\documentclass[australian,english,aps]{revtex4-2}
\usepackage[T1]{fontenc}
\usepackage[latin9]{inputenc}
\usepackage{xcolor}
\usepackage{textcomp}
\usepackage{amsmath}
\usepackage{amssymb}
\usepackage{esint}

\makeatletter
\usepackage{babel}

\makeatother

\usepackage{babel}
\begin{document}
\title{Mesoscopic and macroscopic quantum correlations in photonic, atomic
and optomechanical systems}
\author{Run Yan Teh$^{1}$, Laura Rosales-Z\'arate$^{2}$ , Peter D. Drummond$^{1}$
and M. D. Reid$^{1}$}
\address{$^{1}$Centre for Quantum Science and Technology Theory, Swinburne
University of Technology, Melbourne 3122, Australia}
\address{\selectlanguage{australian}%
$^{2}$Centro de Investigaciones en \'Optica A.C., Le\'on, Guanajuato
37150, M\'exico}
\selectlanguage{english}%
\begin{abstract}
This paper reviews the progress that has been made in our knowledge
of quantum correlations at the mesoscopic and macroscopic level. We
begin by summarizing the Einstein-Podolsky-Rosen (EPR) argument and
the Bell correlations that cannot be explained by local hidden variable
theories. It was originally an open question as to whether (and how)
such quantum correlations could occur on a macroscopic scale, since
this would seem to counter the correspondence principle. The purpose
of this review is to examine how this question has been answered over
the decades since the original papers of EPR and Bell. We first review
work relating to higher spin measurements which revealed that macroscopic
quantum states could exhibit Bell correlations. This covers higher
dimensional, multi-particle and continuous-variable EPR and Bell states
where measurements on a single system give a spectrum of outcomes,
and also multipartite states where measurements are made at multiple
separated sites. It appeared that the macroscopic quantum observations
were for an increasingly limited span of measurement settings and
required a fine resolution of outcomes. Motivated by this, we next
review correlations for macroscopic superposition states, and examine
predictions for the violation of Leggett-Garg inequalities for dynamical
quantum systems. These results reveal Bell correlations for coarse-grained
measurements which need only distinguish between macroscopically distinct
states, thus bringing into question the validity of certain forms
of macroscopic realism. Finally, we review progress for massive
systems, including Bose-Einstein condensates and optomechanical oscillators,
where EPR-type correlations have been observed between massive systems.
Experiments are summarized, which support the predictions of quantum
mechanics in mesoscopic regimes.
\end{abstract}
\maketitle

\section{Introduction}

In recent decades, there has been huge progress made in the manipulation
of quantum systems for the purpose of applications in the field of
quantum information and quantum technologies. A large part of that
progress is a direct result of our knowledge of quantum correlated
systems. In this review, we summarize the status of what has been
learned about macroscopic quantum correlations. Quantum correlations
are defined as those correlations described by quantum mechanics that
cannot also be described by classical theory or classical-like theories.
Here, we restrict the meaning further, to imply those correlations
whose source is a quantum entangled state, with emphasis given to
the type of quantum correlations considered by Einstein, Podolsky
and Rosen \citep{Einstein:1935} and Bell \citep{Bell_Physics1964}.
Our approach is to present the progress as a historical timeline of
discoveries relating to mesoscopic and macroscopic quantum correlations.

Arguably, the study of quantum correlations began with the Einstein-Podolsky-Rosen
(EPR) argument \citep{Einstein:1935}. We begin therefore with a summary
of the EPR paradox, and the theorem given by Bell \citep{Bell_Physics1964,Bell_1966_RMP,Bell_1971_FoundQM,Bell_1976_Epistemol},
who proved how such correlations for two separated spin $1/2$ systems
cannot be explained by any local hidden variable (LHV) theory.  This
apparently resolved the paradox by implying that the premise of local
realism on which the EPR argument was based was fundamentally invalid.
The property of the correlations that violate Bell's LHV theories
is termed Bell nonlocality \citep{Brunner_RMP2014}. A brief summary
is given in Section II.

Originally, it was an open question as to whether such quantum correlations
could manifest on a mesoscopic, or macroscopic, scale. It is the answer
to this question that we analyze in the review. The anticipated answer
may well have been no, in order to ensure compatibility with the correspondence
principle. First, ``macroscopic scale'' could refer to systems of
large size, i.e. to systems possessing a large number of particles.
It was shown by Mermin that quantum mechanics predicts the failure
of local hidden variable theories for two separated systems of higher
spin \citep{Mermin:1990aa}. This translates to the prediction of
failure of local hidden variable theories for separated systems possessing
multiple particles \citep{Drummond:1983}, or to systems of higher
dimensionality.

Another way to increase the system size is to consider three or more
separated particles, or systems. Measurements can then be made locally
on each of many systems. Svetlichny analyzed three systems to show
how to confirm a genuine mesoscopic form of multipartite Bell nonlocality,
that cannot be explained as correlations arising from a bipartite
Bell nonlocality shared among just two systems \citep{Svetlichny_PRD1987}.
Greenberger, Horne and Zeilinger (GHZ) showed the extreme paradox
associated with some types of tripartite quantum states \citep{Greenberger_1990_AJP},
this being later extended to $N$ particles by Mermin \citep{mermin1990extreme},
who demonstrated the possibility of an increasing amount of violation
of a Bell inequality for systems of increasing size. The exact nature
of the increasing size needed to be considered carefully however.
For a genuine multipartite nonlocality, it was revealed that the violations
would be constant with increasing size. The macroscopic quantum observations
appeared to be restricted to an increasingly limited span of measurement
settings and also appeared to require a fine resolution of measurement
outcomes. Later work countered some of these conclusions, for different
systems and measurements. In Sections III, IV and VII, we review results
for Bell nonlocality for higher dimensional systems, multi-particle
systems, and multipartite systems.

One way to achieve macroscopic correlations is to amplify the size
or dimensionality of each system involved in the correlations. A natural
limit is provided where measurements give continuous variable (CV)
outcomes. For the CV systems involving the measurement of quadrature
field amplitudes, the number of photons at each site is amplified
by introducing local oscillator fields that allow assignment of the
correlations in terms of macroscopic Schwinger spin observables.
We thus review quantum correlations in CV systems. It is feasible
to generate EPR correlations and entanglement for fields, but the
generation of Bell-nonlocal states that falsify local hidden variable
theories for continuous-variable measurements is a more difficult
task. This can be done, however, and we give a summary of continuous-variable
quantum correlations in Section V.

Leggett argued that macroscopic quantum mechanics can only be rigorously
tested using states that are superpositions of macroscopically distinct
states, in the spirit of the Schr\"odinger cat paradox \citep{Schrodinger:1935_Naturwiss,Frowis_RMP2018}.
In Sections V and VI, we thus review the quantum correlations associated
with entangled cat states, which includes superpositions of two-mode
number states (NOON states), the GHZ states, and entangled coherent
states. The hybrid state involving a microscopic qubit entangled with
a macroscopic qubit is also analyzed, as the prototype of the original
Schr\"odinger cat state modeling a measurement apparatus \citep{Brune_PRL1996}.
We summarize interpretations, which gives insight into the measurement
problem, in Section VI. The impact of a coarse graining of measurement
outcomes is reviewed in Sections VII and VIII, and for some states
and measurements is found not to be fundamentally limiting.

Leggett and Garg proposed to test macroscopic realism in a setting
where measurements are made at different times, and where measurements
need only distinguish between two macroscopically distinct states
\citep{Leggett:1985}. Motivated by this, we review in Sections VII
and VIII the dynamics creating macroscopic superposition states (cat
states), and the macroscopic correlations in time predicted to violate
the Leggett-Garg inequalities. For suitably adapted systems, Bell
correlations can be predicted using macroscopic measurements that
distinguish only between two macroscopically distinct coherent states,
which act as macroscopic qubits. This allows macroscopic quantum
correlations under coarse-graining of measurement outcomes at a macroscopic
level, although there is sensitivity to decoherence (losses) and to
the imprecision of measurement settings. In Section VIII, we summarize
the progress toward tests of Leggett-Garg's macro-realism. Experiments
realizing these correlations are more limited, but developments are
summarized.

An alternative perspective is that for a system size to be truly macroscopic,
the systems must have large mass. In Sections IX, we review the status
of quantum correlations in atomic and Bose-Einstein condensate (BEC)
systems, where experiments have demonstrated mesoscopic EPR correlations,
both for atoms within a condensate and for separated condensates.
In Section X, we give a summary of the significant progress made in
achieving quantum correlations in optomechanical systems, including
for EPR-type and Bell experiments. Overall, however, there has not
yet been reported a rigorous violation of a Bell inequality where
the hidden variables involved are for spatially separated objects
of significant mass. The final section of this review gives a conclusion.

\section{Bell correlations for small systems}

Bell considered the singlet state of two spin $1/2$ particles, 
\begin{equation}
|\psi_{Bell}\rangle=\frac{1}{\sqrt{2}}\{|\uparrow\rangle_{A}|\downarrow\rangle_{B}-|\downarrow\rangle_{A}|\uparrow\rangle_{B}\}\label{eq:bell-state}
\end{equation}
where $|\uparrow\rangle$ and $|\downarrow\rangle$ are the spin $1/2$
eigenstates for the spin $Z$ component $\hat{S}_{z}$ of the spin
vector $\hat{S}=(\hat{S}_{x},\hat{S}_{y},\hat{S}_{z})$ \citep{Bell_Physics1964}.
The particles are labelled $A$ and $B$ and become spatially separated.
We will use superscripts to denote that the observables apply to the
system $A$ or $B$. For the singlet state, the outcomes of the two
spin components $\hat{S}_{z}^{(A)}$ and $\hat{S}_{z}^{(B)}$ are
anti-correlated. We also introduce the Pauli spin observables $\hat{\sigma}=(\hat{\sigma}_{x},\hat{\sigma}_{y},\hat{\sigma}_{z})$
where $\frac{\hbar}{2}\hat{\sigma}=\hat{S}$, for which the outcomes
of the measurements of the spin components are $\pm1$. The rotated
spin components are 
\begin{eqnarray}
 &  & \hat{\sigma}_{\theta}^{(A)}=\cos\theta\thinspace\hat{\sigma}_{z}^{(A)}+\sin\theta\thinspace\hat{\sigma}_{x}^{(A)},\ \hat{\sigma}_{\phi}^{(B)}=\cos\phi\thinspace\hat{\sigma}_{z}^{(B)}+\sin\phi\thinspace\hat{\sigma}_{x}^{(B)}\ ,\label{eq:2,:rotation}
\end{eqnarray}
which can be measured on $A$ and $B$ respectively, by a Stern-Gerlach
apparatus at each site. A transformation into a different spin basis
reveals the anti-correlation between the outcomes at each site, for
measurements of the spins $\hat{\sigma}_{\theta}^{(A)}$ and $\hat{\sigma}_{\phi}^{(B)}$
where $\theta=\phi$.

Most experiments to date use the photonic version of the Bell states,
the rotations (\ref{eq:2,:rotation}) being realized at each site
by beam splitters with a variable transmission (or beam splitters
with phase shifts), polarizing beam splitters (PBS), or polarizers
(refer \citep{Clauser:1978_ReportPP,Brunner_RMP2014}). Each particle
is a photon, and the Bell state (\ref{eq:bell-state}) is mapped onto
a four-mode state where $|\uparrow\rangle_{A}=|1\rangle_{a+}|0\rangle_{a-}$,
$|\downarrow\rangle_{A}=|0\rangle_{a+}|1\rangle_{a-}$, $|\uparrow\rangle_{B}=|1\rangle_{b+}|0\rangle_{b-}$
and $|\downarrow\rangle_{B}=|0\rangle_{b+}|1\rangle_{b-}$, as originally
described in Ref. \citep{Reid1986}. The modes designated $a_{\pm}$
and $b_{\pm}$ are two modes at the sites $A$ and $B$, respectively.
The $a_{\pm}$ ($b_{\pm}$) may be realized as orthogonally polarized
modes of the photon incident on a polarizer at $A$ ($B$). The horizontally
and vertically polarized states are written as $|\uparrow\rangle_{A}=|1\rangle_{a+}|0\rangle_{a-}\equiv|H\rangle_{A}$
and $|\downarrow\rangle_{A}=|0\rangle_{a+}|1\rangle_{a-}\equiv|V\rangle_{A}$,
and similarly for system $B$. As well as $|\psi_{Bell}\rangle$,
we also consider the positively correlated Bell state $|\psi_{Bell,+}\rangle=\frac{1}{\sqrt{2}}\{|\uparrow\rangle_{A}|\uparrow\rangle_{B}+|\downarrow\rangle_{A}|\downarrow\rangle_{B}\}$.
The Bell state is then written conveniently as 
\begin{equation}
|\psi_{Bell,+}\rangle=\frac{1}{\sqrt{2}}\{|H,H\rangle+|V,V\rangle\}\label{eq:bell+}
\end{equation}
where $|H,H\rangle\equiv|H\rangle_{A}|H\rangle_{B}$ and $|V,V\rangle\equiv|V\rangle_{A}|V\rangle_{B}$.
The photon emerges from the polarizer as for a beam splitter. The
photon is detected to be either in a $+$ position or a $-$ position,
which we identify as the spin result ``up'' and the spin result
``down'', and which indicates the direction of polarization of the
emerging photon. The polarizer (or beam splitter) provides the mode
transformations
\begin{align}
\hat{c}_{+} & =\cos\frac{\theta}{2}\hat{a}_{+}+\sin\frac{\theta}{2}\hat{a}_{-},\ \hat{c}_{-}=\sin\frac{\theta}{2}\hat{a}_{+}-\cos\frac{\theta}{2}\hat{a}_{-}\nonumber \\
\hat{d}_{+} & =\cos\frac{\phi}{2}\hat{b}_{+}+\sin\frac{\phi}{2}\hat{b}_{-},\hat{d}_{-}=\sin\frac{\phi}{2}\hat{b}_{+}-\cos\frac{\phi}{2}\hat{b}_{-}\ .\label{eq:rotc}
\end{align}
Here $\hat{a}_{\pm}$ and $\hat{b}_{\pm}$ are the boson destruction
operators for the modes $a_{\pm}$ and $b_{\pm}$, respectively. With
only one photon incident on the polarizer $A$ ($B$), the spin observable
$\hat{\sigma}_{\theta}^{(A)}$ ($\hat{\sigma}_{\phi}^{(B)}$) is given
by the mode number difference $\hat{c}_{+}^{\dagger}\hat{c}_{+}-\hat{c}_{-}^{\dagger}\hat{c}_{-}$
($\hat{d}_{+}^{\dagger}\hat{d}_{+}-\hat{d}_{-}^{\dagger}\hat{d}_{-}$).

Calculation reveals that for systems $A$ and $B$ prepared in the
Bell state (\ref{eq:bell-state}), the expectation value for the Pauli
spin product is
\begin{equation}
E(\theta,\phi)\equiv\langle\sigma_{\theta}^{A}\sigma_{\phi}^{B}\rangle=-\cos(\theta-\phi)\ .\label{eq:exp-prod}
\end{equation}
A similar result is obtained for the photonic example, where for $|\psi_{Bell,+}\rangle$
one obtains $E(\theta,\phi)=\langle(\hat{c}_{+}^{\dagger}\hat{c}_{+}-\hat{c}_{-}^{\dagger}\hat{c}_{-})(\hat{d}_{+}^{\dagger}\hat{d}_{+}-\hat{d}_{-}^{\dagger}\hat{d}_{-})\rangle=\cos(\theta-\phi)$.\textcolor{red}{}

Bell's first proof assumed a deterministic local realistic theory,
where there are definite values $\lambda_{\theta}^{(A)}$ and $\lambda_{\phi}^{(B)}$
for the spin components $\sigma_{\theta}^{(A)}$ and $\sigma_{\phi}^{(B)}$
of both the particles \citep{Bell_Physics1964,Bell_1966_RMP}.\textcolor{red}{{}
}These values predetermine the result of the spin component if measured.
Here, $\lambda_{\theta}^{(A)}=\pm1$, $\lambda_{\phi}^{(B)}=\pm1$,
though the anti-correlation of the Bell state (\ref{eq:bell-state})
would imply $\lambda_{\theta}^{(A)}=-\lambda_{\phi}^{(B)}$. Bell's
proof was later generalized to cover more general local realistic
theories where there may be local stochastic interactions due to,
for example, the measurement apparatus \citep{Bell_1971_FoundQM,Bell_1976_Epistemol,Clauser:1978_ReportPP,Bell_2001_JohnBell,Bell_2004,Wiseman_2014}.
Bell's more general work accounted for all theories that are local
realistic and locally causal. Bell \citep{Bell_1971_FoundQM} and
Clauser, Horne, Shimony and Holt (CHSH) \citep{Clauser:1969_PRL23}
considered the predictions of all local hidden variable theories (LHV)
for which the joint spin product satisfies 
\begin{equation}
E(\theta,\phi)=\int\rho(\lambda)\ E^{(A)}(\theta,\lambda)E^{(B)}(\phi,\lambda)\ d\lambda\ .\label{eq:lhv}
\end{equation}
This LHV constraint may also be expressed in terms of the measurable
joint probabilities $P_{++}(\theta,\phi)$ as 
\begin{equation}
P_{++}(\theta,\phi)=\int\rho(\lambda)\ P_{+}^{(A)}(\theta,\lambda)P_{+}^{(B)}(\phi,\lambda)\ d\lambda\ .\label{eq:lhv-2}
\end{equation}
Here $P_{++}(\theta,\phi)$ is the joint probability of the outcome
$+$ at both sites, with measurement settings $\theta$ and $\phi$
respectively. The $\rho(\lambda)$ is a distribution over a set of
hidden variables $\{\lambda\}$ and $E^{(A)}(\theta,\lambda)$ is
the expectation value of the measurement $\sigma_{\theta}^{(A)}$
at site $A$, given the measurement setting $\theta$ at that site,
for the hidden variable state $\{\lambda\}$. Similarly, $P_{+}^{(A)}(\theta,\lambda)$
is the probability for outcome $+$ at site $A$, given $\{\lambda\}$.
The $E^{(A)}(\theta,\lambda)$ and $P_{+}^{(A)}(\theta,\lambda)$
are dependent on the parameters $\lambda$ that describe the state
to be measured at $A$, and are dependent on the measurement setting
$\theta$, but are assumed independent of the setting $\phi$ chosen
at the space-like separated site $B$. This is justified based on
causality, and assuming independence of choices of the settings at
each site (locality). The $E^{(B)}(\phi,\lambda)$ and $P_{+}^{(B)}(\phi,\lambda)$
are defined similarly. The restriction (\ref{eq:lhv}) leads to a
bound on the correlations given by the Bell-Clauser-Horne-Shimony-Holt
(CHSH) inequality \citep{Clauser:1969_PRL23} $|B|\leq2$ where 
\begin{equation}
B\equiv E(\theta,\phi)-E(\theta,\phi')+E(\theta',\phi)+E(\theta',\phi')\ .\label{eq:bell-ineq}
\end{equation}
This is readily seen, by noting the algebraic constraint $|B|\leq2$
on $B$ for any deterministic local hidden variable theory, where
one assumes predetermined spin variables $A_{1}$, $A_{2}$, $B_{1}$
and $B_{2}$ for measurements $\sigma_{\theta}^{(A)}$, $\sigma_{\theta'}^{(A)}$,
$\sigma_{\phi}^{(B)},$and $\sigma_{\phi'}^{(B)}$, the values of
the variables being either $+1$ or $-1$.\textcolor{red}{{} }Since
a probability distribution compatible with the general local hidden
variable theory (\ref{eq:lhv}) is convex, the same bound will apply
to the more general local hidden variable theory. A convex set is
fully determined by the extremal points, as any other points in the
convex set can be expressed as convex combinations of these extremal
points. Proofs are given in Scarani \citep{scarani2012device} and
Brunner et al \citep{Brunner_RMP2014}.

For the choices of angle \textcolor{black}{$\theta=0$, $\phi=\pi/4$,
$\theta'=\pi/2$, $\phi'=3\pi/4$}, the quantum prediction of (\ref{eq:exp-prod})
for $|\psi_{Bell}\rangle$ violates the bound giving $B=-2\sqrt{2}$.
Similarly, for state $|\psi_{Bell,+}\rangle$ the value is $B=2\sqrt{2}$.
In this way, rather dramatically, it is shown that the predictions
of quantum mechanics cannot be compatible with any local realistic
(or local causal) theory that embodies the very simple and reasonable
premises, (\ref{eq:lhv}) and (\ref{eq:lhv-2}).

Bell's original paper addressed the EPR paradox, of 1935 \citep{Einstein:1935}.
The assumption (\ref{eq:lhv}) reduces to that of EPR's local realism,
when $\theta=\phi$. For any $\theta=\phi$, we note the anti-correlation
between the spin outcomes at each site is maximum, as for the Bell
state (\ref{eq:bell-state}). EPR argued in their paper that if one
can predict with certainty the result of a measurement on a system
without disturbing that system, then the result of the measurement
was predetermined and describable by a hidden variable. In Bohm's
version of the argument \citep{bohm2013quantum}, the fact that one
can infer the outcome for spin $\hat{\sigma}_{\theta}^{(A)}$ of system
$A$ by measuring the spin $\hat{\sigma}_{\theta}^{(B)}$ of the space-like
separated system $B$ implies the condition of EPR, since this measurement
is justified to be noninvasive to system $A$. The EPR argument then
implies that for any $\theta$, there exists a hidden variable $\lambda_{\theta}^{(A)}$
to predetermine the value of the measurement of $\hat{\sigma}_{\theta}^{(A)}$.
Since this description is not consistent with any local quantum state
for system $A$, EPR would conclude quantum mechanics to be incomplete.
Bell's paper thus showed that any hidden variable theory consistent
with the assumption of local realism could not be compatible with
the predictions of quantum mechanics.

Early evidence of a Bell state was given by Bleuler and Brandt \citep{Bleuler_PhysRev1948}
and Hanna \citep{HANNA:1948aa} using Geiger counters, and by Wu and
Shaknov \citep{Wu_PR1950} who used scintillation counters that have
higher detection efficiency. These authors measured the correlation
between two gamma photons generated by positron-electron annihilation.
The annihilation produces the two photons in a state $|\psi\rangle=\frac{1}{\sqrt{2}}\left(|HV\rangle-|VH\rangle\right)$.
The coincidence measurements with polarizer angle settings $\theta$
and $\phi$ at $0$ and $\pi/2$ gave an enhanced counting rate, in
agreement with the theory proposed by Wheeler \citep{Wheeler:1946aa}
and Pryce and Ward \citep{PRYCE:1947aa}. This suggests correlations
along the lines of an EPR paradox. An EPR paradox occurs when there
is a maximum correlation between the elements of two pairs of non-commuting
observables, such as \{$\hat{\sigma}_{z}^{(A)},\hat{\sigma}_{z}^{(B)}$\}
and \{$\hat{\sigma}_{x}^{(A)},\hat{\sigma}_{x}^{(B)}$\}. While the
original EPR argument considered position and momentum \citep{Einstein:1935},
Bohm's version considered non-commuting spin observables \citep{bohm2013quantum}.
In the ideal case, this correlation leads to a contradiction between
the assumption of local realism, and the completeness of quantum mechanics.
Bohm and Aharonov \citep{Bohm_PR1957}\textcolor{red}{} later explained
how the results of the Wu-Shaknov experiment were consistent with
the Bell state $|\psi\rangle=\frac{1}{\sqrt{2}}\left(|\uparrow\downarrow\rangle-|\downarrow\uparrow\rangle\right)$,
and not for a non-separable classical mixture of the states $|\uparrow\rangle|\downarrow\rangle$
and $|\downarrow\rangle|\uparrow\rangle$. The Wu-Shaknov experiment
thus gave early evidence of quantum correlation, in the form of Bohm's
EPR paradox and entanglement.

An early test of Bell's theorem was performed by Freedman and Clauser
using the correlation in polarization between photons emitted as a
pair in an atomic cascade \citep{Clauser:1972_PRL28}, following the
proposal of Clauser, Horne, Shimony and Holt \citep{Clauser:1969_PRL23}.
This was followed by increasingly rigorous experimental tests including
those by Aspect, Dalibard and Roger \citep{Aspect:1982_PRL49} and
Aspect, Grangier and Roger \citep{Aspect:1981_PRL47,Aspect_1982_PRL49},
which supported the quantum predictions. Later, it was suggested by
Reid and Walls \citep{Reid1986} and Shih and Alley \citep{Shih_1988_PRL}
to use the correlated photon pairs generated in parametric down conversion,
modeled by the two-mode Hamiltonian $H=i\kappa E(\hat{a}^{\dagger}\hat{b}^{\dagger}-\hat{a}\hat{b})$.
The first such test was performed by Ou and Mandel, using the two-mode
version and the equivalent of beam splitters to generate four modes
\citep{Ou_Mandel1988_PRL}. The four mode variants based on the interaction
Hamiltonian 
\begin{equation}
H=i\kappa E(\hat{a}_{+}^{\dagger}\hat{b}_{+}^{\dagger}+\hat{a}_{-}^{\dagger}\hat{b}_{-}^{\dagger}-\hat{a}_{+}\hat{b}_{+}-\hat{a}_{-}\hat{b}_{-})\label{eq:four-mode-ham}
\end{equation}
were outlined in Reid and Walls \citep{Reid1986} and Horne, Shimony
and Zeilinger \citep{Horne_PRL1989}, with proposals and an experimental
violation given by Rarity and Tapster \citep{RarityTapster1990_PRA,RarityTapster1990_PRL}.
A set of experiments based on parametric down conversion \citep{Kwiat_PRL_1995,Tittel_1998_PRL,Weihs_1998_PRL,Tapster_PRL_1994}
further confirmed quantum predictions. Loopholes due to poor detection
efficiencies and lack of spatial separation (needed to justify locality)
have since been overcome \citep{Giustina2013_Nature,Giustina_2015_PRL,Christensen_2013_PRL,Shalm_2015_PRL,Hensen_2015_Nature,Hensen2016_SciRep,Rosenfeld_PRL_2017,li2018test},
with more recent work focusing on multiple sources of loopholes \citep{Abellan2015_PRL,BigBellExp}.
The experiments support quantum mechanics, giving a violation of Bell's
inequality. Bell correlations have found significant applications
in quantum information, including for device-independent entanglement
detection \citep{mayers1998quantum,barrett2005no,acin2006bell} and
the certification of random numbers \citep{Pironio:2010aa}.

\section{Bell correlations for higher spin, multi-particle and higher dimensional
systems}

Originally, it may have been speculated that the violation of Bell
inequalities could only be possible for microscopic systems. The correspondence
principle suggests Bell inequalities to be valid, for large systems.
The extent to which this is true is an ongoing investigation, and
is the topic of this review.

\subsection{Mermin's result for higher spin}

The work of Mermin gave an advance in this direction, by showing that
Bell's theorem applies to higher spin systems \citep{Mermin_PRD1980}.
Mermin considered the higher spin state that is the generalization
of the singlet Bell state (\ref{eq:bell-state}), 
\begin{equation}
|\psi\rangle=\frac{1}{\left(2j+1\right)^{1/2}}\sum_{m=-j}^{j}\left(-1\right)^{j-m}|m,-m\rangle_{\hat{n},\hat{n}}\,\label{eq:Mermin_spin_state}
\end{equation}
where the state has zero total spin. Here, $|m_{1},m_{2}\rangle_{\hat{n}_{1},\hat{n}_{2}}$
denotes the state for two spin-$j$ particles labelled $i=1,2$, which
have a spin $m_{i}$ along the axis $\hat{n}_{i}$. Defining $m_{i}\left(a\right)$
to be the spin observable of particle $i$ measured along the direction
$a$, Mermin started out with the inequality $\left|m_{1}\left(a\right)+m_{1}\left(b\right)\right|\geq-\left(m_{1}\left(a\right)+m_{1}\left(b\right)\right)$
and the relation $m_{1}\left(c\right)=-m_{2}\left(c\right)$, to derive
an inequality of the form 
\begin{align}
j\langle\left|m_{1}\left(a\right)-m_{2}\left(b\right)\right|\rangle & \geq\langle m_{1}\left(a\right)m_{2}\left(c\right)\rangle+\langle m_{1}\left(b\right)m_{2}\left(c\right)\rangle\,.\label{eq:Mermin_inequality1}
\end{align}
Inequality (\ref{eq:Mermin_inequality1}) is satisfied if deterministic
local realism is valid, where definite values for any component of
either of the two correlated spins always exist. This assumption justifies
the relation $m_{1}\left(c\right)=-m_{2}\left(c\right)$ for the correlations
of (\ref{eq:Mermin_spin_state}). Thus, a violation of the inequality
implies the contradiction between local realism and quantum theory.

Mermin showed that violations of (\ref{eq:Mermin_inequality1}) are
possible for the state Eq. (\ref{eq:Mermin_spin_state}). The averages
in the inequality Eq. (\ref{eq:Mermin_inequality1}) were calculated
using Eq. (\ref{eq:Mermin_spin_state}), which leads to the inequality
\begin{align}
\frac{1}{2j+1}\sum_{m,m^{'}}\left|m-m^{'}\right|\left|\langle m|e^{-2i\theta S_{y}}|m^{'}\rangle\right| & \geq\frac{2}{3}\left(j+1\right)\sin\theta\,.\label{eq:Mermin_inequality2-1}
\end{align}
Here, $\pi-2\theta$ is the angle between the spin directions $a$
and $b$, while the angle between $a$ and $c$ is identical to the
angle between $b$ and $c$, and is given by $\pi/2+\theta$. Mermin
found that there exists a range for $\theta$, given by the condition
$0<\sin\theta<1/2j$, such that inequality (\ref{eq:Mermin_inequality2-1})
will always be violated for a spin $j$. This gives the surprising
result that there is a violation of local realism for large spin $j$.
However, it was noted that for spin $1/2$, the inequality (\ref{eq:Mermin_inequality2-1})
is violated for any $\theta$. By contrast, in the limit of $j\rightarrow\infty$,
the range for $\theta$ to show violation is of order $1/j$. This
gives the possibility of local realism as a description of the quantum
mechanical state in the classical limit. Since the control of the
angle $\theta$ is limited by experimental precision, consistency
is given with the correspondence principle.

The higher spin system admits the $2j+1$ outcomes $-j$, ..$j$ at
each site, for systems $A$ and $B$, and thus is an example of a
\emph{higher dimensional} system of dimension $d=2j+1$. The higher
dimensionality can be achieved in different ways, with different mappings
onto physical systems.

\subsection{Multi-particle violations}

The significance of the higher dimensions for macroscopic quantum
mechanics is given by the work of Drummond, who considered the four-mode
bosonic state \citep{Drummond_1983_PRL} 
\begin{align}
|\Psi_{N}\rangle & =\frac{1}{N!\sqrt{N+1}}\left(\hat{a}_{+}^{\dagger}\hat{b}_{+}^{\dagger}+\hat{a}_{-}^{\dagger}\hat{b}_{-}^{\dagger}\right)^{N}|0000\rangle\,\label{eq:N-boson}
\end{align}
with $N$ bosons at each site $A$ and $B$, modeled after $|\psi_{Bell,+}\rangle$
for $N=1$. This state maps onto the higher spin Bell state (\ref{eq:Mermin_spin_state})
considered by Mermin, thereby highlighting how the higher spin results
lead to mesoscopic Bell violations involving multiple particles at
each site. Here, $|0000\rangle$ symbolizes the vacuum state of all
four modes. A polarizer (or polarizing beam splitter) is placed at
each site, aligned at an angle $\theta$ or $\phi$, the mode transformations
being given as (\ref{eq:rotc}). An incoming particle can be detected
at either the $+$ or $-$ mode at the output of the polarizer, similar
to the up or down state of a spin $1/2$ particle. At each site,
$N$ indistinguishable bosons are incident on the polarizer, and the
possible outcomes are $N$, $N-1$, ..., $1$, $0$ bosons detected
at the $+$ output mode. The configuration therefore maps to a two
systems of higher spin, with dimension $d=N+1$. This work confirmed
violations of Bell inequalities for the $N$-particle state (\ref{eq:N-boson})
for arbitrarily large $N$.

The violations given in \citep{Drummond_1983_PRL} were formulated
in terms of the Clauser-Horne Bell inequalities \citep{Clauser_PRD_1974}.
Here, the outcome at $A$ ($B$) is assigned the value $+1$ if \emph{all}
$N$ bosons are detected in the $+$ mode i.e. in the mode $c_{+}$
(or $d_{+}$), and otherwise are assigned the value $0$. The Clauser-Horne
(CH) inequality can be expressed as $-1\leq CH\leq0$, where 
\begin{align}
CH & \equiv P_{++}(\theta,\phi)+P_{++}(\theta',\phi')+P_{++}(\theta,\phi')-P_{++}(\theta,\phi')-P_{+}^{(A)}(\theta')+P_{+}^{(B)}(\phi)\ .\label{eq:CH}
\end{align}
$P_{++}(\theta,\phi)$ is the joint probability for detecting $+1$
at both sites with polarizer angle settings $\theta$ and $\phi$,
and $P_{+}^{(A)}(\theta)$ ($P_{+}^{(B)}(\phi)$) are the marginal
probabilities for detecting $+1$ at the site $A$ ($B$) only. Where
all bosons can be detected, the marginal probabilities become equivalent
to the one-sided joint probabilities $P_{+}^{(A)}(\theta,-)$ ($P_{+}^{(B)}(-,\phi)$)
for detecting $+1$ at site $A$ ($B$)\emph{ and} a total of $N$
bosons at site $B$ ($A$). The fair sampling assumption (also called
the no-enhancement axiom) justifies that the marginal probabilities
can be measured as the one-sided joint probabilities, in situations
of limited detection efficiency. In the original paper \citep{Drummond_1983_PRL},
the detection probabilities were expressed as proportional to the
higher-order normally-ordered moments 
\begin{align}
P_{++}(\theta,\phi) & \propto\langle\psi|(\left(\hat{c}_{+}^{\dagger}\right)^{N}\left(\hat{d}_{+}^{\dagger}\right)^{N}\hat{c}_{+}^{N}\hat{d}_{+}^{N}|\psi\rangle\,.\label{eq:prob-prop}
\end{align}
For the state (\ref{eq:N-boson}), quantum mechanics predicts that
$P_{+}^{(A)}(\theta)$ and $P_{+}^{(A)}(\theta,-)$ are both independent
of angle choice $\theta$ (and similarly that $P_{+}^{(B)}(\phi)$
and $P_{+}^{(B)}(-,\phi)$ are independent of $\phi$). The fair sampling
assumption was used (along with the symmetry of the marginal probabilities)
to justify an expression for the left-side of (\ref{eq:CH}) in terms
of the measurable ratio of joint to one-sided probabilities: $g(\theta,\phi)=P_{++}(\theta,\phi)/(P_{+}^{(A)}(\theta')+P_{+}^{(B)}(\phi))\equiv P_{++}(\theta,\phi)/P_{+}^{(A)}(\theta,-)$.
The quantum prediction 
\begin{equation}
g(\theta,\phi)=\cos^{2N}(\theta-\phi)\label{eq:ratio-g}
\end{equation}
for the $2N$-boson state (\ref{eq:N-boson}) gives a violation of
the CH inequality (\ref{eq:CH}) for suitable angle choices, for any
$N$. The selected angles are expressed as \textcolor{black}{$\theta=0$,
$\phi=\varphi$, $\theta'=2\varphi$, $\phi'=3\varphi$}, and the
value of $\varphi$ optimized. As for the higher spin case, the range
of angle $\theta$ for which violation is possible reduces with increasing
$N$. The use of the Clauser-Horne approach ensured that the result
was not restricted to the assumption of local deterministic theories.
Rather, all local hidden variable theories were ruled out, for all
$N$. Drummond also considered the predictions where $J<N$ bosons
are detected at the $+$ mode, at each site \citep{Drummond_1983_PRL}.
The violations decreased with decreasing $J$. The reduced values
of $J$ corresponded to the loss of information about the particle
outcomes and the effect is similar to that expected for decoherence
and detection efficiency losses. Overall, the work of \citep{Drummond_1983_PRL}
was consistent with Mermin's prediction that the range of parameter
space for which violations can be observed is reduced with increasing
$N$.

The generation of the multi-particle bosonic Bell states (\ref{eq:N-boson})
can be achieved from four-mode parametric down conversion (\ref{eq:four-mode-ham}),
as was explained in \citep{Reid_2002_PRA}. The solution for the
parametric Hamiltonian is $|\psi\rangle=\sum_{N=0}^{\infty}c_{N}|\Psi_{N}\rangle\,,$
where $|\Psi_{N}\rangle$ is the quantum state in Eq. (\ref{eq:N-boson}),
$c_{N}=\sqrt{N+1}\tanh^{N}r/\cosh^{2}r$, and $r=\kappa t$. This
implies the state $|\Psi_{N}\rangle$ can be generated, conditioned
on the detection of a total of $N$ photons, at each site.

An experimental realization of a higher spin Bell experiment was given
by Howell et al \citep{Howell_PRL2002}. They used the method of Lamas-Linares
et al \citep{lamas2001stimulated} to generate a four-mode entangled
state from parametric down conversion. \textcolor{red}{}Here the
two photon polarization entangled modes are considered as spin$-1$
particles. In order to generate these spin$-1$ particles, polarization-entangled
four photon states are created from a pulsed type-II parametric down-conversion
process. The detection is performed using a post-selection measurement,
in order to consider the second order term $|\Psi_{2}\rangle$ of
the down-converted field which is given by: 
\begin{equation}
\frac{1}{\sqrt{3}}\left(\vert2H,2V\rangle-\vert HV,VH\rangle+\vert2V,2H\rangle\right).\label{eq:high-spin-howell}
\end{equation}
Here $\vert2V,2H\rangle$ means that if Alice measures two vertical
photons, Bob measures two horizontal photons. The possible outcomes
for Alice (Bob) are three: $\vert2H\rangle$, $\vert HV\rangle$ ($\vert VH\rangle$)
and $\vert2V\rangle$, which are denoted as $\vert1\rangle$, $\vert0\rangle$
and $\vert-1\rangle$ respectively, and all of them have the same
probability. The assignment of values is the following: $+1$ for
both measurements results $\vert1\rangle$ and $\vert-1\rangle$,
while $-1$ for a $\vert0\rangle$ result. Using this assignment,
it is possible to measure the probabilities for the CHSH Bell-type
inequality given in (\ref{eq:bell-ineq}). A violation of the CHSH
Bell-type inequality of $2.27\pm0.02$ is obtained, for polarizer
analyzer settings corresponding to $\theta=-16^{\circ},$$\phi^{\prime}=14^{\circ},$$\theta^{\prime}=4^{\circ}$
and $\phi=6^{\circ}.$\textcolor{teal}{}

\subsection{Higher dimensional Bell inequalities}

The higher spin Bell inequalities were the first examples of Bell
inequalities for higher dimensions. The violation of local realism
for two higher dimensional (higher spin) systems $A$ and $B$ has
been studied extensively since Mermin's original paper \citep{mermin1982joint,ogren1983some,garg1982bell,garg1983local,ardehali1991hidden,peres1992finite,gisin1992maximal,durt2001violations,Gisin_1992_PhysLettA,Peres:1999aa,Kaszlikowski_2000_PRL,Collins_PRL2002,Fu_PRL_2004,Lee_PRA_2009,Weiss_NJP_2016,Dalton:2021aa}.
It was pointed out that the violation of local realism could be obtained
for a broader range of settings $\theta$ if different Bell inequalities
were used \citep{ogren1983some,mermin1982joint}. Garg and Mermin
further considered higher spin systems in 1984 and used a geometrical
method to derive Bell inequalities for $j=1$, $3/2$ and $5/2$ \citep{Garg:1984aa}.
Gisin and Peres discovered perhaps surprisingly that for a pair of
spin-$j$ particles in a singlet state, the Bell violation can be
as large as for a pair of spin-$1/2$ particles \citep{Gisin_1992_PhysLettA}.
\textcolor{red}{}Kaszlikowski et al \citep{Kaszlikowski_2000_PRL}
numerically investigated violations of local realism for qudits using
bipartite $d-$dimensional mixed states, suggesting that in fact the
maximal violation increases monotonically with $d$. They argued that
the limitation for violations was the restricted use of measurements.

In 2002, Collins, Gisin, Linden, Massar and Popescu (CGLMP) \citep{Collins_PRL2002}
constructed a family of Bell inequalities for bipartite systems of
arbitrary dimension $d$, which for certain quantum states gave a
violation as $d\rightarrow\infty$. They first considered two parties
with two possible measurements $A_{1},A_{2},B_{1},B_{2}$ each. Each
measurement in turn has $d$ possible outcomes: $0,...,d-1$. Starting
from a combination of these measurements, and noting that $\left(B_{1}-A_{1}\right)+\left(A_{2}-B_{1}\right)+\left(B_{2}-A_{2}\right)+\left(A_{1}-B_{2}\right)=0$,
they write down a Bell expression 
\begin{eqnarray*}
I & \equiv & P\left(A_{1}=B_{1}\right)+P\left(B_{1}=A_{2}+1\right)+P\left(A_{2}=B_{2}\right)+P\left(B_{2}=A_{1}\right)\,
\end{eqnarray*}
where $P\left(A_{i}=B_{j}+k\right)\equiv\sum_{m=0}^{d-1}P\left(A_{i}=m,B_{j}=m+k\text{mod}d\right)$.
Since only $3$ of the probability distributions are allowed due to
the constraint, $I$ that satisfies local realism obeys the inequality
$I\leq3$. The violation of this inequality implies nonlocal correlations.
Collins et al then considered generalized expressions for $I$, deriving
a family of inequalities $I_{d}\leq2$ which are satisfied by all
local realistic theories, yet for nonlocal theories can obtain higher
values. The restriction is not to deterministic local hidden variable
theories. In fact, the task of deriving Bell inequalities has been
turned into a geometric one \citep{Froissart:1981aa,Garg:1984aa,Peres:1999aa,Pitowsky:1989aa,Werner_PRA2001}.
In this approach, deterministic probability distributions form the
extremal points in the convex set of probability distributions that
are compatible with local variable model. Bell inequalities are the
(hyper)-planes that bound the convex set and probability distributions
that violate these inequalities are points that lie outside of these
convex sets. To derive CGLMP inequalities (following Acin et al. \citep{Acin_PRL2005}),
one may start with the deterministic model, for which $\left[\left(B_{1}-A_{1}\right)+\left(A_{2}-B_{1}\right)+\left(B_{2}-A_{2}\right)+\left(A_{1}-B_{2}\right)-1\right]=d-1$
where $\left[x\right]_{d}=x\text{mod}d$. From the inequality $\left[x\right]_{d}+\left[y\right]_{d}\geq\left[x+y\right]_{d}$,
one finds $\left[B_{1}-A_{1}\right]_{d}+\left[A_{2}-B_{1}\right]_{d}+\left[B_{2}-A_{2}\right]_{d}+\left[A_{1}-B_{2}\right]_{d}\geq\left[\left(B_{1}-A_{1}\right)+\left(A_{2}-B_{1}\right)+\left(B_{2}-A_{2}\right)+\left(A_{1}-B_{2}\right)-1\right]_{d}$.
Equality holds in the case of deterministic distributions and nonlocal
correlation is demonstrated if $\left[B_{1}-A_{1}\right]_{d}+\left[A_{2}-B_{1}\right]_{d}+\left[B_{2}-A_{2}\right]_{d}+\left[A_{1}-B_{2}\right]_{d}>\left[\left(B_{1}-A_{1}\right)+\left(A_{2}-B_{1}\right)+\left(B_{2}-A_{2}\right)+\left(A_{1}-B_{2}\right)-1\right]_{d}=d-1$.
This leads to 
\begin{align*}
 & \langle\left[B_{1}-A_{1}\right]_{d}\rangle+\langle\left[A_{2}-B_{1}\right]_{d}\rangle+\langle\left[B_{2}-A_{2}\right]_{d}\rangle+\langle\left[A_{1}-B_{2}\right]_{d}\rangle>d-1
\end{align*}
where $\langle\left[x\right]_{d}\rangle=\sum_{k=0}^{d-1}kP\left(\left[x\right]_{d}=k\right)$.
It was shown that the maximally entangled states $|\psi\rangle=(1/\sqrt{d})\sum_{k=0}^{d-1}|k,k\rangle$
violate Bell CGLMP inequalities \citep{durt2001violations,Collins_PRL2002}.
Here, $k=0,..,d-1$ are the possible outcomes for the measurement
made on the subsystems labelled $i=1,2$; $|k_{1},k_{2}\rangle$ denotes
a state with outcome $k_{i}$ at the site $i$. The maximally entangled
state has outcomes that are fully correlated. The CGLMP inequalities
have been violated experimentally up to $d=12$ for photons entangled
in orbital angular momentum \citep{Dada2011_NatPhys} and up to $d=16$
for polarization-entangled photon pairs \citep{Lo_2016_ScientRep}.

Further work by Fu derived a CHSH inequality for bipartite systems
of arbitrary dimension $d$ \citep{Fu_PRL_2004}. \textcolor{black}{The
natural and reasonable guess that a maximally entangled state should
maximally violate CGLMP inequalities was shown not to be true b}y
Acin et al. \citep{Acin2002Quantum} for two qutrit and also two $d$-dimensional
systems of up to $d=8$. The result was generalized by Chen et al
\citep{Chen2006_PRA}, using the inequalities of Fu. Lee and Jaksch
later derived optimal Bell inequalities \citep{Lee_PRA_2009}, defined
as Bell inequalities that are tight and maximally violated by maximally
entangled states. The bipartite settings are identical to those of
Fu (2004), where each party has two possible measurements and each
measurement has $d$ possible outcomes.\textcolor{red}{}\textcolor{teal}{}

\section{Quantum correlations of multipartite systems}

The original set-up of Bell was bipartite, meaning two separated systems
where local measurements are made on each system by independent observers,
or parties. A different question was addressed by Svetlichny, who
analyzed whether Bell nonlocality can genuinely exist between three
or more separated systems \citep{Svetlichny_PRD1987}. Svetlichny
showed that for certain quantum states, the tripartite correlations
cannot be explained as arising from classical mixtures of two-party
nonlocal states, i.e. from bipartite states displaying failure of
bipartite LHV models (\ref{eq:lhv}) or (\ref{eq:lhv-2}). Such states
are called genuinely tripartite Bell nonlocal. In fact, quantum mechanics
predicts that Bell nonlocality can exist genuinely shared over an
arbitrarily large number $N$ of separated sites. Such states are
called $N$-partite nonlocal, and are said to exhibit genuine $N$-partite
Bell nonlocality.

\subsection{Multipartite Bell nonlocality with qubits}

It might have been expected that the level of violation of local realism
would be less for the three-party than for the two-party case. It
was shown that, in some sense, the reverse is true \citep{Greenberger1989_Bell,Greenberger_1990_AJP,Clifton_1991_FoundP}.
The Greenberger-Horne-Zeilinger (GHZ) state generalized to $N$ parties
is a Schr\"odinger-cat extreme superposition state of the form \citep{mermin1990extreme}
\begin{equation}
|\psi\rangle=\frac{1}{\sqrt{2}}(|\uparrow\rangle^{\otimes N}-|\downarrow\rangle^{\otimes N})\ .\label{eq:ghz}
\end{equation}
Here there are $N$ spatially separated spin-$1/2$ systems, labelled
by $k=1,2..,N$. We use the notation $|\uparrow\rangle^{\otimes N}=\prod_{k=1}^{N}|\uparrow\rangle_{k}=|\uparrow\uparrow..\uparrow\rangle$
where $|\uparrow\rangle_{k}$ is the eigenstate for the Pauli spin
$\sigma_{z}^{(k)}$ of the $k$-th system. For $N=3$, this is the
GHZ state examined in \citep{Mermin:1990aa}, based on the work of
\citep{Greenberger1989_Bell,Greenberger_1990_AJP,Clifton_1991_FoundP}.
According to quantum mechanics, the measurement of $\sigma_{x}^{(1)}\sigma_{y}^{(2)}\sigma_{y}^{(3)}$
(and the permutations with respect to $k$) always gives $1$, whereas
$\sigma_{x}^{(1)}\sigma_{x}^{(2)}\sigma_{x}^{(3)}$ always gives $-1$.\textcolor{red}{{}
}By contrast, an LHV theory would always predict $1$ for $\sigma_{x}^{(1)}\sigma_{x}^{(2)}\sigma_{x}^{(3)}$,
\emph{if} it had been measured that $\sigma_{x}^{(1)}\sigma_{y}^{(2)}\sigma_{y}^{(3)}$
is always $1$. The difference between the quantum and LHV predictions
for $\sigma_{x}^{(1)}\sigma_{x}^{(2)}\sigma_{x}^{(3)}$ can thus be
detected in one run of an experiment, once the values of $\sigma_{x}^{(1)}\sigma_{y}^{(2)}\sigma_{y}^{(3)}$
are established experimentally. This is an extreme ``all or nothing''
violation of EPR's local realism. The same logic applies to the state
(\ref{eq:ghz}) with an arbitrarily large (odd) value of $N$. Such
a state also exhibits an all or nothing violation of local realism.

It was subsequently shown by Mermin that the GHZ state exhibits a
violation of a Bell inequality by an amount that exponentially increases
with $N$ \citep{mermin1990extreme}. Mermin considered the state
$|\psi\rangle=\frac{1}{\sqrt{2}}\left(|\uparrow\uparrow...\uparrow\rangle+i|\downarrow\downarrow...\downarrow\rangle\right)$
and measurements of the spin of each particle either along the $x$
or $y$ axis. The product of all permutations of these measurements
constitutes the operator $\hat{A}=\prod_{k=1}^{N}\hat{A}_{k}^{s_{k}}$
where $s_{k}\in\{+,-\}$ and 
\begin{equation}
\hat{A}_{k}^{\pm}=\left(\hat{\sigma}_{x}^{(k)}\pm i\hat{\sigma}_{y}^{(k)}\right)e^{-i\theta_{k}}\ .\label{eq:mermin-spin}
\end{equation}
Here, $\theta_{k}$ is a phase shift chosen independently at each
site that allows for a rotation of the spin axes of the measurement
on system $k$. This phase shift was selected zero in Mermin's paper.
One may also consider a complex function $F_{k}^{\pm}=X_{k}\pm iY_{k}$
where $X_{k}$, $Y_{k}$ are real, representing in a local hidden
variable theory (LHV) the outcomes of the spin measurements. The average
of $\hat{A}$ is a complex number with real and imaginary parts, given
as $A=\left\langle \hat{A}\right\rangle =\mathrm{Re}A+i\mathrm{Im}A$
where the LHV prediction is of the form $\left\langle \hat{A}\right\rangle \equiv\langle\prod_{j=1}^{M}F_{j}^{s_{j}}\rangle$.
It is found that if LHV theories are valid, then the expectation value
for the imaginary part of $A$ satisfies the Bell-Mermin inequality
\begin{align}
\langle\text{Im}A\rangle_{local} & \leq2^{N/2}\,,N\,\text{even}\nonumber \\
\langle\text{Im}A\rangle_{local} & \leq2^{\left(N-1\right)/2}\,,N\,\text{odd}\,,\label{eq:im}
\end{align}
while the quantum prediction is $\langle\text{Im}A\rangle=2^{\left(N-1\right)}$.
This gives the exponential increase of violation with $N$ noted by
Mermin. For $N=3$, $\theta_{k}=0$ the Bell-Mermin inequality becomes:
$\langle\sigma_{y}^{(1)}\sigma_{x}^{(2)}\sigma_{x}^{(3)}\rangle+\langle\sigma_{x}^{(1)}\sigma_{y}^{(2)}\sigma_{x}^{(3)}\rangle\ \ \ \ +\langle\sigma_{x}^{(1)}\sigma_{x}^{(2)}\sigma_{y}^{(3)}\rangle-\langle\sigma_{y}^{(1)}\sigma_{y}^{(2)}\sigma_{y}^{(3)}\rangle\leq2$.
The similar Bell-Mermin inequality 
\begin{eqnarray}
\langle\sigma_{y}^{(1)}\sigma_{y}^{(2)}\sigma_{x}^{(3)}\rangle+\langle\sigma_{x}^{(1)}\sigma_{y}^{(2)}\sigma_{y}^{(3)}\rangle+\langle\sigma_{y}^{(1)}\sigma_{x}^{(2)}\sigma_{y}^{(3)}\rangle-\langle\sigma_{x}^{(1)}\sigma_{x}^{(2)}\sigma_{x}^{(3)}\rangle & \leq & 2\label{eq:mermin-ineq-1}
\end{eqnarray}
also applies, if one takes instead $X_{k}=\sigma_{y}^{(k)}$ and $Y_{k}=\sigma_{x}^{(k)}$.
This gives a violation for the state (\ref{eq:ghz}), the left side
being $4$.

Ardehali \citep{Ardehali_PRA1992} considered a different state $|\phi\rangle=\frac{1}{\sqrt{2}}\left(|\uparrow\uparrow...\uparrow\rangle-|\downarrow\downarrow...\downarrow\rangle\right)$
and more general measurements, allowing $\theta_{k}$ to be nonzero
for the system $k=N$. In this case, LHV theories imply 
\begin{align}
\mathrm{Re}\hat{A}+\mathrm{Im}\hat{A} & \leq2^{N/2}\,,N\,\text{even}\nonumber \\
\mathrm{Re}\hat{A}+\mathrm{Im}\hat{A} & \leq2^{\left(N+1\right)/2}\,,N\,\text{odd}\,,\label{eq:im-re}
\end{align}
whereas the quantum prediction is $\mathrm{Re}\hat{A}+\mathrm{Im}\hat{A}=2^{N-1/2}$.
The inequalities obtained by Ardehali also work for the state $|\phi\rangle=\frac{1}{\sqrt{2}}\left(|\downarrow\uparrow\uparrow\uparrow...\uparrow\rangle\pm|\uparrow\downarrow\downarrow\downarrow...\downarrow\rangle\right)$
or $|\phi\rangle=\frac{1}{\sqrt{2}}\left(|\downarrow\downarrow\uparrow\uparrow...\uparrow\rangle\pm|\uparrow\uparrow\downarrow\downarrow...\downarrow\rangle\right)$
or $|\phi\rangle=\frac{1}{\sqrt{2}}\left(|\downarrow\downarrow\downarrow\uparrow...\uparrow\rangle\pm|\uparrow\uparrow\uparrow\downarrow...\downarrow\rangle\right)$
or any other state with distinct permutations of $\uparrow$ and $\downarrow$;
and also the state $|\phi\rangle=\frac{1}{\sqrt{2}}\left(|\downarrow\uparrow\uparrow...\uparrow\rangle\pm i|\uparrow\downarrow\downarrow...\downarrow\rangle\right)$
or any other state with distinct permutations of $\uparrow$ and $\downarrow$.
Results were further generalized by Belinski and Klyshko \citep{Belinski_PhysUsp1993,Klyshko:1993aa,Gisin:1998aa,Zukowski_2002_PRL}.
 In summary, introducing the operator

\begin{equation}
\hat{V}=\begin{cases}
\mathrm{Re}\hat{A}+\mathrm{Im}\hat{A}, & N\ \mathrm{is\ even}\\
\sqrt{2}\,\mathrm{Im}\hat{A}, & N\ \mathrm{is\ odd}
\end{cases}\label{eq:V_Op}
\end{equation}
leads to the formulation of the Mermin-Ardehali-Belinski\u{\i}-Klyshko
(MABK) Bell inequalities which gives the following inequality for
any LHV theory

\begin{equation}
V\equiv\vert\langle\hat{V}\rangle\vert\le2^{N/2}\ .\label{eq:V}
\end{equation}
The inequality is violated by quantum mechanics for the GHZ states
and the measurement choices $\theta_{k}$ given in \citep{Mermin_PRD1980,Ardehali_PRA1992,Belinski_PhysUsp1993}.
The quantum prediction $V_{QM}$ is:

\begin{equation}
V=2^{N-1/2}\ .\label{eq:qpred}
\end{equation}
The MABK inequality for $N=2$ reduces to the Clauser-Horne-Shimony-Holt
(CHSH) Bell inequality \citep{Clauser:1969_PRL23}, and for $N=3$
is given by (\ref{eq:mermin-ineq-1}), and the similar inequality
obtained by exchanging $x$ with $y$. As Mermin pointed out, as the
number of particles $N$ increases, the violation of the inequalities
increases exponentially. Other papers investigating $N$-partite correlations
include \citep{roy1991tests,Zukowski_PRA1997,gisin1998bell,Werner_PRA2001}.\foreignlanguage{australian}{\textcolor{blue}{}}

Experimental evidence of the nonlocality of the GHZ correlations was
given by Pan et al \citep{Pan:2000aa}. They produced GHZ states
using the experimental set-up of Bouwmeester et al. \citep{Bouwmeester_PRL1999},
based on the proposal by Zeilinger et al. \citep{Zeilinger1997_ThreeParticle}.
The proposal allows the generation of $3$ entangled photons (GHZ
state) from $2$ pairs of polarization entangled photons. In order
to demonstrate the creation of GHZ state, they followed \citep{Greenberger1989_Bell,Greenberger_1990_AJP,Clifton_1991_FoundP,Mermin:1990aa}
and measured the values $\langle\sigma_{x}^{(1)}\sigma_{y}^{(2)}\sigma_{y}^{(3)}\rangle$,
$\langle\sigma_{y}^{(1)}\sigma_{x}^{(2)}\sigma_{y}^{(3)}\rangle$,
$\langle\sigma_{y}^{(1)}\sigma_{y}^{(2)}\sigma_{x}^{(3)}\rangle$,
$\langle\sigma_{x}^{(1)}\sigma_{x}^{(2)}\sigma_{x}^{(3)}\rangle$
to check the predictions of a GHZ state. The results are consistent
with the violation of the MABK inequality (\ref{eq:mermin-ineq-1}).

A different type of multipartite entangled state is the W state,
written for $N=3$ as \citep{Dur_PRA2000}
\begin{equation}
|W\rangle=\frac{1}{\sqrt{3}}(|\uparrow\rangle|\downarrow\rangle|\downarrow\rangle+|\downarrow\rangle|\uparrow\rangle|\downarrow\rangle+|\downarrow\rangle|\downarrow\rangle|\uparrow\rangle)\ .\label{eq:W}
\end{equation}
The $W$state with $N=4$ was shown to be genuinely multipartite entangled
in experiments performed by Papp et al \citep{papp2009characterization}.
The W states were generated using a photon source and a sequence of
beam splitters and can also violate the Mermin inequalities \citep{Cereceda_PRA2002,Swain2019_QIP}.\textcolor{red}{{}
}For higher $N$, graph states are multi-qubit states including GHZ
and cluster states with applications in quantum computing. Bell inequalities
giving tests of local realism for graph states were developed by Guhne
et al \citep{guhne2005bell} and Toth et al \citep{toth2006two},
using the method of stabilizing operators. These authors showed that
for certain families of graph states, Bell inequalities can be constructed
such that the violation of local realism increases exponentially with
$N$. \foreignlanguage{australian}{More generally, quantum networks
may have several independent sources of entangled states distributed
in the network. Rosset et al \citep{rosset2016nonlinear} and Tavakoli
\citep{tavakoli2016bell} derived nonlinear Bell inequalities which
can detect the nonlocality of the correlations distributed to distant
observers on such networks .\textcolor{red}{}}

Experimentally, four-photon entangled states where generated by Pan
et al. (2001) \citep{Pan_PRL2001} and Zhao et al. \citep{Zhao2003_PRL}.
Using higher pump powers and with improved photo-detection efficiency,
six \citep{Lu:2007aa}, eight \citep{Yao:2012aa}, ten \citep{Gao:2010aa}
and 18 \citep{Wang_PRL2018,Chao:2019aa} qubit entangled photonic
states have also been created. A multipartite entangled state of $14$
atoms was created in an ion trap \citep{Monz_PRL2011}. The violation
of the MABK inequality was reported for $N=4$ by Zhao et al \citep{Zhao2003_PRL},
who generated a GHZ entanglement, the experimental result being $V=4.43$.
Other reports of violations of the Mermin inequality for $N=3$ include
Chen et al. \citep{Chen_PRL2006}, Hamel et al \citep{Hamel:2014aa},
Patel et al \citep{Patel_ScienceAdvances2016} and Li et al \citep{Li:2019aa}.

The signatures of multipartite entanglement do not always involve
the MABK inequalities, however. Lu et al confirmed six-photon entanglement
for graph states using an entanglement witness. Monz et al generated
atomic GHZ states \citep{Monz_PRL2011}. Here, the system was initialized
into a state $|1...1\rangle$, where each ion is in the electronic
ground state $|1\rangle\equiv S_{1/2}\left(m=-1/2\right)$. An entanglement
interaction \citep{Sorensen_PRA2000,Benhelm:2008aa} then produced
GHZ states of the form $\left(|0...0\rangle+|1...1\rangle\right)/\sqrt{2}$.
The density matrix elements inferred in the measurement process gave
evidence of the coherence and fidelity of the GHZ state, and the multi-particle
entanglement was inferred by violation of inequalities deduced for
separable states \citep{guhne2010separability}. More recently, the
experiments of Wei et al \citep{Wei_PRA2020} and Song et al generated
$18$-qubit GHZ states using superconducting qubits \citep{Chao:2019aa},
with Song et al creating multicomponent superpositions of atomic coherent
states, involving up to 20 transmon qubits. Omran et al \citep{omran2019generation}
created an atomic Schr\"odinger cat state in the form of an $N$-particle
GHZ state generated in an array of Rydberg atoms, where $N\sim20$.
\textcolor{red}{}In the experiment , two-level Rydberg atoms are
prepared in an $N$-partite GHZ superposition state 
\begin{equation}
|GHZ_{N}\rangle=\frac{1}{\sqrt{2}}\left(|0101...\rangle+|1010...\rangle\right)\ \label{eq:ghzryd}
\end{equation}
where $|0\rangle$ is the atomic ground state and $|1\rangle$ is
the excited Rydberg state. Lasers manipulate an ensemble of $N$ atoms
prepared in the ground state into the GHZ state. The signature of
the GHZ state is the GHZ-state fidelity $F>0.5$, this being a witness
of the multipartite entanglement of the $N$-partite GHZ state.

In fact, the violations of local realism for $N$-partite systems
is possible for a broad range of states. The initial question of whether
all bipartite entangled states can show Bell nonlocality was addressed
by Gisin \citep{gisin1991bell}. \textcolor{blue}{}For any multipartite
entangled state, this result was generalized in 2012 by Yu et al.
\citep{yu2012all}. These authors proved that all entangled pure states
violate a CHSH-type Bell inequality involving two dichotomic measurements
at each site. The proof is based on Hardy's paradox, which is an all-or-nothing
illustration of the violation of local realism, similar to the GHZ
paradox.\textcolor{blue}{}

\subsection{Genuine multipartite Bell nonlocality}

The $N$-partite GHZ states (\ref{eq:ghz}) are seen to illustrate
a contradiction with the correspondence principle, that classical
statistics follows for large systems. This is because violation of
LHV theories is possible for an arbitrarily large $N$, as shown by
the Bell-MABK inequalities (\ref{eq:V}). Furthermore, the amount
of violation exponentially increases with $N$. On the other hand,
the Bell-MABK inequalities are derived assuming locality between
every one of the $N$ systems i.e. between all pairs of the $N$ particles.
The MABK inequalities if violated therefore imply a Bell nonlocality
to exist at least between a pair of particles. It might be argued
that it is then not surprising that the violation increases with an
increasing number of such pairs.

This criticism can be overcome. The GHZ states and the multipartite
generalizations illustrate a genuine $N$-partite Bell nonlocality
of the type initially considered by Svetlichny \citep{Svetlichny_PRD1987}.
In fact, a stricter set of inequalities can be derived which if violated
will imply a genuine $N$-partite Bell nonlocality, meaning that the
nonlocality cannot be described as arising from a mixing of states
which allow genuine $k$-partite Bell nonlocality, where $k<N$. The
nonlocality is mutually shared between \emph{all} $N$ particles.
The approach is to relax the assumptions made in the derivation of
the $N$-partite Bell inequality. There is not the requirement that
there is locality assumed between all of the $N$ subsystems. Rather,
one considers all possible bipartitions $A_{j}-B_{j}$ of the $N$
spatially separated subsystems, where $A_{j}$ is a set of specific
subsystems, and $B_{j}$ denotes the complementary set. Hidden variables
states are considered where locality is assumed between the $A_{j}$
and $B_{j}$, but not between the subsystems of $A_{j}$ and $B_{j}$.
If one supposes the system to be modeled by a convex mixture of all
such bilocal descriptions, then the correlations are constrained
by Bell inequalities \citep{Collins2002_PRL_Bell,Seevinck_PRL2002}.
The Bell inequality described by Collins, Gisin, Popescu, Roberts
and Scarani (CGPRS) is \citep{Collins2002_PRL_Bell} 
\begin{eqnarray}
V_{{\cal S}}\equiv\mathrm{Re}A+\mathrm{Im}A & \leq & 2^{N-1}\ .\label{eq:merminsteerstat-1}
\end{eqnarray}
Svetlichny's inequality is a version for $N=3$ \citep{Svetlichny_PRD1987}.
The quantum prediction (\ref{eq:qpred}) maximizes to predict a violation,
for even $N$, by a \emph{constant }amount: $\frac{V_{QM}}{V_{S}}=\sqrt{2}$
\citep{Werner_PRA2001}. \foreignlanguage{australian}{It should be
noted that since Svetlichny's work, further improved definitions of
genuine multipartite nonlocality have been constructed by Gallego
et al and Bancal et al \citep{gallego2012operational,bancal2013definitions,dutta2020operational}.
These take into account no-signalling and the time ordering of measurements
between the observers, and enable genuine $N$-partite nonlocality
to be detected for a broader set of entangled states.}

Experiments for $N=3$ verifying the violation of Svetlichny's inequality
were performed by Lavioe et al \citep{Lavoie_2009} and Lu et al
\citep{Lu_PRA2011} using three-photon GHZ states with correlated
polarization. Lavioe et al followed the approach of Bouwmeester et
al. \citep{Bouwmeester_PRL1999} to generate the GHZ state. The experiment
setup by Lu et al. was based on the proposal by Rarity and Tapster
\citep{Rarity_PRA1999}, involving entangled photon pairs and a weak
coherent state. They shone an infrared pulse with wavelength $780$nm
into a crystal that up converts into an ultraviolet pulse with wavelength
$390$nm. This pulse was split into two beams, where one beam was
subsequently sent into another nonlinear crystal to produce a polarization
entangled state $\left(|H_{2}H_{3}\rangle+|V_{2}V_{3}\rangle\right)/\sqrt{2}$,
while the other beam was prepared in a state $\left(|H_{1}\rangle+|V_{1}\rangle\right)/\sqrt{2}$.
Beam $1$ and $2$ are sent into a polarizing beam splitter such that
the total state is a GHZ state $\left(|HHH\rangle+|VVV\rangle\right)/\sqrt{2}$.
The ion trap experiments of Barreiro et al \citep{Barreiro:2013aa}
reported multipartite device-independent entanglement for up to $N=6$
ions. When there are two measurement settings with two possible outcomes
for each setting, the device-independent entanglement witnesses used
by Barreiro et al. are equivalent to the $n$-partite Svetlichny inequalities.
In this experiment, up to $6$ ions were conclusively shown to have
genuine multipartite quantum nonlocal correlations. Higher numbers
of ions have the problem of cross-talks in the measurement process
(local measurements on individual ions affect neighboring ions).\textcolor{red}{{}
}\textcolor{black}{Subsequent reports of violations of Svetlichny-}CGPRS\textcolor{black}{{}
inequal}ities  for three qubit GHZ-photon states include those of
Erven et al. (2014) \citep{Erven:2014aa}, Hamel et al. \citep{Hamel:2014aa}
and Patel et al\foreignlanguage{australian}{ }\citep{Patel_ScienceAdvances2016}\foreignlanguage{australian}{.
Recently, violations of Svetlichny inequalities were verified using
the IBM quantum computer for W and GHZ states }\citep{Swain2019_QIP}\foreignlanguage{australian}{.
}

\subsection{Multipartite Bell nonlocality with qudits}

After the original GHZ papers, further studies demonstrated that the
``all or nothing'' violation of local realism can apply even where
there is a higher dimensionality at each of the sites. The paper by
Reid and Munro \foreignlanguage{australian}{\citep{reid1992macroscopic}
}generalized the three-party GHZ contradiction with local realism,
for $N$ particles at each of three sites, $j=1,2,3$. They considered
the multipartite extension of the state (\ref{eq:N-boson}), given
by
\begin{equation}
|\psi\rangle=\frac{\left(\hat{a}_{1+}^{\dagger}\hat{a}_{2+}^{\dagger}\hat{a}_{3+}^{\dagger}+\hat{a}_{1-}^{\dagger}\hat{a}_{2-}^{\dagger}\hat{a}_{3-}^{\dagger}\right)^{N}\vert0\rangle}{N!\left[\sum_{r=0}^{N}r!(N-r)\right]^{1/2}}.\label{eq:QSRM}
\end{equation}
Here, the state $|\psi\rangle=\frac{1}{\sqrt{2}}\left(\hat{a}_{1+}^{\dagger}\hat{a}_{2+}^{\dagger}\hat{a}_{3+}^{\dagger}+\hat{a}_{1-}^{\dagger}\hat{a}_{2-}^{\dagger}\hat{a}_{3-}^{\dagger}\right)\vert0\rangle$
with $N=1$ is a GHZ state similar to the spin version (\ref{eq:ghz}),
since we may map the two-state system $|1\rangle_{+}|0\rangle_{-}$,
$|0\rangle_{+}|1\rangle_{-}$ onto spin qubits $|\uparrow\rangle$,
$|\downarrow\rangle$ for each mode given by $a_{j}$, $j=1,2,3$.
The $\hat{a}_{j+}^{\dagger}$, $\hat{a}_{j-}^{\dagger}$ are boson
operators for six orthogonal field modes, and the $\pm$ denotes the
orthogonal modes/ polarizations at the same energy. The detected outputs
for the analyzers (polarizers or beam splitters) at each site correspond
to the following transformed modes: $\hat{d}_{j\pm}\left(\phi_{j}\right)=\frac{1}{\sqrt{2}}\left(\pm\hat{a}_{j+}+e^{i\phi_{j}}\hat{a}_{j-}\right)$,
similar to (\ref{eq:rotc}). The choices for the settings are $\phi=0$
and $\phi=\pi/2$, which gives measurements $\hat{\sigma}_{x}^{(j)}$
or $\hat{\sigma}_{y}^{(j)}$ for site $j$. Incident on each of the
three analyzers ($j=1,2,3$) are $N$ bosons, some of which are detected
in polarization mode $+$ and the rest as $-$. Assuming the number
of bosons with the final polarization $+$ (or $-$) is measurable,
one may treat the bosons as though distinguishable particles and determine
the product $S_{jx}^{N}$ (or $S_{jy}^{N}$) of the individual spin
outcomes $+1$ or $-1$ associated with the $N$ incident bosons at
site $j$. The spin products under consideration are $S_{1y}^{N}S_{2y}^{N}S_{3x}^{N},$
$S_{1y}^{N}S_{2x}^{N}S_{3y}^{N},$ $S_{1x}^{N}S_{2y}^{N}S_{3y}^{N}$
and $S_{1x}^{N}S_{2x}^{N}S_{3x}^{N}$. The expectations values of
these products is calculated by rewriting the state $|\psi\rangle$
given in Eq. (\ref{eq:QSRM}), in the transformed modes $\hat{d}_{j\pm}^{x}$
and $\hat{d}_{j\pm}^{y}$. The expectation values for $N$ odd for
the products $S_{1x}^{N}S_{2y}^{N}S_{3y}^{N},$ $S_{1y}^{N}S_{2x}^{N}S_{3y}^{N},$
and $S_{1y}^{N}S_{2y}^{N}S_{3x}^{N}$ is always $-1,$ while for $S_{1x}^{N}S_{2x}^{N}S_{3x}^{N}$
the state is transformed in the $\hat{d}_{j\pm}^{x}$ modes, obtaining
that the expectation value for this product is $+1$ for all $N$.
However, on calculating the classical predictions for this expectation
value, the result is always $-1$, in disagreement with the previous
result of $+1$. This is the ``all or nothing'' distinction between
quantum and classical predictions, applied to arbitrary large values
of odd $N,$ where one has a macroscopic state. The authors further
showed how one may extend the approach to consider violation of the
Mermin inequality for $N$ particles at each site. GHZ correlations
for multi-dimensional system without inequalities were also considered
by Cabello \citep{cabello2001multiparty}.

Higher-dimensional multipartite Bell inequalities have since been
studied extensively \citep{Peres:1999aa,cabello2002bell,ChenPRA_2006,Chen2009_PRA,Arnault_2012_JPA,Son:2006_PRL}.\textcolor{red}{{}
}Cabello extended the MABK inequality to $n$ spin $s$ particles,
and showed that higher dimensional GHZ states maximally violate these
inequalities \citep{cabello2002bell}. Cabello demonstrated that the
violation for an arbitrary but fixed $s$ increases exponentially
with $N$, thus extending the observation of Mermin \citep{mermin1990extreme}
to higher spin. Cabello also demonstrated that for arbitrary but fixed
$N$, the violation does not decrease with $s$, thus generalizing
the result of Gisin and Peres \citep{Gisin_1992_PhysLettA} for $N=2$.
Son, Lee and Kim derived generalized MABK inequalities for multipartite
systems of arbitrary dimension i.e. for arbitrary $d$ and $N$ \citep{Son:2006_PRL}.
They showed that the higher dimensional extensions of the GHZ states
given as
\begin{equation}
\frac{1}{\sqrt{d}}\{\sum_{k=1}^{d-1}|k,k,k,..\rangle\}\ \label{eq:multi}
\end{equation}
may violate these inequalities. Here $|k,k,k,..\rangle\equiv|k\rangle_{1}..|k\rangle_{j}..|k\rangle_{N}$,
the $|k\rangle_{j}$ being an orthogonal basis set for states at the
site $j$. However, it was known that the MABK inequalities were not
tight, in the sense that there exist entangled states that would not
violate the MABK inequalities. For $N=3$, Chen et al presented two-setting
inequalities that were shown numerically to be violated for all entangled
states \citep{chen2004gisin}. In 2009, Chen and Deng extended this
result, to derive a Bell inequality based on the CHSH correlation
functions for $N$ $d$-dimensional systems, which was shown tight
for a range of systems, including up to $d=10$ for $N=3$. \textcolor{blue}{}

Multipartite and higher dimensional Bell tests provide a way to overcome
losses and noise, although this can also be achieved with multiple
settings. It could be argued that the violations associated with the
$N$-partite GHZ qubit states which have dichotomic outcomes ($d=2$)
do not satisfy the requirement of a macroscopic Bell violation. For
the Mermin inequality of type (\ref{eq:mermin-ineq-1}), for example,
the products of the Pauli spins at each site are either $+1$ or $-1$.
 Therefore, each spin system must be measured exactly, or else this
product changes sign. A microscopic resolution of measurement outcomes
is required. The Bell violations are thus lost when there is enough
decoherence of the system so that one of the spins changes sign. This
is made apparent by the fact that the violations are readily destroyed
by reduced detection efficiencies. On the other hand, \emph{multi-setting}
Bell tests allow a choice of more than two measurement settings at
each site \citep{gisin1999bell,collins2004relevant} and are known
to enhance the tests of local realism, allowing violations to be obtained
for greater levels of decoherence, as realized by losses associated
with lower detection efficiencies. One prediction is for a failure
of local realism with a detection efficiency as low as 43\% at one
detector, for a non-maximally entangled state, using three measurement
settings \citep{cabello2007minimum,brunner2007detection}. For highly
entangled states however, violations of local realism are possible
at 50\% efficiency using combinations of multi-settings and/ or multiple
sites \citep{brunner2007detection,larsson2001strict,cabello2008necessary,reid2013two,Kiesewetter2015Violations}.
The work of Durt, Kaszlikowski and Zukowski \citep{durt2001violations}
showed the possibility that higher dimensional Bell tests are more
robust with noise (refer Section VII).

\section{Continuous-variable quantum correlations}

\subsection{EPR correlations}

The original EPR argument was based on the correlations of position
and momenta, the outcomes of which are continuous variables \citep{Einstein:1935}.
The EPR state is a two-party state, where each subsystem $i=1,2$
has a position $q_{i}$ and momentum $p_{i}$. The EPR state is simultaneously
an eigenstate of position difference $q_{1}-q_{2}$ and momenta sum
$p_{1}+p_{2}$. This means that $(\Delta(q_{1}-q_{2}))^{2}\rightarrow\infty$
and $(\Delta(p_{1}+p_{2}))^{2}\rightarrow\infty$, where here we use
the notation $(\Delta x)^{2}$ as the variance of $x$.

The correlations can be realized for fields using the conjugate quadrature
phase amplitude observables $X_{i}$ and $P_{i}$ of two modes $i=1,2$,
where (for a rotating frame) $\hat{X}_{i}=\hat{a}_{i}+\hat{a_{i}}^{\dagger}$
and $\hat{P}_{i}=(\hat{a}_{i}-\hat{a_{i}}^{\dagger})/i$ \citep{Reid:1988,Reid:1989}.
This was shown by Reid \citep{Reid:1989} for the output of the non-degenerate
parametric amplifier, which is equivalent to the two-mode squeezed
vacuum state. Here, $\hat{a_{i}}$ and $\hat{a}_{i}^{\dagger}$ are
boson creation and destruction operators for the single mode field
labelled by $i$. The non-degenerate parametric amplifier is described
by the interaction Hamiltonian 
\begin{equation}
H_{I}=i\kappa E(\hat{a}_{1}^{\dagger}\hat{a}_{2}^{\dagger}-\hat{a}_{1}\hat{a}_{2})\ \label{eq:ham-da-two-mode}
\end{equation}
where $\kappa$ is the coupling strength between the modes and $E$
the pump intensity. The unitary interaction generates a two-mode squeezed
vacuum state of type $|TMSS\rangle=\text{sech}r\sum_{n=0}^{\infty}\tanh^{n}r\ |n\rangle_{1}|n\rangle_{2}$,
where here $|n\rangle_{i}$ is the number state for mode $i$. Following
\citep{Reid:1989}, one may solve (\ref{eq:ham-da-two-mode}) directly
to confirm EPR correlations. We find $\dot{\hat{a}}_{1}=\kappa E\hat{a}_{2}^{\dagger}$
and $\dot{\hat{a}}_{2}=\kappa E\hat{a}_{1}^{\dagger}$, implying $\dot{\hat{X}}_{1}=\kappa EX\hat{X}_{2}$,
$\dot{\hat{X}}_{2}=\kappa E\hat{X}_{1}$ from which it is clear that
the solution for $\hat{X}_{\pm}=\hat{X}_{1}\pm\hat{X}_{2}$ is $\hat{X}_{\pm}(t)=\hat{X}_{\pm}(0)e^{\pm\kappa Et}$.
Similarly, defining $\hat{P}_{\pm}=\hat{P}_{1}\pm\hat{P}_{2}$, the
solutions are $\hat{P}_{\pm}(t)=\hat{P}_{\pm}(0)e^{\mp\kappa Et}$
\citep{rosales2015decoherence}. Using that $(\Delta\hat{X}_{\pm}(0))^{2}=2$,
this gives 
\begin{equation}
(\Delta(\hat{X}_{1}-\hat{X}_{2}))^{2}=2e^{-2r},\ (\Delta(\hat{P}_{1}+\hat{P}_{2}))^{2}=2e^{-2r}\label{eq:epr-solns}
\end{equation}
where $r=\kappa Et$ is the two-mode squeezing parameter, leading
to EPR correlations for $r>0$, as $r\rightarrow\infty.$

The EPR correlations can also be realized using a beam splitter, with
either one or two vacuum squeezed inputs \citep{Reid:1989,DiGuglielmo_PRA2007,Eberle2011_PRA}.
In this case, the two incoming modes $a_{1}^{(in)}$, $a_{2}^{(in)}$
are transformed into the two output modes $a_{1}$, $a_{2}$ according
to $\hat{a}_{1}=\cos\theta\hat{a}_{1}^{(in)}+\sin\theta\hat{a}_{2}^{(in)}$,
$\hat{a}_{2}=\sin\theta\hat{a}_{1}^{(in)}-\cos\theta\hat{a}_{2}^{(in)}$.
The $\hat{a}_{1}$, $\hat{a}_{1}^{(in)}$ and $\hat{a}_{2}$, $\hat{a}_{2}^{(in)}$
are the destruction boson operators for the modes. Let us assume the
input $a_{2}^{(in)}$ to be squeezed along the $X$ quadrature so
that $\Delta X_{2}^{(in)}=e^{-r_{2}}$, and the input $a_{1}^{(in)}$
to be squeezed along $P$ so that $\Delta P_{1}^{(in)}=e^{-r_{1}}$.
Here, $r_{i}>0$ are the squeezing parameters. The solutions for the
outputs of a 50/50 beam splitter are
\begin{equation}
(\Delta(\hat{X}_{1}-\hat{X}_{2}))^{2}=2e^{-2r_{2}},\ (\Delta(\hat{P}_{1}+\hat{P}_{2}))^{2}=2e^{-2r_{1}}\ ,\label{eq:epr-bs}
\end{equation}
which leads to ideal EPR correlations for large $r_{1}$ and $r_{2}$.
More complete details are given elsewhere \citep{teh2021full}. If
$r_{2}=0$ so that $a_{2}^{(in)}$ is not squeezed ($\Delta\hat{X}_{2}^{(in)}=\Delta\hat{P}_{2}^{(in)}=1$),
but if $a_{1}^{(in)}$ remains squeezed with $r_{1}>0$, then the
correlations are no longer ideally EPR correlated. Nonetheless, as
explained below, we will see that the system may be regarded as EPR
correlated, since the correlations result in an EPR paradox. Later
studies revealed that number states incident on beam splitters can
also generate entanglement between the output modes \citep{kim2002entanglement}.

More generally, ideal EPR correlations manifest for two conjugate
(non-commuting) observables of a single system $A$ ($i=1$), when
both of those observables can be estimated with perfect accuracy by
a measurement on a remote space-like separated system $B$. The measurement
at $B$ will be different in each case. Let us denote the observables
as $O_{x}$ and $O_{p}$, and take the special case where the commutator
is of the form $[O_{x},O_{p}]=C$, where $C$ is a constant. For position
and momentum, $C=i\hbar$ and for quadratures, $C=[\hat{X}_{i},\hat{P}_{i}]=2i$.
This implies an uncertainty $\Delta\hat{X}_{i}\Delta\hat{P}_{i}\geq1$.
Following \citep{Reid:1989}, EPR correlations allow an inferred estimate
for outcomes $X_{i}$ and $P_{i}$ of $\hat{X}_{i}$ and $\hat{P}_{i}$
that approaches zero in each case. Taking the estimate $X_{1,est}$
of $X_{1}$ at $A$ to be the optimal linear combination of quadrature
amplitudes measurable at $B$, this estimate is quantified by the
inference variance defined as $\Delta_{inf}^{2}\hat{X}_{1}\equiv(\Delta(\hat{X}_{1}-X_{1,est}))^{2}$,
where we use the notation $(\Delta_{inf}\hat{X}_{1})^{2}=\Delta_{inf}^{2}\hat{X}_{1}$
to simplify use of brackets. The EPR correlations are observed for
$O_{x}$ and $O_{p}$ when the product reduces below that of the uncertainty
bound, $\Delta_{inf}O_{x}\Delta_{inf}O_{p}<\frac{|\langle C\rangle|}{2}$,
which for $X$ and $P$ reduces to the EPR condition \citep{Reid:1989}
\begin{equation}
E_{1|2}\equiv\epsilon=\Delta_{inf}\hat{X}_{1}\Delta_{inf}\hat{P}_{1}<1\ .\label{eq:epr-crit}
\end{equation}
Specifically, for parametric down conversion, the best linear estimate
of $X_{1}$ is $X_{1,est}=gX_{2}$. Similarly, the best linear estimate
of $P_{1}$ is $P_{1,est}=g'P_{2}$ where $g$ and $g'$ are real
numbers. Therefore 
\begin{equation}
E_{1|2}\equiv\Delta(\hat{X}_{1}-g'X_{2})\Delta(\hat{P}_{1}+gP_{2})<1\label{eq:steer-gh}
\end{equation}
is a condition sufficient to demonstrate correlations of the EPR paradox.
The result for the two-mode squeezed state generated by down conversion
is known to be \citep{Reid:1989,rosales2015decoherence} 
\begin{equation}
\Delta_{inf}^{2}\hat{X}_{1}=\Delta_{inf}^{2}\hat{P}_{1}=\frac{1}{\cosh2r}\ \label{eq:epr-solns-1}
\end{equation}
using $g=g'=\tanh2r$ where $r=\kappa Et$. More details are given
in \citep{teh2021full}. For the beam-splitter configuration described
above with two squeezed inputs, the result is given by the expression
(\ref{eq:epr-solns-1}) but where $r$ denotes the squeeze parameter
of the input fields. With only one squeezed input, we find $\Delta_{inf}\hat{X}_{1}\Delta_{inf}\hat{P}_{1}=\frac{1}{\cosh r}$
with $g=g'=\frac{1-e^{-2r}}{1+e^{-2r}}$.

A further generalization of the EPR paradox is to consider the local
hidden variable models, as in (\ref{eq:lhv}) and (\ref{eq:lhv-2}).
Here, one considers two separated systems labelled $i=1,2$ which
in (\ref{eq:lhv}) and (\ref{eq:lhv-2}) were labelled $A$ and $B$,
respectively. The structure (\ref{eq:lhv-2}) can be taken as a generalized
definition of local realism, or of local causality. One may then \emph{also}
consider whether the expressions $P_{+}^{(A)}(\phi,\lambda)$ for
$A$ are consistent with a local \emph{quantum} state description,
which would be given by a density operator $\rho_{A,\lambda}$. We
write 
\begin{equation}
P_{++}(\theta,\phi)=\int\rho(\lambda)\ P_{+,Q}^{(A)}(\theta,\lambda)P_{+}^{(B)}(\phi,\lambda)\ d\lambda\label{eq:lhv-2-1}
\end{equation}
where the subscript $Q$ denotes this extra condition on the local
hidden state (LHS) at $A$. One is asking whether, within the framework
of local hidden variables (based on a generalized premise of local
realism, or local causality), the elements of reality (interpretable
as the $\lambda$) are consistent with a local quantum state description
at $A$. If no, one may interpret the result as a generalized EPR
paradox, since the assumption of local realism / causality is shown
to be incompatible with local hidden variable states for $A$ that
are consistent with a local quantum state. The failure of the condition
given as (\ref{eq:lhv-2-1}) is referred to as ``steering'', or
``EPR steering''. Steering is the term used by Schr\"odinger in
his response to EPR's original 1935 paper \citep{Schrodinger:1935a,Schrodinger:1936}.
The connection between the expression (\ref{eq:lhv-2}), ``steering''
and quantum tasks was given by Wiseman, Jones and Doherty \citep{Wiseman_PRL2007,saunders2010experimental}.
The failure of (\ref{eq:lhv-2-1}) gives the condition for a steering
of system $A$.

The EPR condition (\ref{eq:epr-crit}) involving $E_{A|B}$ is sufficient
for demonstration of EPR steering of system $A$ \citep{Cavalcanti2009},
and gives a one-sided device-independent condition for entanglement
\citep{Branciard_PRA2012,Opanchuk_PRA2014}. The condition has been
shown necessary and sufficient for two-mode steering where the systems
$A$ and $B$ are Gaussian single-mode systems \citep{Jones_PRA2007}.
This implies restricting to Gaussian states and Gaussian measurements
\citep{weedbrook2012gaussian,dAuria2009full}. Links between EPR-variance
criteria for entanglement and steering and the criteria of Simon and
Duan et al \citep{Simon_PRL2000,duan2000inseparability} derived in
the context of Gaussian \textcolor{black}{states have been formalised
by Marian and Marian \citep{Marian_JOPA2018,Marian_PRA2021}.}

In order for the EPR correlations to be observed, the inferred uncertainties
are compared relative to the value which is given as the commutator
$C$. The correlation can then hardly be called ``macroscopic''.
On the other hand, the method of detection in optical physics is to
amplify the quantum noise level, using local oscillator fields which
in quantum mechanics are modeled (approximately) as coherent states
$|\alpha\rangle$ with large amplitude $\alpha$. This implies that
the detection involves large numbers of photons incident on detectors,
in contrast with the detection of the Bell correlations described
in Section III. A careful analysis shows that in some experiments,
the amplification occurs \emph{prior} to the choice of a phase angle
which determines the measurement setting i.e. whether $X$ or $P$
is to be measured. This is the case for polarization entanglement
experiments \citep{Bowen_PRL2002}, where the local oscillator is
combined with the signal field ahead of impinging on the polarizer
beam splitter, the setting of which determines whether $X$ or $P$
is measured at the final detector. The continuous variable experiment
can then be mapped onto an equivalent macroscopic spin EPR experiment,
where the observables that are measured are (once the local oscillator
mode is accounted for) the Schwinger spins \citep{Reid_PRA2018,Reid_PRL2000,Reid_PRA2000},
\begin{equation}
\hat{S}_{X}^{(i)}=\frac{1}{2}(\hat{c_{i}}^{\dagger}\hat{a_{i}}+\hat{c_{i}}\hat{a_{i}}^{\dagger}),\ \hat{S}_{Y}^{(i)}=\frac{1}{2i}(\hat{c}_{i}^{\dagger}\hat{a_{i}}-\hat{c}_{i}\hat{a}_{i}^{\dagger}),\ \hat{S}_{Z}^{(i)}=\frac{1}{2}(\hat{c}_{i}^{\dagger}\hat{c}_{i}-\hat{a}_{i}^{\dagger}\hat{a}_{i})\ .\label{eq:schwinger}
\end{equation}
Here, $\hat{a_{i}}$, $\hat{a}_{i}^{\dagger}$ are the boson operators
for the field mode labelled $i$ that is being measured. The operators
$\hat{c}_{i}$, $\hat{c}_{i}^{\dagger}$ are the boson operators for
the local oscillator fields associated with each mode $i$, which
are model-led as an intense coherent state of amplitude $E=\alpha$
(taken to be real). In this limit, $\hat{S}_{X}\rightarrow\frac{E}{2}\hat{X}$
and $\hat{S}_{Y}\rightarrow\frac{E}{2}\hat{P}$ where $\hat{X}$ and
$\hat{P}$ are the quadrature amplitudes of the field $i$. Then we
see that because $E=\alpha$ is large, $\langle\hat{c}_{i}^{\dagger}\hat{c}_{i}\rangle\gg\langle\hat{a_{i}}^{\dagger}\hat{a_{i}}\rangle$
and the values of the spins can also be large i.e. macroscopic. The
combined system at each site comprises a single mode $\hat{a}_{i}$
and a second very intense field $\hat{c_{i}}$ and hence this system
prior to the choice of measurement setting $X$ or $P$ is macroscopic.
In this way, one can argue that the continuous variable (CV) correlations
are ``macroscopic''. The analysis of the CV experiment gives an
example of amplification due to measurement, and the analogy with
the Schr\"odinger cat gedanken experiment has been given in \citep{Reid2019_PRL}.

Schr\"odinger noted in his response to the EPR argument of 1935 that
the paradoxical correlations occur when the states are ``entangled''
\citep{Schrodinger:1935_Naturwiss}. This led to the concept of entanglement
defined within quantum theory: two systems $A$ and $B$ are entangled
if the density operator $\rho$ for the combined systems cannot be
expressed in the separable form $\rho=\sum_{R}P_{R}\rho_{R}^{(A)}\rho_{R}^{(B)}$,
where here the system $A$ ($B$) is identified with system $i=1$
($2)$. The verification of entanglement between two single modes
generally requires a less strict bound than for EPR steering. As an
example, a criterion for entanglement between systems $1$ and $2$
is given by 
\begin{equation}
\Delta_{prod}=\Delta(\hat{X}_{1}-g'\hat{X}_{2})\Delta(\hat{P}_{1}+g\hat{P}_{2})<1+gg'\ \label{eq:ent-cond}
\end{equation}
as derived by Giovannetti et al \citep{giovannetti2003characterizing}
in their Eq (5), with $a_{1}=b_{1}=1$, $a_{2}=g$, $b_{2}=g'$. From
this, we see that the output modes in the beam splitter configurations
above are entangled for all $r$. The product entanglement criterion
$\Delta_{prod}=\Delta\left(\hat{X}_{1}-\hat{X}_{2}\right)\Delta\left(\hat{P}_{1}+\hat{P}_{2}\right)<2$
derived earlier by Tan \citep{Tan1999ConfirmingPhysRevA.60.2752}
is a special case of (\ref{eq:ent-cond}) for $g=g'=1$. On noting
that for any real numbers $x$ and $y$, $x^{2}+y^{2}\geq2xy$, we
see that the sum criterion 
\begin{align}
\Delta_{sum} & =\left[\Delta^{2}\left(\hat{X}_{1}-g'\hat{X}_{2}\right)+\Delta^{2}\left(\hat{P}_{1}+g\hat{P}_{2}\right)\right]<2(1+gg')\ \label{eq:Simon_Duan-1}
\end{align}
derived by Duan et al and Simon \citep{duan2000inseparability,Simon_PRL2000}\textcolor{red}{{}
}for $g=g'$, is also a special case of (\ref{eq:ent-cond}).\textcolor{red}{{}
} Schr\"odinger's historical responses to the EPR argument motivated
the classifications of entanglement, steering, Bell nonlocality. More
details are given in the reviews \citep{Reid:2009_RMP81,Brunner_RMP2014,Uola2020_RMP}.

There have been many examples of realizations of the continuous variable
(CV) EPR quantum correlations, beginning with the demonstration of
the CV EPR paradox by Ou et al \citep{Ou:1992}. Since then, there
have been reports of both entanglement and EPR steering in optical
CV systems, including those referred to in the review \citep{Reid:2009_RMP81}
(see also \citep{Walborn2011_PRL,Schneeloch2013_PRL,Wang:2010_OptExpress,Yan2012_PRA}).
Experiments have also demonstrated EPR correlations between pairs
of photons, using the EPR criterion (\ref{eq:epr-crit}) \citep{Lee2016_PRL,HowellPRL_2004}.
We see from the expressions (\ref{eq:epr-solns}), (\ref{eq:epr-crit})
and (\ref{eq:ent-cond}) that the EPR entanglement and steering conditions
are also conditions for squeezing, since a reduction in the noise
below that governed by the uncertainty principle is required. The
most significant results for two-mode squeezing, EPR entanglement
and EPR steering have been reported in a set of optical experiments,
which achieve as low as $\epsilon\sim0.1$\textcolor{red}{{} }\citep{Hage2010_PRA,Eberle2011_PRA,Samblowski_arxiv2010,Samblowski2011_AIP,Steinlechner2013_PRA,Eberle:2013_OptExp}.

It is also possible to construct EPR conditions for spin operators,
defined as (\ref{eq:schwinger}), which have discrete outcomes. For
spin $1/2$, this corresponds to Bohm's version of the EPR paradox
\citep{bohm2013quantum}. Conditions to realize Bohm's EPR paradox
for spin systems have been put forward in \citep{Reid:2009_RMP81,Cavalcanti:2009aa}.
These apply the uncertainty relations involving spin commutation
relations. In fact, the macroscopic realizations of EPR correlations
obtained using the Stokes spin observables (defined similarly to (\ref{eq:schwinger}))
may be interpreted as a higher spin version of Bohm's paradox \citep{Bowen_PRL2002}.
The relation $\Delta\hat{S}_{X}\Delta\hat{S}_{Y}\geq\frac{|\langle\hat{S}_{Z}\rangle|}{2}$
for the Schwinger relations (\ref{eq:schwinger}) is used, in which
case $c_{i}$ are intense local oscillator fields and the value of
the $|\langle\hat{S}_{Z}\rangle|$ becomes large, giving a macroscopic
level of quantum noise. In the experiment of Bowen et al \citep{Bowen_PRL2002},
the correlated source is generated by non-degenerate down conversion
modeled as (\ref{eq:ham-da-two-mode}). The modes are combined with
strong fields $c_{i}$ using beam splitters and then passed through
polarizers at each site. This achieves the scenario where an EPR correlation
is expressed in terms of spin operators, with a macroscopic level
of quantum noise. 

\subsection{Bell tests}

An important question is whether one may demonstrate Bell nonlocality
for continuous variable measurements. This would rule out all local
hidden variable theories, which is a stronger result than confirming
either entanglement or EPR steering. As we have seen, the CV measurements
using optical homodyne methods can be expressed in terms of the Schwinger
operators, for certain arrangements at least, which would allow more
macroscopic tests of LHV theories.

Bell violations for continuous variable measurements were proposed
by Bell. Bell considered a quantum mechanical state with a Wigner
function \citep{Bell_2001_JohnBell} 
\begin{align}
W\left(q_{1},q_{2},p_{1},p_{2}\right) & =Ke^{-q^{2}}e^{-p^{2}}\delta\left(p_{1}+p_{2}\right)\left[\left(q^{2}+p^{2}\right)^{2}-5q^{2}+p^{2}+\frac{11}{4}\right]\label{eq:wigner-1}
\end{align}
where $K$ is a constant, $q=q_{1}-q_{2}$, and $p=\left(p_{1}-p_{2}\right)/2$.
This Wigner function is negative in certain regions, for example at
$p=0,\,q=1$, and was used by Bell to show a violation of a Bell inequality.
This is carried out by calculating the sign correlation function between
particle $1$ at time $t_{1}$ after free evolution, and particle
$2$ at time $t_{2}$.

It was understood by Bell that a positive Wigner function would provide
a local hidden variable theory for the measurements of $q_{i}$ and
$p_{i}$. Leonhardt and Vaccaro \citep{Leonhardt:1995aa} obtained
the wave-function 
\begin{align}
\tilde{\psi}\left(p_{1},p_{2}\right) & \propto\left[(p_{1}-p_{2})^{2}-4\right]e^{-\frac{1}{8}\left(p_{1}-p_{2}\right)^{2}}\delta\left(p_{1}+p_{2}\right)\,\label{eq:reduced_wigner-1}
\end{align}
in the momentum representation from the Wigner function Eq. (\ref{eq:wigner-1}).
They noted that the two features in Eq. (\ref{eq:reduced_wigner-1})
giving rise to a nonlocal correlation are the Dirac delta function
$\delta\left(p_{1}+p_{2}\right)$ and the admission of negative values
in the Wigner function. Leonhardt and Vaccaro \citep{Leonhardt:1995aa}
proposed mixing a squeezed vacuum state with a superposition of Fock
state to produce a state with these two features. A squeezed vacuum
has a momentum wave-function of the form $\tilde{W}_{sq}\left(p\right)\propto\delta\left(p\right)$
in the strong squeezing limit. In order to introduce negativity into
the Wigner function, a superposition of $|0\rangle$ and $|2\rangle$
is mixed with the squeezed vacuum using a $50:50$ beam splitter.
The momentum wave-function of the output state from the beam splitter
resembles that of the wave-function Eq. (\ref{eq:reduced_wigner-1})
and violates a Bell inequality. The optical equivalence of free particle
evolution in the Bell work is the homodyne detection of the rotated
quadratures $q_{\theta}\equiv q\cos\theta-p\sin\theta$ with the angle
$\theta=\text{arctan}t$, where $t$ is the time as in Bell's work.
The approach was to create dichotomic outcomes from the continuous
variable outcomes by binning according to the sign of $q_{\theta}$,
so that a traditional Bell inequality based on dichotomic observables
could be implemented.

Gilchrist, Deuar and Reid showed how local hidden variable theories
can be excluded for correlations based on pair-coherent states and
superpositions of squeezed two-mode cat states \citep{Gilchrist_PRL1998,gilchrist1999contradiction}.
Cat states are the superpositions $|\psi\rangle\sim|\alpha\rangle+e^{i\theta}|-\alpha\rangle$
of two macroscopically distinct coherent states, $|\alpha\rangle$
and $|-\alpha\rangle$ \citep{Yurke_PRL1986}. These states have been
generated experimentally in optical \citep{Ourjoumtsev:2007aa} and
microwave systems \citep{Kirchmair:2013aa,Vlastakis_Science2013}.
Collapse and revivals reported for matter waves suggest similar states
to be generated in a Bose-Einstein condensate (BEC) \citep{Greiner:2002aa}.
These authors studied entangled cat states based on two spatially
separated modes $A$ and $B$. In particular, pair-coherent states
are the continuous superpositions of coherent states with a fixed
amplitude but arbitrary phase $\zeta$ \citep{agarwal1988nonclassical}
\begin{align}
|\Psi\rangle_{m} & =\mathcal{N}\intop_{0}^{2\pi}e^{-im\zeta}|\alpha_{0}e^{i\zeta}\rangle_{A}|\alpha_{0}e^{-i\zeta}\rangle_{B}\,d\zeta\,.\label{eq:pair-coherent_state}
\end{align}
Here $\mathcal{N}$ is a normalization constant, $m$ is the particle/
photon number difference between modes $A$ and $B$, and $\alpha_{0}$
is the amplitude of the coherent state. Gilchrist et al \citep{gilchrist1999contradiction}
also considered a squeezed entangled cat state. The entangled cat
state is defined as 
\begin{align}
|cat\rangle & =N\left(|\alpha_{0}\rangle_{A}|\beta_{0}\rangle_{B}+e^{i\theta}|-\alpha_{0}\rangle_{A}|-\beta_{0}\rangle_{B}\right)\label{eq:cat-state}
\end{align}
where $N$ is the normalization constant and $\alpha_{0}$, $\beta_{0}$
are the amplitudes of the coherent states for each mode. The squeezed
two-mode cat state is generated under the evolution of the squeezing
interaction Hamiltonian $H_{I}$ (\ref{eq:ham-da-two-mode}) such
that the density operator of the system is $\rho_{sc}=exp\left(-\frac{iH_{I}t}{\hbar}\right)|cat\rangle\langle cat|exp\left(\frac{iH_{I}t}{\hbar}\right)\,$.
Rotated quadrature phase amplitudes are defined, as $\hat{X}_{\theta}=\hat{X}\cos\theta+\hat{P}\sin\theta$,
and binning was used, to classify the outcomes as either $+1$ or
$-1$ depending on the sign of $\hat{X}_{\theta}$. Similar to the
approach of \citep{Leonhardt:1995aa}, violations are then evident
as violations of the Clauser-Horne-Shimomy-Holt (CHSH) or Clauser-Horne
(CH) Bell inequalities. It is interesting that the bin can distinguish
the dead or alive aspect of the cat-like state, determined by the
sign of the amplitude. In this case, however, in the limit where $\alpha$
is large, the violations vanish or become increasingly small.

Recent work by Kumar, Saxena and Arvind shows how the CH inequalities
based on polarization can be used to confirm the nonlocality of both
the entangled cat state $|cat\rangle$ and the pair coherent state
\citep{Kumar2021Continuous}. These authors considered modes of definite
polarization, where intensity correlations are detected after passing
through polarizers, after some mode transformations. The outcomes
are binned according to whether photons are detected at the output
location, or not. Violations of the CH inequality are found possible,
for coherent amplitudes of order $\sqrt{2}$, where the two coherent
states ($|\alpha_{0}\rangle$ and $|-\alpha_{0}\rangle$) for each
mode are distinctly separated. \textcolor{teal}{}\textcolor{blue}{}

Banaszek and Wodkiewicz considered the two-mode entangled cat states
$|cat\rangle$ given by (\ref{eq:cat-state}) and showed how to achieve
Bell violations for these states, in a way that does not decay for
larger $\alpha$ and $\beta$, using phase space distributions \citep{Banaszek:1999b_Nonlocality}.
The work was extended by Jeong et al. \citep{jeong2003quantum} and
Milman et al \citep{Milman_EPJD2005}.\textcolor{blue}{{} }Banaszek
and Wodkiewicz demonstrated how the phase-space $Q$ and Wigner functions
can be used to infer nonlocal correlations. These quasi-probability
distributions are functions of continuous variables, defined by a
complex amplitude for each mode. The approach considered a two-mode
cat-state that is displaced before measurements. If detectors detect
no photons the event is assigned $0$, and $1$ otherwise. The joint
probability of no-count events in both detectors is $Q_{AB}\left(\alpha,\beta\right)$
where, $Q_{AB}(\alpha,\beta)=\langle\hat{Q}_{A}\left(\alpha\right)\hat{Q}_{B}\left(\beta\right)\rangle$
and $\hat{Q}_{A}\left(\alpha\right)=\hat{D}\left(\alpha\right)|0\rangle\langle0|\hat{D}^{\dagger}\left(\alpha\right)=|\alpha\rangle\langle\alpha|$.
Here, $\hat{D}\left(\alpha\right)$ and $\hat{D}\left(\beta\right)$
are the standard displacement operators with amplitude $\alpha$ and
$\beta$ for modes labelled $A$ and $B$ respectively. The Clauser-Horne
(CH) Bell inequality can be written as $-1\leq CH\leq0$, where 
\begin{align}
CH & \equiv Q_{AB}\left(0,0\right)+Q_{AB}\left(\alpha,0\right)+Q_{AB}\left(0,\beta\right)-Q_{AB}\left(\alpha,\beta\right)-Q_{A}\left(0\right)-Q_{B}\left(0\right)\,.\label{eq:CH-Q}
\end{align}
The realization that $Q$ is the Husimi $Q$-function, defined as
$Q(\alpha)=\langle\alpha|\rho|\alpha\rangle/\pi$ \citep{Husimi1940},
thus implies that $Q$-functions can display nonlocal quantum correlations.
The authors then considered another measurement where the number of
photons detected can be resolved. A parity value $P$ of $+1$ ($-1$)
is assigned to even (odd) number of detected photons. The operators
corresponding to these measurements are $\hat{\Pi}^{(+)}\left(\alpha\right)=\hat{D}\left(\alpha\right)\sum_{k=0}^{\infty}|2k\rangle\langle2k|\hat{D}^{\dagger}\left(\alpha\right)$
and $\hat{\Pi}^{(-)}\left(\alpha\right)=\hat{D}\left(\alpha\right)\sum_{k=0}^{\infty}|2k+1\rangle\langle2k+1|\hat{D}^{\dagger}\left(\alpha\right).$
The correlation function is shown to be 
\begin{align}
\hat{\Pi}_{AB}\left(\alpha,\beta\right) & =(\hat{\Pi}_{A}^{(+)}\left(\alpha\right)-\hat{\Pi}_{A}^{(-)}\left(\alpha\right))\otimes(\hat{\Pi}_{B}^{(+)}\left(\beta\right)-\hat{\Pi}_{B}^{(-)}\left(\beta\right))\nonumber \\
 & =\hat{D}_{A}\left(\alpha\right)\hat{D}_{B}\left(\beta\right)\left(-1\right)^{\hat{n}_{a}+\hat{n}_{b}}\hat{D}_{A}^{\dagger}\left(\alpha\right)\hat{D}_{B}^{\dagger}\left(\beta\right)\,\label{eq:corr-bw}
\end{align}
where $\langle\hat{\Pi}_{AB}(\alpha,\beta)\rangle=\pi^{2}W(\alpha,\beta)/4$
corresponds to a scaled two-mode Wigner function $W(\alpha,\beta)$
\citep{Wigner1932Quantum}. The $\langle\hat{\Pi}_{AB}(\alpha,\beta)\rangle$
corresponds to the average parity product $\hat{P}_{A}\hat{P}_{B}$
after the displacements. It follows that the outcomes $\Pi_{AB}(\alpha,\beta)$
are bounded by $\pm1$, implying the Bell inequality $|B|\leq2$ where
\begin{align}
B & \equiv\langle\hat{\Pi}_{AB}\left(\alpha',\beta'\right)\rangle+\langle\hat{\Pi}_{AB}\left(\alpha,\beta'\right)\rangle+\langle\hat{\Pi}_{AB}\left(\alpha',\beta\right)\rangle-\langle\hat{\Pi}_{AB}\left(\alpha,\beta\right)\rangle\,.\label{eq:Bell-BW}
\end{align}
Thus, Wigner functions can also be used to demonstrate nonlocal correlations.
Violations are predicted for the entangled cat state, which for optimally
selected values of the state and of the displacements $\alpha$ and
$\beta$ allow the maximum violation $B=2\sqrt{2}$ for $\alpha_{0},\beta_{0}\rightarrow\infty$
\citep{Milman_EPJD2005}.

Milman et al \citep{Milman_EPJD2005} proposed the generation of the
two-mode cat states (\ref{eq:cat-state}) in a cavity QED setting.
The setting consists of a Rydberg atom and two superconducting cavities.
Firstly, the atom is prepared in a superposition state $\left(|g\rangle+|e\rangle\right)/\sqrt{2}$,
which is then passed through two cavities that are both in a coherent
state $|\lambda\rangle$. The atom interacts with the cavity field
in such a way that a phase shift $\phi$ is induced, depending on
the state of the atom: $|e\rangle|\lambda\rangle\rightarrow e^{i\phi}|e\rangle|\lambda e^{i\phi}\rangle$
and $|g\rangle|\lambda\rangle\rightarrow e^{-i\phi}|g\rangle|\lambda e^{-i\phi}\rangle$.
By choosing the interaction time and detuning, the phase $\phi$ can
be fixed and they focused on the case $\phi=\pi/2$, so that the atom-cavity
state after the interaction is given by $\left(-|e\rangle|\alpha_{0}\rangle|\alpha_{0}\rangle+|g\rangle|-\alpha_{0}\rangle|-\alpha_{0}\rangle\right)/\sqrt{2}$
with $\alpha_{0}=i\lambda$. Finally, a $\pi/2$ pulse is applied
to the atom, transforming $|e\rangle\rightarrow\left(-|e\rangle+|g\rangle\right)/\sqrt{2}$
and $|g\rangle\rightarrow\left(|e\rangle+|g\rangle\right)/\sqrt{2}$.
This brings the atom-cavity state to be: 
\begin{align}
|\phi\rangle & =\frac{1}{2}\Bigl(|e\rangle\left(|\alpha_{0}\rangle|\alpha_{0}\rangle+|-\alpha_{0}\rangle|-\alpha_{0}\rangle\right)-|g\rangle\left(|\alpha_{0}\rangle|\alpha_{0}\rangle-|-\alpha_{0}\rangle|-\alpha_{0}\rangle\right)\Bigl)\,.\label{eq:state}
\end{align}
A detection of the state $|e\rangle$ prepares the two-mode cat state
$|\psi^{+}\rangle\propto|\alpha_{0}\rangle|\alpha_{0}\rangle+|-\alpha_{0}\rangle|-\alpha_{0}\rangle$
with an even photon number (even parity), while a detection of the
state $|g\rangle$ prepares the two-mode cat state $|\phi^{-}\rangle\propto|\alpha_{0}\rangle|\alpha_{0}\rangle-|-\alpha_{0}\rangle|-\alpha_{0}\rangle$
with an odd photon number (odd parity). In order to measure the correlation
function $\langle\hat{\Pi}_{AB}\left(\alpha,\beta\right)\rangle=Tr\bigl[\rho D_{A}\left(\alpha\right)D_{B}\left(\beta\right)\left(-1\right)^{\hat{n}_{a}+\hat{n}_{b}}D_{A}^{\dagger}\left(\alpha\right)D_{B}^{\dagger}\left(\beta\right)\bigr]$,
the cavity states $A$ and $B$ are displaced by amplitudes $-\alpha$
and $-\beta$ respectively. A second atom is then passed through both
cavities, where the subsequent measurement on the atom gives the measurement
outcome of $\langle\hat{\Pi}_{AB}\left(\alpha,\beta\right)\rangle$.

Wang et al \citep{Wang_Science2016} have experimentally created a
two-mode cat state for microwave fields. The two-mode cat state generation
process is based on the modified proposal by Leghtas et al \citep{Leghtas_PRA2013},
who showed how to map an arbitrary two-mode qubit state onto a two-mode
cat-state superposition in a cavity. As above, Wang et al prepared
the atom in a superposition state, with the two cavities being in
the vacuum state. Instead of passing the atom through two cavities
that are both in a coherent state $|2\lambda\rangle$, the cavity
states are only transformed to coherent states if the atom is in the
$|g\rangle$ state, while both cavities remain in the vacuum state
if the atom is in the $|e\rangle$ state. The state of the whole system
up to this point is given by $\left(|g\rangle|2\lambda\rangle|2\lambda\rangle+|e\rangle|0\rangle|0\rangle\right)/\sqrt{2}$.
Another conditional interaction puts the atom in the $|e\rangle$
state back to the $|g\rangle$ state, while leaving the $|g\rangle$
state unchanged, thus disentangling the atomic state from the cavity
states. Finally, a displacement operator with a displacement amplitude
$-\lambda$ is applied to both cavities, which prepares the whole
system in the cat state, as 
\begin{equation}
|g\rangle\left(|\lambda\rangle|\lambda\rangle+|-\lambda\rangle|-\lambda\rangle\right)/\sqrt{2}\,.\label{eq:wang}
\end{equation}
Following these proposals, Wang et al \citep{Wang_Science2016} experimentally
measured the joint parity and observed a violation of a Bell inequality,
based on the observation of the four points of the Wigner function,
as in (\ref{eq:Bell-BW}).

We note that the measurements involved in the above schemes require
to distinguish between a $0$ or $1$ photon count, or else to determine
the parity of the cat state, which refers to whether the cat state
has an odd or even photon number. These distinctions are not macroscopic.
If one infers the correlations using the phase-space analysis by measuring
$W(\alpha,\beta)$., the measurements do not distinguish between amplitudes
(say, $\alpha_{0}$ and $-\alpha_{0}$ for the cat state (\ref{eq:cat-state}))
that are macroscopically separated in phase space. While one can infer
failure of LHV theories for a macroscopic state with $\alpha_{0}$
large, it could be argued that this is not a macroscopic correlation
in the sense of a fully macroscopic measurement.

\selectlanguage{australian}%
Further different approaches have been taken. Ketterer et al \citep{Ketterer_PRA2015}\foreignlanguage{english}{
considered testing the CHSH inequality without a prior choice of binning
procedure and with no prior knowledge of the Hilbert space dimension
of the system, following from work by Horodecki \citep{horodecki2003mean}.
They proposed modular variables to turn unbounded observables into
bounded ones and then to check for violation of the CHSH inequality.
Arora and Asadian also propose a macroscopic Bell scheme by transforming
an unbounded momentum operator $\hat{p}$ into a bounded one \citep{Arora_PRA2015}.
They consider the observable 
\begin{align}
\hat{X} & \equiv\cos\left(\frac{\hat{p}L}{\hbar}\right)\label{eq:AAcos}
\end{align}
so that the value is bounded by $\pm1$, and define two states $|\psi_{0}\rangle$and
$|\psi_{1}\rangle$, which are superposition of position states. The
position superposition state can be implemented with a grating with
$N=2M$ slits. A linear combination of these two states can be formed
where $|\psi_{\pm}\rangle=\frac{1}{\sqrt{2}}\left(|\psi_{0}\rangle\pm|\psi_{1}\rangle\right)$.
These states are such that the bounded operator $\hat{X}$ have the
following expectation values: $\langle\psi_{\pm}|\hat{X}|\psi_{\pm}\rangle=\pm\frac{N-1}{N}$.
In other words, depending on the state $|\psi_{+}\rangle$ or $|\psi_{-}\rangle$,
the expectation value of $\hat{X}$ either returns a positive or negative
value, for large $N$, which are justified as macroscopically distinguishable.
In order to implement different measurement settings as required in
CHSH inequality, they considered a unitary operator $U$ such that
$U\left(\phi\right)|\psi_{0}\rangle=e^{i\phi/2}|\psi_{0}\rangle$
and $U\left(\phi\right)|\psi_{1}\rangle=e^{-i\phi/2}|\psi_{1}\rangle$.
For a state $|\Psi\rangle=(|\psi_{+}\rangle_{1}|\psi_{-}\rangle_{2}-|\psi_{-}\rangle_{1}|\psi_{+}\rangle_{2})/2$,
the correlation function is found to be 
\begin{equation}
\langle\hat{X}\left(\phi\right)\otimes\hat{X}\left(\theta\right)\rangle=-\left(\frac{N-1}{N}\right)^{2}\cos\left(\phi-\theta\right)\,,\label{eq:aa-xx}
\end{equation}
giving the Bell parameter $|\langle\hat{B}\rangle|=\left(\frac{N-1}{N}\right)^{2}2\sqrt{2}$.
The authors suggest a physical implementation based on entangled beams
in polarization. A similar scheme has been proposed by Huang et al
\citep{huang2020re}.}

\selectlanguage{english}%
Recent work by Thearle et al \citep{Thearle_PRL2018} infers an experimental
violation of a Bell inequality using intensities and CV measurements.
They justify that the violations are continuous variable, since in
their scheme the photon correlation intensity functions are measured
via continuous-variable quadrature-phase amplitude measurements. This
work is based on the earlier proposals, which evaluate Bell correlations
in terms of normalized intensity correlations \citep{Huntington2001_CV,Ralph2000_Proposal}.
Reid and Walls originally considered four-mode states generated from
parametric down conversion (\ref{eq:four-mode-ham}), with two polarization
modes $a_{\pm}$ and $b_{\pm}$ at the respective sites $A$ and $B$.
They considered bounded observables at each site according to a normalized
intensity difference \citep{Reid1986}. At $A$ the normalized intensity
is defined as $(I_{+}^{(A)}(\theta)-I_{-}^{(A)}(\theta))/(I_{+}^{(A)}(\theta)+I_{-}^{(A)}(\theta))$,
where $I_{\pm}^{(A)}(\theta)$ are the outcomes for $\hat{I}_{\pm}^{(A)}(\theta)=\hat{c}_{\pm}^{\dagger}\hat{c}_{\pm}$
given in (\ref{eq:rotc}). Similar intensity differences are defined
at $B$ in terms of the angle $\phi$. The normalization motivates
the application of CHSH-type Bell inequalities in terms of intensity
correlations, where more than a single photon might be detected. Thearle
et al. generated a $4$-mode entangled state from optical parametric
oscillation (OPO), and the output is sent to two parties. Each party
is free to apply the mixer that allows the polarization along any
axis to be measured. Explicitly, the correlation function of interest
has the form 
\begin{equation}
R^{ij}\left(\theta,\phi\right)=\langle R_{A}^{i}\left(\theta\right)R_{B}^{j}\left(\phi\right)\rangle=\langle\hat{a}_{i}^{\dagger}\hat{a}_{i}\hat{b}_{j}^{\dagger}\hat{b}_{j}\rangle\label{eq:int}
\end{equation}
where $\theta,\phi$ are the measurement settings for party $A$ and
$B$ respectively, while $i,j\in\left\{ +,-\right\} $ characterized
the detector that detects the mode $\pm$. In the scheme, the operators
$\hat{a}_{i}^{\dagger}\hat{a}_{i}$ are measured in terms of quadrature
phase amplitude observables, which required additional assumptions
about vacuum corrections.\textcolor{blue}{{} }It is important to
measure the vacuum intensity and this is carried out in the experiment
by randomly swapping between measuring the quadratures and vacuum
intensity. They inferred a Bell violation with $|\langle B\rangle|=2.31>2$.

The work of Zukowski, Wiesniak and Laskowski \citep{zukowski2016bell}
demonstrates how to obtain violation of Bell inequalities for the
four-mode intensity outputs of the parametric down conversion process
given by (\ref{eq:four-mode-ham}), directly. In a series of papers
\citep{zukowski2016entanglement,zukowski2016bell,zukowski2017normalized},
Zukowski et al rigorously formalized the renormalisation approach,
by giving a careful treatment of the vacuum state, for which the normalized
intensities become undefined. This led to new predictions of violation
of Bell inequalities at higher intensities. \textcolor{teal}{}\textcolor{blue}{}
A similar renormalisation procedure also using the Moore-Penrose operator
was developed by He et al \citep{he2012entanglement} to justify the
use of EPR entanglement criteria involving normalized atomic detection
counts, which were applied to give predictions for EPR correlations
in a Bose-Einstein condensate \citep{he2011einstein,he2012entanglement}.

Lee and Jaksch (2009) extended their approach of obtaining an optimal
Bell inequality to continuous variable systems. They considered the
two-mode squeezed state $|TMSS\rangle=\text{sech}r\sum_{n=0}^{\infty}\tanh^{n}r|n\rangle|n\rangle$.
The measurement basis is obtained by carrying out quantum Fourier
transform on the basis state $|n\rangle$. This gives the phase state:
$|\theta,k\rangle=(1/\sqrt{s+1})\sum_{n=0}^{s}exp(in\theta_{k})|n\rangle$.
The correlation operator $\hat{E}\left(\theta,\phi\right)=\Pi\left(\theta\right)\otimes\Pi\left(\phi\right)$,
where $\Pi\left(\theta\right)=\sum_{k=0}^{s}\left(-1\right)^{k}|\theta,k\rangle\langle\theta,k|$.
They find the Bell correlation function $\mathcal{B}_{QM}$ arbitrarily
close to $2\sqrt{2}$ when the squeezing strength $r\rightarrow\infty$.

Many of the above methods rely on a binning, or renormalisation, of
outcomes in order to obtain Bell violations for continuous variables
measurements. This provides an observable with outcomes bounded by
$\pm1$, so that one may take advantage of the CHSH or CH Bell inequalities.
Cavalcanti, Foster, Reid and Drummond (CFRD) showed that the falsification
of local hidden variable theories could be obtained \emph{without}
this binning, directly from the continuous variable spectrum \citep{cavalcanti2007bell}.
These authors adapted the approach of Mermin \citep{mermin1990extreme}
for continuous variable outcomes. They considered the function 
\begin{equation}
F_{j}^{\pm}=X_{j}\pm iY_{j}\label{eq:f}
\end{equation}
of measurement outcomes $X_{j},Y_{j}$ at each site $j$, where the
quantum observable for these measurements is $\hat{X_{j}}$, $\hat{Y_{j}}$
respectively. For a local hidden variable (LHV) theory, it is always
true that $|\langle\prod_{j=1}^{N}F_{j}^{s_{j}}\rangle|^{2}\leq\int d\lambda P(\lambda)\prod_{j=1}^{N}|\langle F_{j}^{s_{j}}\rangle_{\lambda}|^{2}.$
Since $|\langle F_{j}^{\pm}\rangle_{\lambda}|^{2}=\langle X_{j}\rangle_{\lambda}^{2}+\langle Y_{j}\rangle_{\lambda}^{2}$,
it follows from the non-negativity of variances that for any LHV state,
$|\langle F_{j}^{\pm}\rangle_{\lambda}|^{2}\leq\langle X_{j}^{2}\rangle_{\lambda}+\langle Y_{j}^{2}\rangle_{\lambda}$.
This leads to the bound 
\begin{equation}
|\langle\prod_{j=1}^{N}F_{j}^{s_{j}}\rangle|\leq\left\langle \prod_{j=1}^{N}(X_{j}^{2}+Y_{j}^{2})\right\rangle ^{1/2}\,,\label{eq:main_ineq}
\end{equation}
giving the CFRD Bell inequality.

Cavalcanti et al \citep{cavalcanti2011unified} pointed out that if
one constrains the local hidden variables for some of the sites, say
labelled $r=1,..,T$, to be consistent with quantum predictions, then
there is a further restriction given by the uncertainty relation.
They consider quantum uncertainty relations of the form $(\Delta\hat{X}_{j})^{2}+(\Delta\hat{Y}_{j})^{2}\geq C_{j}$,
where $C_{j}$ may depend on the operators associated with $x_{j}$
and $y_{j}$. This will imply for \emph{quantum }states that $|\langle F_{j}^{\pm}\rangle_{\lambda}|^{2}\leq\langle X_{j}^{2}\rangle_{\lambda}+\langle Y_{j}^{2}\rangle_{\lambda}-C_{j}$,
leading to the unified nonlocality inequalities 
\begin{equation}
|\langle\prod_{j=1}^{N}F_{j}^{s_{j}}\rangle|\leq\left\langle \prod_{j=1}^{T}(X_{j}^{2}+Y_{j}^{2}-C_{j})\prod_{j=T+1}^{N}(X_{j}^{2}+Y_{j}^{2})\right\rangle ^{1/2}.\label{eq:unified}
\end{equation}
If one takes observables $X$ and $Y$ to be spin, as in Mermin's
original approach given by (\ref{eq:mermin-spin}), then the inequalities
\ref{eq:main_ineq}) becomes the MABK inequalities, if $T=0$. For
$T=N$, a set of entanglement criteria is derived, corresponding to
the criteria of Roy \citep{roy2005multipartiteseparability}.

The original work of CFRD considered $X$ and $Y$ to be continuous-variable
outcomes of quadrature phase amplitudes, $X$ and $P$ \citep{cavalcanti2007bell}.
Symbolizing $\hat{a}^{+}=\hat{a}^{\dagger}$ and $\hat{a}^{-}=\hat{a}$,
the nonlocality inequalities (\ref{eq:unified}) will be violated
when 
\begin{equation}
|\langle\hat{a}_{1}^{s_{1}}\cdots\hat{a}_{N}^{s_{N}}\rangle|>\langle\prod_{j=1}^{T}\hat{n}_{j}\prod_{j=T+1}^{N}(\hat{n}_{j}+1/2)\rangle^{1/2}\,.\label{eq:ataineq}
\end{equation}
For $T=0$, these are the CV Bell inequalities derived by CFRD \citep{cavalcanti2007bell}.
These authors showed that certain GHZ states with $N>9$ will violate
the inequalities. If $T=N$, one arrives at inequalities for entanglement
given by Hillery and Zubairy \citep{Hillery2006Entanglement}, whereas
for intermediate $T$, the inequalities are steering inequalities
\citep{cavalcanti2011unified}.

Further work by Salles et al \citep{salles2010bell} and He et al
\citep{he2010bell} studied the CFRD Bell inequalities. Salles et
al showed violations of generalizations of the inequalities for three
settings using spin qubits and GHZ states, for $N\geq3$. The approach
was adapted by Shchukin and Vogel \citep{shchukin2008quaternions}
who derived multi-setting inequalities closely related to the algebra
of quaternions and octonions. Violations were predicted for GHZ states
with $N\geq3$ \citep{Kiesewetter2015Violations}. The unified approach
to deriving Bell and steering inequalities was applied to spin qubits
by Cavalcanti et al \citep{cavalcanti2011unified}, who derived MABK
steering inequalities. Criteria were also developed by Jebaratnam
et al \citep{Jebaratnam_PRA2018}. Li et al have experimentally investigated
the violation of Mermin steering inequalities for qubits where $N=3$
\citep{Li:2019aa}.

\section{Quantum correlations of cat states}

The Schr\"odinger cat gedanken experiment analyses the correlated
state of type \citep{Schrodinger:1935_Naturwiss,Frowis_RMP2018} 
\begin{equation}
|\psi\rangle_{cat}=\frac{1}{\sqrt{2}}(|\alpha\rangle_{A}|\uparrow\rangle_{B}+e^{i\theta}|-\alpha\rangle_{A}|\downarrow\rangle_{B})\label{eq:cat}
\end{equation}
where, for large $\alpha$, a microscopic system (the spin) is coupled
to a macroscopic system (modeled as a field mode). Here, $|\alpha\rangle$
is a coherent state. Where $A$ and $B$ refer to different systems,
this state is entangled. In Schr\"odinger's original argument, a
microscopic system $B$ is prepared in superposition state, and is
then coupled to a macroscopic system $A$ which represents the measurement
device. In the above analogy, system $B$ is originally in the superposition
$(|\uparrow\rangle+|\downarrow\rangle)/\sqrt{2}$ and system $A$
is prepared in a coherent state $|\alpha\rangle$, where $\alpha$
is large. A coupling $H$ between the systems creates the state after
an interaction time. This models the measurement procedure, since
the result of a measurement of the coherent-state amplitude (whether
$\alpha$ or $-\alpha$) is correlated with the result of the spin
measurement $\hat{\sigma}_{z}$ on $B$.

Correlated cat states such as (\ref{eq:cat}) have been generated
experimentally. The experiment of Monroe et al \citep{Monroe:1996aa}
created a cat state similar to (\ref{eq:cat}) where the correlation
is between the internal degree of freedom of an atom and its spatial
degree of freedom. The atom is trapped in a harmonic potential and
is prepared in the state $|\downarrow,0\rangle$. Here, $|\downarrow\rangle$
is the internal state and $|0\rangle$ is the motional ground state
of the atom. A $\pi/2$ pulse puts the atom in a superposition and
a displacement beam excites the motional state to a coherent state
$|\alpha e^{-i\phi/2}\rangle$, conditioned on the internal state
being $|\uparrow\rangle$. This brings the state to $\left(|\downarrow,0\rangle+|\uparrow,\alpha e^{-i\phi/2}\rangle\right)/\sqrt{2}$.
After a $\pi$ pulse and further displacement, the cat state $\left(|\uparrow,\alpha e^{i\phi/2}\rangle+|\downarrow,\alpha e^{-i\phi/2}\rangle\right)/\sqrt{2}$
is created.

On the other hand, the experiment of Brune et al \citep{Brune_PRL1996}
creates a cat state of type (\ref{eq:cat}) by entangling Rydberg
atoms one at a time with a microwave field. The Rydberg rubidium atom
is prepared in a superposition of two atomic states $e$ and $g$,
and then passed through a high $Q$ microwave cavity prepared in a
coherent state $|\alpha\rangle$ of a few photons. A dispersive interaction
between the atom and cavity generates the entangled cat state. The
superposition was inferred by observation of coherence using two-atom
correlations, based on a proposal by Davidovich et al \citep{Davidovich_PRA1996}.
The experiment modeled the coupling of a system to a meter, in which
the decay of the coherence was found to increase with the separation
$\phi$ in phase space of the two coherent states, in agreement with
predictions \citep{Zurek_PRD1981,Zurek_PRD1982,Caldeira:1983aa,Walls_PRA1985,Yurke_PRL1986,Brune_PRA1992}.
The experiment thus models the rapid decay of a macroscopic pointer
from the macroscopic superposition (\ref{eq:cat}) of two distinct
coherent states into a classical mixture of the two states.

The entangled cat state (\ref{eq:cat}) is paradoxical, since the
concept of realism is challenged by the correlations. The correlations
are readily shown to be of the EPR-type. We find it useful to specify
the entangled cat state 
\begin{equation}
|\psi\rangle_{cat}=\frac{1}{\sqrt{2}}(|\alpha\rangle_{A}|\uparrow\rangle_{B}+i|-\alpha\rangle_{A}|\downarrow\rangle_{B})\,.\label{eq:cat-1}
\end{equation}
The measurement of the spin $\hat{S}_{z}^{(B)}$ of the system $B$
indicates the sign of the amplitude $\hat{X}_{A}$ of the field mode
$A$ (taking $\alpha$ to be real). If we consider the inference of
$X_{A}$ based on the measurement of $\hat{S}_{z}^{(B)}$, then as
$\alpha\rightarrow\infty$, the system $A$ is projected into either
$|\alpha\rangle$ or $|-\alpha\rangle$. Hence $\Delta_{inf}\hat{X}=1$.
This can be verified by a complete calculation of the conditional
distributions for $X$ conditioned on the outcome $\pm1$ for spin
at $B$. On the other hand, if one measures $\hat{S}_{x}^{(B)}$ at
$B$, then we consider 
\begin{eqnarray}
|\psi\rangle_{cat} & = & (|\alpha\rangle+i|-\alpha\rangle)|\uparrow\rangle_{x}+(|\alpha\rangle-i|-\alpha\rangle)|\downarrow\rangle_{x}\label{eq:rotcat}
\end{eqnarray}
which implies the conditional variance for $\hat{P}$ given an outcome
$\pm1$ for $\hat{S}_{x}^{(B)}$. The result for the spin $\hat{S}_{x}$
projects the system into one of the cat-states $(|\alpha\rangle\pm i|-\alpha\rangle)/\sqrt{2}$
considered by Yurke and Stoler \citep{Yurke_PRL1986}, these states
having the same variance $(\Delta\hat{P}_{A})^{2}$ in $\hat{P}_{A}$.
This gives \citep{Reid:2017aa,Cavalcanti_PRA2008,rosales2015decoherence}
\begin{equation}
\Delta_{inf}^{2}\hat{P}=(\Delta\hat{P}_{A})^{2}=1-2\alpha^{2}e^{-4\alpha^{2}}\,.\label{eq:infp}
\end{equation}
Hence, the EPR-paradox and EPR steering condition (\ref{eq:epr-crit})
(for steering of $A$) is satisfied, with $\epsilon<1$. If one assumes
an EPR-type local realism, which is the assumption that the system
is described by a local realistic theory, then the localized state
of the system cannot be directly modeled as a quantum state.

An EPR-steering paradox has been explained for other two-mode correlated
cat states, such as the NOON states given by \citep{Dowling:2008aa}
\begin{equation}
\frac{1}{\sqrt{2}}(|N\rangle_{A}|0\rangle_{B}+|0\rangle_{A}|N\rangle_{B})\,.\label{eq:noon}
\end{equation}
Here $|N\rangle_{I}$ is the number state for the mode labelled $I$.
These states have been generated experimentally \citep{Walther:2004aa,Mitchell:2004aa,Slussarenko:2017aa,Afek879}\foreignlanguage{australian}{.
EPR correlations can be detected using an EPR criterion of the type
(\ref{eq:epr-crit}) but based on number and phase measurements, as
shown in }\citep{Teh_PRA2016}\foreignlanguage{australian}{. Similar
NOON-type states are generated in macroscopic or multimode versions
of the Hong-Ou-Mandel effect \citep{walborn2003multimode,Stobinska_PRA2012,Iskhakov_NJP2013,Yu_Spasibko2014Interference}.}\textcolor{red}{{}
}EPR correlations are also evident in the correlated cat states given
by (\ref{eq:cat-state}) \citep{thenabadu2020bipartite}.

The important question of whether one can illustrate failure of local
realism directly for an entangled cat state was answered by the analyses
that gave violations of Bell LHV theories for the state (\ref{eq:cat-state}).
These analyses are summarized in the previous section. Relevant to
the two-mode cat state (\ref{eq:cat-state}) is the analysis of Banaszek
and Wodkiewicz, which allows violations of Bell inequalities that
do not decay with $\alpha$. This shows that the cat states are not
compatible with the predictions of any LHV model, indicating that
the premise of local realism/ local causality underlying the EPR argument
is itself invalid. A similar result applies to the Schr\"odinger
cat state (\ref{eq:cat}).\textcolor{red}{{} }Wodkiewicz \citep{Wodkiewicz:2000aa}
investigated the nonlocal Bell correlations for the hybrid cat state
(\ref{eq:cat}), based on the Banaszek and Wodkiewicz approach. The
correlation is expressed in terms of the Wigner function and shows
the entanglement and interference of the spin and the cat system.
Violations of the Banaszek-Wodkiewicz Bell inequality for the hybrid
cat state are obtained for certain spin orientations, showing the
nonlocality of this state.\textcolor{blue}{}\textcolor{red}{}

Bell correlations for the NOON states (\ref{eq:noon}) were demonstrated
by Wildfeuer, Lund and Dowling \citep{Wildfeuer_PRA2007}. They first
applied the Bell scheme of Banaszek and Wodkiewicz to NOON states,
i.e. to coherently displace the state before detection. They found
a maximal violation of the CH-Bell inequality for $N=1$, with the
degree of violation reducing with $N$, making experimental observation
difficult for $N\geq3$. Similar conclusions were reached in the parity
measurements based on the Wigner function. In this case, the CHSH
inequality is violated for $N=1$ but not for $N>1$. However, for
different Bell inequalities, the authors found Bell violations for
a NOON state that are independent of $N$. The argument is that the
CH and CHSH inequalities involve only four joint probability distributions
and are not sufficiently sensitive to detect the nonlocal correlations
in a NOON state. Bell inequalities with more joint probability distributions
are needed. They considered four Bell-type inequalities derived by
Janssens et al. \citep{Janssens:2004aa} that have six joint probability
distributions. Written in terms of the Q-functions, these inequalities
are given by
\begin{align}
J_{1} & =Q\left(\alpha\right)+Q\left(\beta\right)+Q\left(\gamma\right)+Q\left(\delta\right)-Q\left(\alpha,\beta\right)-Q\left(\alpha,\gamma\right)-Q\left(\alpha,\delta\right)-Q\left(\beta,\gamma\right)-Q\left(\beta,\delta\right)-Q\left(\gamma,\delta\right)\leq1\nonumber \\
J_{2} & =2Q\left(\alpha\right)+2Q\left(\beta\right)+2Q\left(\gamma\right)+2Q\left(\delta\right)-Q\left(\alpha,\beta\right)-Q\left(\alpha,\gamma\right)-Q\left(\alpha,\delta\right)-Q\left(\beta,\gamma\right)-Q\left(\beta,\delta\right)-Q\left(\gamma,\delta\right)\leq3\nonumber \\
J_{3} & =Q\left(\alpha\right)-Q\left(\alpha,\beta\right)-Q\left(\alpha,\gamma\right)-Q\left(\alpha,\delta\right)+Q\left(\beta,\gamma\right)+Q\left(\beta,\delta\right)+Q\left(\gamma,\delta\right)\geq0\nonumber \\
J_{4} & =Q\left(\alpha\right)+Q\left(\beta\right)+Q\left(\gamma\right)-2Q\left(\delta\right)-Q\left(\alpha,\beta\right)-Q\left(\alpha,\gamma\right)+Q\left(\alpha,\delta\right)-Q\left(\beta,\gamma\right)+Q\left(\beta,\delta\right)+Q\left(\gamma,\delta\right)\leq1\,.\label{eq:q-noon}
\end{align}
The authors numerically determined the parameters $\alpha$, $\beta$,
$\gamma$, $\delta$ that optimally violate each of these inequalities.
$J_{1}$, $J_{2}$, and $J_{4}$ are found to be violated by a constant
amount for all $N$, while $J_{3}$ shows a decrease in violation
as a function of $N$.

\section{Coarse-graining and decoherence}

The argument could be put forward that the quantum correlations considered
so far are not genuinely macroscopic because, although describing
macroscopic systems, they require measurements that are microscopically
or mesoscopically resolving i.e. they require measurements that distinguish
at some point between states that are only microscopically or mesoscopically
distinct.\textcolor{red}{{} }In fact, it was originally an open question
whether, in a Bell test with $N$ particles at each site, violations
were possible with the loss of information about the result of just
\emph{one} of the particles. For some Bell inequalities and states,
the answer was negative, as for the MABK inequalities with GHZ states
(refer Section IV). More generally, this has been shown to not be
the case.

Peres pointed out that additional noise in $x$ and $p$ of order
the quantum noise level ($\sim\hbar$) would damp out any Bell violations
arising from the $x$ and $p$ measurements \citep{peres1995quantum}.
This is apparent because the addition of such noise transforms any
negative Wigner function $W(x_{i},p_{i})$ (for two modes $i=1,2$)
into a positive distribution, given by the $Q$ function \citep{Husimi1940}.
There then exists a local hidden variable theory which models the
measured $x_{i}$ and $p_{i}$ moments as arising from a joint probability
distribution, $W$. From this point of view, the measurements must
be highly resolved if one is to observe Bell correlations using continuous
variable (CV) $\hat{x}$ and $\hat{p}$ measurements. There will be
a threshold value of noise, as measured by a standard deviation $\sigma_{t}$
(of order the quantum limit), beyond which Bell violations will not
be possible.

CV Bell violations can be observed in the presence of significant
coarse-graining, however, if measured in an \emph{absolute} sense.
As an example, examination of the two-mode Schwinger operators given
by (\ref{eq:schwinger}) indicates that Bell violations are possible
for large absolute noise values in the spins $\hat{S}_{X}$ and $\hat{S}_{Y}$.
These observables are measured when continuous variable (CV) measurements
are carried out by optical or atomic homodyne. Consider a CV measurement,
where $\sigma_{t}$ is the threshold value of noise in $X$ and $P$
beyond which there can be no violation of a CV Bell inequality. In
terms of the Schwinger operators, this threshold corresponds to an
amplified noise value of $\sim E\sigma_{t}$. The $E\sigma_{t}$ reflects
the allowed noise in the two-mode Schwinger photon number difference
given by (\ref{eq:schwinger}), but is nonetheless \emph{small} relative
to the total mode occupation numbers, of order $\sim E^{2}$. The
relative threshold noise level scales as $1/\sqrt{N}$ where $N$
is the total field number.

The work reviewed in Sections III and IV examines Bell violations
using measurements with \emph{discrete} outcomes. Violation of local
realism is possible for arbitrarily large systems (of size $N$) and
for arbitrarily large dimension ($d$). The effect of coarse-grained
measurements on Bell nonclassicality for discrete measurements has
been well studied. After the work of Mermin \citep{Mermin_PRD1980},
Busch developed a mathematical framework that characterizes the sharpness
of observables \citep{busch1986unsharp}. With this formalism, Busch
showed that an EPR experiment with two spin-$1/2$ particles in the
singlet state no longer violates the Bell inequality when the unsharpness
of the observables exceeds certain threshold\textcolor{blue}{}. The
incompatibility with local realism is uncovered by very precise measurement
selections, and was shown sensitive to noise. In this way, the transition
from quantum to classical can be viewed as arising from measurement,
as it becomes increasingly coarse-grained \citep{Kofler_PRL2007}\textcolor{blue}{}.

Durt, Kaszlikowski and Zukowski showed explicitly that the higher
dimensional Bell inequalities may allow a greater robustness to noise
\citep{durt2001violations}. They extended the work of \citep{Kaszlikowski_2000_PRL}
to values of $d=16$. The state of interest is the mixed state
\begin{align}
\rho_{d}\left(F_{d}\right) & =F_{d}\rho_{noise}+\left(1-F_{d}\right)|\Psi_{max}^{d}\rangle\langle\Psi_{max}^{d}|\,\label{eq:mixed_state_Durt-1}
\end{align}
where $F_{d}\leq1$ is a parameter that characterizes the contribution
of the noise state $\rho_{noise}=\frac{\hat{I}}{d^{2}}$. The entangled
state is $|\Psi_{max}^{d}\rangle=\frac{1}{\sqrt{d}}\sum_{m=1}^{d}|m,m\rangle$,
where $m$ is a mode that contains a photon, which can be prepared
using parametric down conversion. The measurement for each party involves
a $2d$-multiport and $d$ phase shifters. A $2d$-multiport is an
optical device that lets a photon enter one of its $d$ input ports,
and that photon has equal probability of exiting from one of the $d$
output ports. By imposing that the quantum prediction can be expressed
in terms of joint probability distributions that satisfy local realism,
the authors obtained $4d^{2}$ linear equations that must be true
for $d^{4}$ local hidden probabilities $P^{HV}$. This is a linear
optimization problem where the minimal noise threshold $F_{d}^{(cr)}$
that allows local realism model to agree with the quantum prediction
is obtained. They found that a larger absolute noise contribution
$F_{d}^{(cr)}$ is allowed for larger $d$.\textcolor{red}{{} }The
behavior for detection efficiencies $\eta$, was also studied. Values
of $\eta<1$ reflect loss (or decoherence) from the system, so that
information can be accessed by another party. The critical detection
efficiency $\eta_{d}^{(cr)}$, below which local realism holds, was
found to decrease very slowly but continuously with $d$. This extended
the result of Mermin and Garg for $d=2$ \citep{garg1982bell}. However,
the conclusions were limited to maximally entangled states with certain
observables only.

The degree of sensitivity to noise and decoherence will depend on
the state and measurements being considered. For a cat state, different
to a maximally entangled state (\ref{eq:multi}), the superposition
is for two macroscopically distinct states only. Here, the impact
of decoherence with increasing size $N$ is more extreme \citep{Caldeira:1983aa,Walls_PRA1985,Yurke_PRL1986,Brune_PRA1992},
as shown by Brune et al \citep{Brune_PRL1996}. In the work of Yurke
and Stoler, the impact of detection efficiency $\eta$ was considered
for the cat state $\frac{1}{\sqrt{2}}(|\alpha\rangle\pm i|-\alpha\rangle)$,
with $\alpha$ real. The quantum coherence was measured by the fringes
in the distribution $P(P)$ of the quadrature phase measurement $\hat{P}$.
The term contributing to the fringes was found to decay as $\sim e^{-2(1-\eta)|\alpha|^{2}}$,
showing an exponentially increased sensitivity with $\alpha$. Similarly,
Kennedy and Walls showed that thermal noise $n_{th}$ destroys the
fringes and there is an increased sensitivity of the decay with $|\alpha|$
\citep{Kennedy_PRA1988}. We expect that the steering signature for
the entangled cat state (\ref{eq:cat-1}) as measured by the conditional
variance (\ref{eq:infp}) will show a similar sensitivity to the efficiency
$\eta$ and thermal noise $n_{th}$ of the steered system. This is
because $\frac{1}{\sqrt{2}}(|\alpha\rangle+i|-\alpha\rangle)$ and
$\frac{1}{\sqrt{2}}(|\alpha\rangle-i|-\alpha\rangle)$ are the states
of the steered system, conditioned on the measurement of the spin
$\hat{S}_{x}^{(B)}$ of the micro-system, as given by (\ref{eq:rotcat}).
This was confirmed for the decoherence of the steering of a cat state
by Rosales-Zarate et al \citep{rosales2015decoherence}. \textcolor{red}{}\textcolor{teal}{}However,
in these analyses, the measurements being considered were limited
to $X$ and $P$.

Jeong, Paternostro and Ralph \citep{Jeong_PRL2009} gave an explicit
example of how Bell violations could be obtained for coarse grained
measurements based on macroscopic superposition states, using more
general measurements involving the displacement operator and homodyne
detection. They considered an entangled thermal cat state \citep{Jeong_PRL2006,Jeong_PRA2007},
with a density operator
\begin{align}
\rho & =\mathcal{N}\left[\rho_{A}^{th}\left(V,d\right)\otimes\rho_{B}^{th}\left(V,d\right)+\sigma_{A}\left(V,d\right)\otimes\sigma_{B}\left(V,d\right)+\sigma_{A}\left(V,-d\right)\otimes\sigma_{B}\left(V,-d\right)+\rho_{A}^{th}\left(V,-d\right)\otimes\rho_{B}^{th}\left(V,-d\right)\right]\,.\label{eq:thermal-cat}
\end{align}
Here, $\rho_{j}^{th}\left(V,d\right)=\intop d^{2}\alpha P^{th}\left(V,d\right)|\alpha\rangle_{j}\langle\alpha|$
is a displaced thermal state for system $j=A,B$ where $P^{th}\left(V,d\right)=\frac{2}{\pi\left(V-1\right)}exp\left(-\frac{2|\alpha-d|^{2}}{V-1}\right)$
with $d$ being the displacement in phase space and $V$ the variance
associated with the distribution; $\sigma_{i}\left(V,d\right)=\intop d^{2}\alpha\,P^{th}\left(V,d\right)|-\alpha\rangle_{i}\langle\alpha|$
and $\mathcal{N}=2\left[1+exp\left(-\frac{4d^{2}}{V}\right)/V^{2}\right]$.
We note that $d$, the effective separation between the two states
of the superposition, can be large. The local measurements comprise
a displacement operator $D_{j}\left(i\theta_{j}/2d\right)=e^{\frac{i\theta_{j}}{2d}a_{j}^{\dagger}+\frac{i\theta_{j}}{2d}a_{j}}$
where $\theta_{j}$ determines the measurement setting for party $j$,
and also involve a unitary operator that represents the Kerr nonlinear
interaction. Finally, a parameter $\eta$ denotes the homodyne detector
efficiency that characterizes the final measurement outcome imprecision.
A Bell-CHSH inequality based on the binned positive or minus outcomes
of the quadrature phase amplitude measurements was evaluated. The
authors gave numerical results for $d=150$ and found strong Bell
violations (approaching $B=2\sqrt{2}$) with a detection efficiency
as low as $\eta=0.05$. It is argued that this observation goes against
the viewpoint that a coarsening of measurement outcomes (i.e. measurement-outcomes
imprecision) contributes to quantum-classical transition.\textcolor{blue}{}

In a later paper, Jeong, Lim and Kim \citep{jeong2014coarsening}
further studied how Bell violations are destroyed by coarse-grained
measurements, in a discrete measurement setting. The authors evaluated
the effect of imprecision in measurement settings and measurement
outcomes on the violation of a CHSH Bell inequality, for an entangled
state with a varying degree of macroscopicity defined by a parameter
$n$. Here, $n$ reflects the dimensionality of the system. Their
work suggested that the coarsening of measurement settings and the
coarsening of the measurement outcomes play different roles in the
quantum-classical transition. The decrease in the violation observed
for imprecise outcomes can be overcome if the entangled state is more
macroscopic (i.e. large $n$), provided the measurement settings are
precise. This is consistent with the results of Durt et al for higher
dimensional systems \citep{durt2001violations}, which showed increasing
robustness of violations against noise for increasing dimensionality
$d$. On the other hand, the authors showed the CHSH inequality is
not violated for a measurement-setting imprecision above a certain
threshold, irrespective of the macroscopicity $n$ of the entangled
state.

The effect of coarse-graining has been analyzed in the context of
other quantum signatures, such as quantum coherence \citep{wang2013precision,park2014gaussian},
quantum entanglement \citep{raeisi2011coarse,correia2019spin}\textcolor{red}{{}
}and entropic uncertainty relations \citep{Veeren_PRA2020}. In particular,
Wang et al examined the impact of coarse-graining and imprecise measurement
on the coherence of the cat state $|\psi\rangle\sim|\alpha\rangle+i|-\alpha\rangle$,
where $\alpha$ is real \citep{wang2013precision}.\textcolor{red}{{}
}The coherence is signified by the revival of a coherent state $|\alpha\rangle$,
after application of the nonlinear unitary dynamics $U_{NL}=e^{i\hat{n}^{2}/2}$,
where $\hat{n}=\hat{a}^{\dagger}\hat{a}$. Here, the measurements
need only distinguish between the macroscopically distinct outcomes
$\alpha_{0}$ and $-\alpha_{0}$, and hence allow macroscopic coarse
graining. Similar to the results of Jeong, Lim and Kim, however, the
authors show that an imprecise transformation $U_{NL}$ will inevitably
affect the outcome. More explicitly, instead of $\pi/2$ phase, the
authors allow $\pi/2+\phi$ where $\phi$ satisfies a Gaussian distribution
with a standard deviation $\sigma$. Wang et al calculated that instead
of obtaining a state $|\alpha\rangle\langle\alpha|$ after applying
the unitary transformation on $|\psi\rangle\langle\psi|$, the measurement-setting
imprecision of the transformation leads to a state 
\begin{align}
C_{\sigma}\left(|\alpha\rangle\langle\alpha|\right) & =e^{-\alpha^{2}}\sum_{n,n^{'}=0}^{\infty}e^{-\frac{1}{2}\sigma^{2}\left(n^{2}-n^{'2}\right)^{2}}\frac{\alpha^{n+n^{'}}}{\sqrt{n!n^{'}!}}|n\rangle\langle n^{'}|\,.\label{eq:cg3}
\end{align}
For large $|\alpha|$, $\sigma\neq0$ will make it hard to distinguish
the two states in the final measurement (even if the measurement is
ideal), and there will be a threshold (depending on $|\alpha|$) where
it is impossible to distinguish the states.

In the next section, we examine results by Thenabadu et al \citep{Thenabadu_PRA2020,thenabadu2020bipartite,reid2021weak}
showing how violations of local realism can be obtained for macroscopic
coarse-grained measurements that need only distinguish between the
two distinct states $|\alpha\rangle$ and $|-\alpha\rangle$, where
$\alpha\rightarrow\infty$. Watts, Halpern and Harrow have also shown
how nonlinear Bell inequalities, which have additional assumptions,
may be violated for macroscopic measurements \citep{Watts2020Nonlinear}.

Furthermore, calculations indicate that the decoherence causing the
fragility of the macroscopic quantum coherence and correlations of
the cat state can be controlled by modifying the environment to which
the cat system is coupled \citep{Frowis_RMP2018}. For the simple
cat state $\frac{1}{\sqrt{2}}(|\alpha\rangle\pm i|-\alpha\rangle)$,
Meccozi and Tombesi showed that coupling to a squeezed reservoir slows
down the otherwise rapid decay of quantum coherence of the cat state
as $\alpha$ increases \citep{Mecozzi_PRL1987_squeeze_environ,Tombesi_JosaB1987}.
Kennedy and Walls and Serafini et al \citep{serafini2005quantifying,serafini2004minimum}
showed a similar effect for the decoherence caused by a thermal reservoir
\citep{Kennedy_PRA1988}. In both cases, it was important to orientate
the squeezed quadrature in suitable direction. Similar results have
been predicted \citep{teh2020overcoming,Munro_PRA1995} for the even
and odd cat states $\frac{1}{\sqrt{2}}(|\alpha\rangle\pm|-\alpha\rangle)$
generated in parametric oscillation above threshold \citep{Wolinsky_PRL1988,Krippner_PRA1994,Gilles_PRA1994,Hach_PRA1994,teh2020dynamics},
and for cat states generated in optomechanical systems \citep{bennett2018rapid}.
The progress made towards creating and manipulating squeezed states
of light in a variety of systems (for example, \citep{Caves_PRD1981,milburn1981production,reid1985squeezing,Movshovich_squeezing_PRL1990,squeezing_purdy_PRX2013,castellanos2008amplification,mehmet2011squeezed,tse2019quantum,wollman2015quantum})
suggests the realisation of more macroscopic cat states to be possible
in the future, which also motivates tests of quantum mechanics \citep{Bassi_RMP2013}.

We have seen how quantum mechanics violates local realism in macroscopic
regimes. Another approach was taken by Navascues and Wunderlich \citep{Navascues:2010aa}.
They considered the bounds placed on bipartite correlations if consistency
with classical mechanics in the macroscopic regime is to be upheld
in the form of a mechanism that they called macroscopic locality.
They first considered a source that produces a pair of particles
which are sent to two spatially separated parties, Alice and Bob.
\textcolor{black}{Each of them can change their measurement settings
by applying one out of $s$ possible interactions on their particle.
}The possible measurement settings are indexed from $1$ to $s$
for Alice, and $s+1$ to $2s$ for Bob. There will be a set of detectors
that determine the measurement outcomes, labelled by $D(a)$ or $D(b)$,
where $D(a)$ is the detector that returns outcome $a$ for Alice,
and similarly $D(b)$ returns outcome $b$ for Bob. The experiment
is characterized by the joint probability $P(a,b)$. To define the
notion of macroscopic locality, they described a macroscopic experiment.
Instead of a pair of particles, $N$ independent pairs of particles\textcolor{teal}{{}
} are produced and Alice and Bob each receive a beam of particles.
In the case of one particle, a chosen measurement setting (one out
of $s$ interactions) will lead to one detector being triggered. \textcolor{teal}{}
In the macroscopic case, the choice of one measurement setting will
lead to the trigger of all detectors, one for each particle. The measurement
outcomes are a distribution of intensities from all the detectors
of either Alice or Bob.  Let the intensities detected by Alice (Bob)
be $\vec{I}_{X}$ ($\vec{I}_{Y}$), where $X$ ($Y$) is one of the
$s$ interactions by Alice (Bob).  In a local hidden variable model,
a joint probability density $P\left(\vec{I}_{1},...,\vec{I}_{2s}\right)$
characterizes the experiment, where $\vec{I}_{j}$ is the hidden variable
state giving the result of the measurement with setting $j$. The
marginal probability density $P\left(\vec{I}_{X},\vec{I}_{Y}\right)d\vec{I}_{X}d\vec{I}_{Y}$
can be estimated. When $N>>1$ is large, the marginal probability
density is demanded to be consistent with classical physics, and hence
admits local hidden variable models. The marginal probability density
then satisfies
\begin{align}
P\left(\vec{I}_{X},\vec{I}_{Y}\right) & =\intop\left(\prod_{Z\neq X,Y}d\vec{I}_{Z}\right)P\left(\vec{I}_{1},...,\vec{I}_{2s}\right)\,.\label{eq:macro-loc}
\end{align}
If this is the case, then the system is said to satisfy \emph{macroscopic
locality}. The set of correlations that satisfies macroscopic locality
is completely characterized by a set $Q_{1}$ \citep{Navascues_PRL2007}.
Navascues and Wunderlich inferred that all quantum correlations (given
by the set of correlations $Q$) are macroscopically local in the
bipartite case ($Q\subset Q^{1}$).\textbf{\textcolor{teal}{}}\textcolor{red}{}\textcolor{teal}{}

\section{Leggett-Garg correlations and macroscopic realism}

In 1985, Leggett and Garg gave an explicit proposal to test the predictions
of macroscopic realism against those of quantum mechanics. In the
proposal, it is only necessary to make measurements distinguishing
between two macroscopically distinct states of a system \citep{Leggett:1985}.
The measurements are thus macroscopic and allow a macroscopic coarse
graining, at least in principle. Leggett and Garg considered systems
which evolve dynamically with time. In their analysis, it was however
necessary to consider a specific definition of macroscopic realism,
which is referred to as macrorealism.

\subsection{Leggett-Garg inequalities and macro-realism}

The question of how to define a truly macroscopic quantum regime was
raised by Leggett \citep{Leggett_PTPS1980,Leggett_2002}. Leggett
suggested that in order to confirm the presence of a quantum superposition
or quantum coherence, some observables/ quantitative measures that
allow the distinction between a superposition and a classical mixture
have to be designed \citep{Leggett_PTPS1980}. In the paper \citep{Leggett_PTPS1980},
Leggett defined disconnectivity as one possible measure. As an example,
an $N$-particle state 
\begin{equation}
\psi=\left(a\phi_{l}+b\phi_{r}\right)^{N}\label{eq:leggett-1}
\end{equation}
is not regarded as a macroscopic quantum state based on the disconnectivity
measure, but an $N$-particle state 
\begin{equation}
\psi_{1}=a\phi_{l}^{N}+b\phi_{r}^{N}\label{eq:leggett-2}
\end{equation}
would be. Leggett also suggested to observe quantum tunneling as evidence
of macroscopic quantum state. In the later paper \citep{Leggett_2002},
Leggett proposed another feature to be included in the notion of 'macroscopic
distinctness'. It is named the extensive difference of the measurement
outcomes. For a superposition of two states, the difference between
the measurement outcome of both states has to be large, relative to
a reference value of an observable which is typical at the atomic
scale. Leggett gave the Bohr magneton as a reference value when the
observable in consideration is the magnetic moment of the state.

In 1985, Leggett and Garg proposed a test of macroscopic realism,
where one makes measurements distinguishing between two macroscopically
distinct states of a system \citep{Leggett:1985}. A review is given
in \citep{Emary:2014}. The assumption of macroscopic realism (MR)
is that the system prior to the measurement is actually in one or
other of the two macroscopically distinct states. One may then assign
a hidden variable $\lambda_{M}$ to the system to denote the outcome
of the measurement, should it be performed. For example, for the cat
state ($\alpha_{0}$ is real)
\begin{equation}
|\psi\rangle\sim|\alpha_{0}\rangle+e^{i\theta}|-\alpha_{0}\rangle\label{eq:cat-single}
\end{equation}
of a system $A$, the measurement $\hat{S}$ defined as the sign of
the quadrature amplitude $\hat{X}_{A}$ distinguishes between the
two states $|\alpha_{0}\rangle$ and $|-\alpha_{0}\rangle$. One assigns
$\lambda_{M}=+1$ if the system is in state giving outcome $+1$,
and $\lambda_{M}=-1$ if the system is in a state giving outcome $-1$.
We note the system \emph{cannot} be regarded as being in either state
$|\alpha_{0}\rangle$ or $|-\alpha_{0}\rangle$. The superposition
$|\psi\rangle$ can be distinguished from the mixture $\rho_{mix}$
of states $|\alpha_{0}\rangle$ and $|-\alpha_{0}\rangle$, for example,
by fringes in the distribution of $\hat{P}_{A}$ \citep{Yurke_PRL1986}.
Macroscopic realism asserts that the system is in a ``state'' with
a definite value for the result for $S$, and does not propose details
about the microscopic nature of that state.

Leggett and Garg's work considered a system dynamically evolving in
such a way that the system at times $t_{i}$ gives outcomes $+1$
or $-1$ for a measurement, these outcomes corresponding to the macroscopically
distinct states. Here, we consider the measurement $\hat{S}_{i}$
of $\hat{S}$ at the time $t_{i}$, and denote the outcomes by $S_{i}\equiv S(t_{i})$.
They made two assumptions, the first being macroscopic realism (MR).
The second assumption, noninvasive measurability (NIM), is that the
value of $\lambda_{M}$ can be measured without a disturbance to the
future (macroscopic) dynamics of the system. The two assumptions (referred
to as macro-realism) imply that certain inequalities will be satisfied.
The Leggett-Garg inequalities involve the two-time correlation functions
$\langle S(t_{i})S(t_{j})\rangle$. One of the inequalities is \citep{Williams_PRL2008,Jordan_PRL2006,Leggett:1985}
\begin{equation}
LG\equiv\langle S(t_{1})S(t_{2})\rangle+\langle S(t_{2})S(t_{3})\rangle-\langle S(t_{1})S(t_{3})\rangle\leq1\label{eq:three-time-lg}
\end{equation}
The inequalities can be violated by macroscopic two-state systems
whose correlation function is given as $\langle S(t_{i})S(t_{j})\rangle=\cos\Omega(t_{i}-t_{j})$
where $\Omega$ is a constant. This is clear, if we choose, for example,
$t_{1}=0$, $t_{2}=\pi/(4\Omega)$ and $t_{3}=\pi/(2\Omega)$. The
no-signaling-in-time inequality gives a test necessary and sufficient
for macrorealism \citep{Clemente_PRL2016,Kofler_PRL2008,Kofler_PRA2013}.
Higher dimensional studies have also been given by Halliwell and Mawby
\citep{Halliwell_PRA2020}. Systems that violate Leggett-Garg inequalities
can be considered to exhibit macroscopic correlations with respect
to time. In these tests, one requires only to make macroscopic coarse-grained
measurements, which need only distinguish between the macroscopically
distinct states, e.g. $|\alpha_{0}\rangle$ and $|-\alpha_{0}\rangle$,
at the given time $t_{i}$.

There have been many predictions and some realizations of violation
of Leggett-Garg inequalities, including \citep{Halliwell_PRA2021,Pan_PRA2020,rosales2018leggett,knee2016strict,Knee:2012aa,Goggin:2011,Palacios-Laloy:2010aa,Robens_PRX2015,Williams_PRL2008,Jordan_PRL2006,Dressel_PRA2014,Dressel_PRL2011,asadian2014probing}
and those referenced in Emary et al \citep{Emary:2014}. In many of
these (e.g. \citep{Dressel_PRL2011,Goggin:2011}), the two states
$+1$ and $-1$ are realized as a photonic qubit similar to the qubit
Bell experiments. In other analyses, the outcomes are binned to create
a dichotomic observable. Kofler and Brukner showed how the Leggett-Garg
inequality could be used to demonstrate non-classical behavior in
high spin systems, and to illustrate the quantum to classical transition
under coarse graining \citep{Kofler_PRL2007}. They tested a Leggett-Garg
inequality for a large spin $j$ system with the Hamiltonian $\hat{H}=\hat{J}^{2}/2I+\omega\hat{J}_{x}$
and an initial maximally mixed state $\rho\left(0\right)=\frac{1}{2j+1}\sum_{m=-j}^{j}|m\rangle\langle m|$.
Here, the spin vector is $\hat{J}=(\hat{J}_{x},\hat{J}_{y},\hat{J}_{z})$
and $|m\rangle$ are the eigenstates of the spin $z$ component $\hat{J}_{z}$.
The parity measurement $\hat{Q}\equiv\sum_{m=-j}^{j}\left(-1\right)^{j-m}|m\rangle\langle m|$
is carried out at different times. In terms of these parity operators,
the Leggett-Garg inequality they considered has the form $K\equiv C_{12}+C_{23}+C_{34}\text{\textminus}C_{14}\text{\ensuremath{\le}}2$,
where $C_{ij}\equiv\langle Q\left(t_{i}\right)Q\left(t_{j}\right)\rangle$.
They found a violation of the Leggett-Garg inequality for arbitrarily
high spin $j$, even for a maximally mixed initial state, provided
that the values $m$ can be resolved perfectly. They then considered
the system initially in a spin-$j$ coherent state 
\begin{align}
|\theta_{0},\phi_{0}\rangle & =\sum_{m}\begin{pmatrix}2j\\
j+m
\end{pmatrix}^{1/2}\cos^{j+m}\frac{\theta_{0}}{2}\sin^{j-m}\frac{\theta_{0}}{2}e^{-im\phi_{0}}|m\rangle\,.\label{eq:cg1-1-1}
\end{align}
The state at time $t$, under the time evolution unitary operator
$U_{t}=e^{-i\omega t\hat{J}_{x}}$, is given by $|\theta,\phi\rangle=U_{t}|\theta_{0},\phi_{0}\rangle$.
The probability that a $\hat{J}_{z}$ measurement at time $t$ for
large spin $j$ gives outcome $m$ is 
\begin{align}
p\left(m,t\right) & =|\langle m|\theta,\phi\rangle|^{2}\approx\frac{1}{\sqrt{2\pi}\sigma}e^{-\left(m-\mu\right)^{2}/2\sigma^{2}}\,,\label{eq:cg2-1-1}
\end{align}
which is a Gaussian. Here, $\mu=j\cos\theta$ and $\sigma=\sqrt{j/2}\sin\theta$.
Sharp measurements will allow this Gaussian distribution to be resolved
and a violation of the Leggett-Garg inequality observed. In order
to introduce a finite resolution, the authors subdivide the $2j+1$
possible outcomes into $\left(2j+1\right)/\Delta m$ coarse-grained
values, where $\Delta m$ determines the resolution of the measurement.
If $\Delta m$ is larger than the standard deviation $\sigma$ in
the distribution of $m$, then the Gaussian function cannot be distinguished.\textcolor{blue}{}

The noninvasive measurability assumption (NIM) is the additional assumption
in the definition of macrorealism, and must be justified in an experiment.
This has been done in different ways, most notably using stationarity
\citep{Formaggio_PRL2016,Zhou_PRL2015}, weak measurements \citep{Goggin:2011,Williams_PRL2008},
or ideal negative-result measurements \citep{Knee:2012aa,Robens_PRX2015}.
The ideal negative-result measurement was proposed originally by Leggett
and Garg \citep{Leggett:1985}, and is based on the assumption that
MR is correct. In some scenarios, this then implies the experimentalist
may measure a result $-1$ without disturbing the system by registering,
for example, an absence of a photon or current, at a given location.
The weak measurement can be shown to have negligible impact on the
quantum system in some limit, yet with results yielding agreement
with the quantum prediction for the ensemble average. The weak measurement
involves postselection and is an ambiguous one, returning results
that can be outside the normal eigenvalue range, which in this case
corresponds to the set $\{-1,1\}$ \citep{Aharonov:1988,Dressel_RMP2014}.
More sophisticated approaches are given in Uola, Vitagliano and Budroni
\citep{Uola_PRA2019}.

Macroscopic tests giving evidence of violations of a Leggett-Garg
inequality have been realized for superconducting experiments (see
e.g. \citep{Palacios-Laloy:2010aa,knee2016strict,White:2016aa}).
The recent experiment of Knee et al reports violations for macroscopic
superconducting qubits, where the noninvasive measurability assumption
is validated by a control experiment in which the macroscopic states
are prepared and the impact of the measurement calibrated \citep{knee2016strict}.
This however involves the assumption that the macroscopic 'state'
of the system is actually the state prepared in the laboratory. Proposals
to test violation of Leggett-garg inequalities in other systems have
been put forward e.g. for optomechanics \citep{asadian2014probing},
atomic states \citep{budroni2015quantum}, and for NOON states and
two-well Bose-Einstein condensates \citep{rosales2018leggett,opanchuk2016quantifying}.
A proposal to demonstrate violations of the inequality (\ref{eq:three-time-lg})
using cat states and a suitable dynamics is given in \citep{thenabadu2019leggett,Thenabadu_PRA2020}.

\subsection{Two-party Leggett-Garg tests}

In this review, we are mainly concerned with macroscopic quantum correlations
connected with spatial separation i.e. with entangled states. However,
it is possible to link the Bell and Leggett-Garg approaches, to obtain
a situation where a two-party Leggett-Garg test is given, corresponding
to Bell violations using macroscopic measurements that only distinguish
between two macroscopically distinguishable states e.g. between $|\alpha\rangle$
and $|-\alpha\rangle$.

Dressel et al \citep{Dressel_PRL2011} extended the Leggett-Garg inequality
to multipartite systems, as well as including ambiguous detections
results (i.e. weak measurements). They derived a two-party generalized
Leggett-Garg inequality. First, a pair of particles are created at
time $t_{0}$. After some time at $t_{1}$, particle $1$ interacts
with an imperfect detector and returns a generalized value $\alpha_{1}\in S$,
where $S$ is a set with $\text{min}S\leq-1$ and $\text{max}S\geq1$
(a weak measurement). At yet a later time $t_{2}$, both particles
$1$ and $2$ are measured with unambiguous detectors (a strong measurement)
with detection results $b_{1},b_{2}\in\{-1,1\}$. The correlation
function considered is $C=\alpha_{1}+\alpha_{1}b_{1}b_{2}-b_{1}b_{2}$,
which has the inequality 
\begin{equation}
-|1-2\text{min}S|\leq\langle C\rangle\leq|2\text{max}S-1|\label{eq:c}
\end{equation}
that must be satisfied for the macroscopic realism model. This inequality
is just one of the many possible correlation functions that can be
formed involving $\alpha_{1}$, $b_{1}$ and $b_{2}$. The setup was
realized in an optical experiment. The degenerate type-II down conversion
process generates entangled photon pairs where the polarization of
these photon pairs are orthogonal to each other. The measurements
that correspond to the correlation function in theory are given by:
$\alpha_{1}\leftrightarrow-\sigma_{z}^{\left(1\right)}$, $b_{1}\leftrightarrow\sigma_{\theta}^{\left(1\right)}$
and $b_{2}\leftrightarrow\sigma_{z}^{\left(2\right)}$. The weak measurement
$\alpha_{1}$ is made by passing the photon beam through a coverslip
before measuring the polarization state $\sigma_{z}^{\left(1\right)}$.
They found violation of the generalized Leggett-Garg inequality.

A proposal to avoid loopholes using a hybrid Leggett-Garg-Bell inequality
was put forward by Dressel and Korotkov \citep{Dressel_PRA2014}.
They combined the generalized Leggett-Garg inequality with Bell locality.
The assumption of non-invasiveness of measurements is replaced by
Bell locality, where a measurement on a system cannot disturb/ influence
the measurements made on the other system that is space-like separated
from it. The hybrid Bell-Leggett-Garg inequality is aimed to circumvent
the non-invasiveness problem, as well as the disjoint sampling loophole
in Bell inequalities, where different experimental settings are required
to check for the Bell inequalities. They considered a similar setup
as in Dressel et al \citep{Dressel_PRL2011}, the difference being
that the weak measurement is also carried out on particle $2$. The
correlation function considered is 
\begin{equation}
C=\alpha_{1}\alpha_{2}+\alpha_{1}b_{2}+\alpha_{2}b_{1}-b_{1}b_{2}\,,\label{eq:c2}
\end{equation}
with the average value of $C$ satisfying the inequality $|\langle C\rangle|\leq2$
in the hybrid Leggett-Garg Bell locality model. Related experiments
investigating violation of hybrid inequalities in the context of weak
measurements were performed by White et al \citep{White:2016aa} for
transmon qubits and Higgens et al \citep{Higgins:2015} using entangled
photons.

It is possible to test macroscopic realism where the noninvasiveness
assumption is replaced by that of a macroscopic Bell locality, if
one considers Bell inequalities derived for macroscopically distinct
qubit states. This allows a situation where all the necessary measurements
are macroscopic, in the sense that one only requires to distinguish
between two macroscopically distinguishable states, for \emph{all
}choices of measurement setting. Thenabadu et al \citep{Thenabadu_PRA2020}
considered macroscopic Bell-CHSH inequalities where the two outcomes
of $\hat{S}$ correspond to detecting one or other of the states $|N\rangle|0\rangle$
or $|0\rangle|N\rangle$, where $|N\rangle$ is a number state. This
replaces the microscopic qubits \{$|1\rangle|0\rangle$, $|0\rangle|1\rangle$\}
with mesoscopic qubits \{$|N\rangle|0\rangle$, $|0\rangle|N\rangle$\},
distinct by $N$ quanta for each mode. The work of \citep{Thenabadu_PRA2020}
considered two space-like separated systems $A$ and $B$. The overall
system is prepared in a four mode NOON-type Bell state, e.g. $\frac{1}{\sqrt{2}}(|N\rangle_{a_{+}}|0\rangle_{a_{-}}|0\rangle_{b_{+}}|N\rangle_{b_{-}}+e^{i\varphi}|0\rangle_{a_{+}}|N\rangle_{a_{-}}|N\rangle_{b_{+}}|0\rangle_{b_{-}})$.
The rotations at each site corresponding in the standard Bell experiments
to polarizer or Stern-Gerlach rotations (\ref{eq:rotc}) are provided
by a nonlinear Josephson interaction. This is given for site $A$
by the Hamiltonian $H_{NL}^{(A)}=\kappa(\hat{a}_{+}^{\dagger}\hat{a}_{-}+\hat{a}_{+}\hat{a}_{-}^{\dagger})+g\hat{a}_{+}^{\dagger2}\hat{a}_{+}^{2}+g\hat{a}_{-}^{\dagger2}\hat{a}_{-}^{2}$
\citep{lipkin1965validity,steel1998quantum}. Here, $\hat{a}_{+}$,
$\hat{a}_{-}$ are the boson operators for the corresponding fields
modeled as single modes $a_{+}$ and $a_{-}$, and $\kappa$ and $g$
are the interaction constants. A similar interaction $H_{NL}^{(B)}$
is defined for site $B$. The solutions for $H_{NL}^{(A)}$ confirm
that to an excellent approximation the state created after a time
$t_{a}$ from an initial state $|N\rangle_{a_{+}}|0\rangle_{a_{-}}$
is 
\begin{equation}
|\psi(t)\rangle\sim\cos\theta\thinspace|N\rangle_{a_{+}}|0\rangle_{a_{-}}+i\sin\theta\thinspace|0\rangle_{a_{+}}|N\rangle_{a_{-}}\label{eq:n-state}
\end{equation}
where $\theta$ is proportional to the interaction time $t_{a}$.
The solution for $H_{NL}^{(B)}$ is similar, with interaction time
$t_{b}$. This implies one can map the microscopic qubit Bell experiment
involving qubits \{$|1\rangle|0\rangle$, $|0\rangle|1\rangle$\}
onto a mesoscopic one involving the qubits \{$|N\rangle|0\rangle$,
$|0\rangle|N\rangle$\}, distinct by $N$ quanta at each site. The
settings $\theta$ and $\phi$ correspond to the interaction times
$t_{a}$ and $t_{b}$. For all relevant choices of settings $\theta$
and $\phi$, the Bell violation can be obtained where the measurement
makes only the distinction between $0$ or $N$ photons at each site.
Leggett and Garg's macroscopic realism is applied at the level of
$N$ quanta, to assert that the system $A$ or $B$ is predetermined
to be in a state giving the outcome of $\sim N$ or $\sim0$. The
noninvasive measurability assumption of Leggett and Garg is justified
in this case as an $N$-scopic Bell locality, that the measurement
on system $B$ cannot induce a change of $\sim N$ to the outcomes
at $A$ (and vice versa). The predictions for the violations were
confirmed numerically for up to $N=100$. This gives a rigorous prediction
for violation of Leggett-Garg's macrorealism at the level of $\sim100$
quanta.

Thenabadu et al continued to confirm violation of a macroscopic Bell-CHSH
inequality, using the macroscopically distinct outcomes provided by
multi-component entangled superpositions of coherent states $|\alpha_{0}\rangle$,
$|\alpha_{0}e^{i\theta}\rangle$, for distinct and fixed $\theta$
where $\alpha_{0}$ is real \citep{Thenabadu_PRA2020}. At each of
two sites, the two outcomes for the measurement of the sign of the
quadrature amplitude $X$ correspond to macroscopically distinct states,
as $\alpha_{0}\rightarrow\infty.$ In this case, the unitary rotation
corresponding to the choice of measurement at each site is achieved
using the interaction \citep{Yurke_PRL1986} 
\begin{equation}
H_{NL}=\Omega\hat{n}^{k}\,.\label{eq:nonlinearH-1}
\end{equation}
This is applied independently at $A$ and $B$, where $k=2$. Here,
\textcolor{black}{$\Omega$ is a constant} and $\hat{n}$ is the field
mode number operator. For certain times $t$, the system in the coherent
state $|\alpha_{0}\rangle$ evolves to a superposition of the two
states, $|\alpha_{0}\rangle$ and $|-\alpha_{0}\rangle$, enabling
the application of the Leggett-Garg premise of macroscopic realism
at those times. The measurement settings $\theta$ and $\phi$ correspond
to those certain times of interaction, $t_{a}$ and $t_{b}$. For
all relevant choices of settings $\theta$ and $\phi$, the Bell violation
can be obtained where the measurements make only the distinction between
a negative or positive value of amplitude $\hat{X}_{A}$ (or $\hat{X}_{B}$),
where these values are increasingly macroscopically separated in phase
space, as $\alpha_{0}\rightarrow\infty$. The noninvasive measurability
assumption of Leggett and Garg is justified in this case as a macroscopic
Bell locality, that the measurement on system $B$ cannot make a macroscopic
change to the outcomes at $A$ (and vice versa). Violations of the
Leggett-Garg-Bell inequalities were predicted in the macroscopic regime,
for arbitrarily large $\alpha_{0}$ \citep{Thenabadu_PRA2020}.

In fact, it is possible to obtain a direct mapping between microscopic
and macroscopic versions of the Bell-Leggett-Garg experiments, involving
spin qubits \{$|\uparrow\rangle$, $|\downarrow\rangle$\} and macroscopic
qubits \{$|\alpha_{0}\rangle$, $|-\alpha_{0}\rangle$\} respectively
\citep{thenabadu2020bipartite,reid2021weak}. For large $\alpha_{0}$,
the two coherent states are orthogonal and one may define Schwinger
spin measurements (\ref{eq:schwinger}) as $\hat{S}_{z}=\frac{1}{2}(|\alpha_{0}\rangle\langle\alpha_{0}\rangle-|-\alpha_{0}\rangle\langle-\alpha_{0}|)$,
$\hat{S}_{x}=\frac{1}{2}(|\alpha_{0}\rangle\langle-\alpha_{0}\rangle+|-\alpha_{0}\rangle\langle\alpha_{0}|)$
and $\hat{S}_{y}=\frac{1}{2i}(|\alpha_{0}\rangle\langle-\alpha_{0}\rangle-|-\alpha_{0}\rangle\langle\alpha_{0}|)$.
Similar observables were considered by Wang et al \citep{Wang_PRA2013}.
The authors Thenabadu and Reid consider two sites prepared in the
entangled cat state (\ref{eq:cat-state}), and propose a macroscopic
Bell violation using the macroscopic qubits $|\alpha_{0}\rangle$
and $|-\alpha_{0}\rangle$, and $|\beta_{0}\rangle$ and $|-\beta_{0}\rangle$,
at each site. In this case, the local interaction that brings about
the unitary rotations for certain crucial values of $\theta$ and
$\phi$ is realized by the nonlinear Hamiltonian (\ref{eq:nonlinearH-1})
with $k=4$ \citep{Yurke_PRL1986,thenabadu2019leggett}. \textcolor{black}{The
systems evolve independently at each site according to the interactions
$H_{NL}^{(A)}$ and $H_{NL}^{(B)}$, for times $t_{a}$ and $t_{b}$
respectively. }At site $A$, for certain interaction times $t_{a}=t_{\theta}$,
a system prepared in a coherent state $|\alpha_{0}\rangle$ evolves
to the superposition \citep{Yurke_PRL1986,thenabadu2019leggett,Thenabadu_PRA2020}\textcolor{red}{}
\begin{equation}
|\alpha_{0}\rangle\rightarrow e^{-iH_{NL}t/\hbar}|\alpha_{0}\rangle=e^{-i\theta}(\cos\theta|\alpha_{0}\rangle+i\sin\theta|-\alpha_{0}\rangle)\label{eq:rot-coherent}
\end{equation}
where $\theta=t_{\theta}/2$. For $k=4$, the result is valid for
interaction times $t_{\theta}=m\pi/8$ where $m$ is a non-negative
integer. Similarly at $B$, the system prepared in $|\beta_{0}\rangle$
evolves to $e^{-i\phi}(\cos\phi|\beta_{0}\rangle+i\sin\phi|-\beta_{0}\rangle)$,
where $\phi=t_{\phi}/2$ after a time $t_{b}=t_{\phi}$. The solutions
imply that for the system prepared initially in the two-mode cat state
(\ref{eq:cat-state}) with $\alpha_{0}=\beta_{0}$, violations of
Bell-CHSH inequalities (\ref{eq:bell-ineq}) will be obtained for
the choice of measurement settings $t_{a}$ and $t_{b}$ corresponding
to $\theta$ and $\phi$ as given for (\ref{eq:bell-ineq}). It is
evident that all the measurements are macroscopic, because the outcomes
of $\hat{X}$ for each measurement setting distinguish the amplitudes
$+\alpha_{0}$ and $-\alpha_{0}$, as associated with the macroscopic
qubit. Violations of the Leggett-Garg-Bell inequalities were thus
predicted in the macroscopic regime, for arbitrarily large separations
of outcomes of order $\sim\alpha_{0}\rightarrow\infty$, for all measurement
settings.

\section{Quantum correlations for atomic systems}

Early experimental investigations of quantum correlations focused
on two separated photonic systems, which could not be called macroscopic
correlations. Another interpretation of the meaning of ``macroscopic''
is that the systems involved possess mass. Quantum correlations have
been realized for atomic systems. An early experiment of Lamehi-Rachti
and Mittig investigated the quantum correlations for pairs of protons
\citep{lamehi1976quantum}, although additional assumptions were necessary
to infer quantum correlations \citep{Clauser:1978_ReportPP}.

\subsection{Quantum correlations between two massive particles}

A two-level atom is an example of a spin $1/2$ system, the two levels
corresponding to the states $|\uparrow\rangle$ and $|\downarrow\rangle$.
A spin for each atom can therefore be assigned in terms of two internal
atomic levels, the spin components being constructed from pseudo-Schwinger
spins, where the $\hat{a}^{\dagger}$ and $\hat{b}^{\dagger}$ operators
refer to the 'creation' of each level. This technique has been used
to detect quantum correlations in atomic systems. In 2001, Rowe et
al reported violations of Bell inequalities for two spin $1/2$ systems
given by the internal levels of two ions, in an ion trap \citep{Rowe2001_Nature}.
Significantly, this experiment overcame detection efficiency loopholes
for violation of a Bell inequality, but the correlations were observed
without the spatial separation required for a rigorous Bell test.
Hensen et al have since demonstrated conclusive loophole-free violations
of Bell inequalities for electron spins in diamond separated by 1.3
km, and Rosenfield et al for two Rb atoms separated by 398 m \citep{Hensen_2015_Nature,Hensen2016_SciRep,Rosenfeld_PRL_2017}.

The two atomic internal states of a single atom could not be called
macroscopically distinct however, and give a weak gravitational interaction.
Other origins of entanglement have been investigated. In 2019, Shin
et al realized Einstein-Podolsky-Rosen-type correlations between spatially
separated propagating atoms \citep{Shin2019Bell}. In their experiment,
spin-entangled pairs of ultra-cold He atoms are created from two colliding
spin-polarized Bose Einstein condensates. The range of settings for
each particle was insufficient to claim a complete test of Bell's
theorem however. Bergschneider et al \citep{bergschneider2019experimental}
similarly demonstrated entanglement between ultra-cold fermions in
coupled wells. Here, the quantum correlation is for the momenta and
positions of the atoms, modeled after the photon experiments of Rarity
and Tapster \citep{RarityTapster1990_PRL} and related theoretical
work that gives mechanisms for achieving Bell violations \citep{Bonneau2018Characterizing,yannouleas2019interference}.
Other experiments have used macroscopic entangled matter-waves to
create interferometers operating beyond the usual classical limits
\citep{lucke2011twin}, and correlated matter-waves to suppress atomic
fluctuations \citep{Bucker2011Twin}. As of yet, there is no violation
of a Bell inequality reported.

\subsection{Multi-atom quantum correlations and depth of entanglement}

Quantum correlations were originally defined along the lines of EPR
and Bell, as existing between separated systems and detected by local
measurements on each system. Such correlations may falsify local hidden
variable theories. Another strategy is to identify quantum correlations
\emph{within} the framework of quantum mechanics, by measuring \emph{collective}
observables. This approach has proved useful for certifying the existence
of quantum correlations within a system of $N$ atoms, where $N$
is large. As a related example, the genuine entanglement of $N$ photonic
systems was confirmed for $N=4$ in the optical experiment of Papp
et al \citep{papp2009characterization}, by measuring collective operators
involving all $N$ systems at one converging site.

Sorenson et al showed that the entanglement of many atoms in a Bose
Einstein condensate can be inferred from the observation of spin squeezing,
as defined for the collective atomic spin operators \citep{sorensen2001many,Sorenson2001EntanglementPhysRevLett.86.4431}.
The total collective spin of $N$ atoms is given as $\hat{J}_{\theta}=\sum_{i=1}^{N}\hat{J}_{\theta}^{(i)}$
where $\hat{J}_{\theta}^{(i)}$ is the spin of the $i$-th atom in
the group of $N$ atoms. The spin squeezing relation is determined
by the uncertainty relation $\Delta\hat{J}_{x}\Delta\hat{J}_{y}\geq\frac{|\langle\hat{J}_{z}\rangle|}{2}$.
Spin squeezing occurs when $\Delta\hat{J}_{x}<|\langle\hat{J}_{z}\rangle|/2$
\citep{wineland1992spin,Kitagawa:1993}. In experiments where $N$
atoms are prepared in the same initial state $|\uparrow\rangle$,
and evolve according to the same Hamiltonian, one may express the
collective system in terms of a spin $J=N/2$. The collective spin
$\hat{J}_{z}$ then gives the population difference between the two
levels. A rotation from one basis to another is realized by a Rabi
rotation, and prepares the atoms in a superposition of the two levels.
Spin squeezing and hence multi-particle entanglement can be created
using interactions modeled as $H=\hat{J}_{x}^{2}$ or $\hat{J}_{x}^{2}-\hat{J}_{y}^{2}$
\citep{wineland1992spin,Kitagawa:1993}. This has confirmed possible
in sophisticated multimode models of Bose Einstein condensates \citep{Steel1998,li2008optimum,li2009spin,Opanchuk2012quantum}.

If there is no entanglement between the $N$ atoms, then separability
of the entire spin $1/2$ system implies that the density matrix factorizes
as 
\begin{equation}
\rho=\sum_{R}P_{R}\prod_{i}^{N}\rho_{R}^{(i)}\,.\label{eq:sepN}
\end{equation}
Here, the $P_{R}$ are probabilities, $\sum_{R}P_{R}=1$, and $\rho_{R}^{(i)}$
is the individual density operator for the $i$-th atom. This assumption
leads to the result that the spin squeezing parameter defined as $\xi_{N}=\frac{\sqrt{2J}\Delta\hat{J}_{x}}{|\langle\hat{J}_{z}\rangle|}$
is constrained \citep{sorensen2001many}. The constraint can be understood
in the following way. It is clear that for a spin $1/2$ system, the
maximum variance is $1/4$, so that $(\Delta\hat{J}_{y})\leq1/2$,
which gives a limit on the amount of spin squeezing: $(\Delta\hat{J}_{x})\geq|\langle\hat{J}_{z}\rangle|$.
For a fully separable system, we then see that for any decomposition
of the density operator, the variance in $\hat{J}_{x}$ has a lower
bound, $(\Delta\hat{J}_{x})^{2}\geq\sum_{R}P_{R}\sum_{i}^{N}(\Delta\hat{J}_{x}^{(i)})_{R}^{2}\geq\sum_{R}P_{R}(\sum_{i}^{N}|\langle\hat{J}_{z}^{(i)}\rangle_{R}|^{2})$.
This follows because for the mixture, $\langle\hat{O}\rangle=\sum_{R}P_{R}\langle\hat{O}\rangle_{R}$
and the overall variance cannot be less than the weighted average
of the variances of its components \citep{hofmann2003violation,duan2000inseparability}.
Here, we denote the average of an operator $\hat{O}$ for a system
in the state $\rho_{R}$ by the subscript $R$: $\langle\hat{O}\rangle_{R}=Tr(\rho_{R}\hat{O})$
and $(\Delta\hat{O})_{R}^{2}=\langle\hat{O}^{2}\rangle_{R}-\langle\hat{O}\rangle_{R}^{2}$.
The Cauchy Schwarz inequality implies $(\sum_{i}^{N}\frac{1}{N}|\langle\hat{J}_{z}^{(i)}\rangle_{R}|^{2})(\sum_{i}^{N}\frac{1}{N})\geq|\sum_{i}^{N}\frac{1}{N}\langle\hat{J}_{z}^{(i)}\rangle_{R}|^{2}.$
Noting that $\langle\hat{J}_{z}\rangle_{R}=\sum_{i}^{N}\langle\hat{J}_{z}^{(i)}\rangle_{R}$,
and that the Cauchy-Schwarz inequality also implies $(\sum_{R}P_{R}|\langle\hat{J}_{z}\rangle_{R}|^{2})(\sum_{R}P_{R})\geq|\sum_{R}P_{R}\langle\hat{J}_{z}\rangle_{R})|^{2}$,
one finally obtains that for a fully separable state, the $N$-atom
system satisfies
\begin{eqnarray}
(\Delta\hat{J}_{x})^{2} & \geq & |\langle\hat{J}_{z}\rangle|^{2}/N\,.\label{eq:spin-squeezing-result}
\end{eqnarray}
Where $\langle J_{z}\rangle=N/2$, this reduces to $(\Delta\hat{J}_{x})^{2}<|\langle\hat{J}_{z}\rangle|^{2}/N=N/4$,
which is the condition for spin squeezing: $\xi_{N}=\frac{\sqrt{2J}\Delta\hat{J}_{x}}{|\langle\hat{J}_{z}\rangle|}<1$
\citep{wineland1992spin,Kitagawa:1993}. Hence, the observation of
spin squeezing implies non-separability i.e. entanglement between
at least one pair of atoms \citep{sorensen2001many}.

The result (\ref{eq:spin-squeezing-result}) was used by Esteve et
al \citep{esteve2008squeezing} and Riedel et al \citep{riedel2010atom}
to deduce entanglement in a Bose Einstein condensate of hundreds of
atoms. However, the question arises as to whether this is truly a
macroscopic effect relating to all $N$ atoms, since logically, the
violation of the spin squeezing inequality can come from the entanglement
of just one pair of atoms. The number of atoms genuinely involved
in the entanglement is referred to as the ``\emph{depth of entanglement}''.

The concept of depth of entanglement was developed by Sorenson and
Molmer \citep{sorensen2001entanglement}. They demonstrated that the
minimum depth of entanglement can be inferred, if the spin squeezing
is sufficiently extreme i.e. if the variance $(\Delta\hat{J}_{x})^{2}$
is reduced below a certain level. For systems of a finite dimensionality
(corresponding to a fixed nonzero spin $J$) and where there is a
nonzero $|\langle\hat{J}_{z}\rangle|$, the amount of squeezing possible
is limited i.e. $(\Delta\hat{J}_{x})^{2}$ cannot be zero. This is
because the variance in $\hat{J}_{y}$ cannot be infinite for finite
$J$. However, as $J$ increases, the lower bound for $(\Delta\hat{J}_{x})^{2}$
approaches zero. Sorenson and Molmer considered a spin $J$ system,
and determined the minimum value of $(\Delta\hat{J}_{x})^{2}$ that
is possible, for a given measured value of $\langle\hat{J}_{z}\rangle$.
The result is a function $F_{J}(\langle\hat{J}_{z}\rangle/J)$. For
each $J$, one can show \citep{sorensen2001entanglement}
\begin{equation}
(\Delta\hat{J}_{x})^{2}/J\geq F_{J}(\langle\hat{J}_{z}\rangle/J)\,.\label{eq:depth-sq}
\end{equation}
The curves $F_{J}$ are convex and monotonically increasing with $\langle\hat{J}_{z}\rangle/J$,
and, for a given $\langle\hat{J}_{z}\rangle/J$, monotonically decreasing
with $J$. This allowed the authors to derive the inequality that
holds for $N$ separable systems of spin $J$: $(\Delta\hat{J}_{x})^{2}/NJ\geq F_{J}(\langle\hat{J}_{z}\rangle/NJ)$.
The inequality provided a calibration: for a given measured variance
$(\Delta\hat{J}_{x})^{2}$, it is possible to determine $J_{0}$ such
that $(\Delta\hat{J}_{x})^{2}/NJ_{0}<F_{J_{0}}(\langle\hat{J}_{z}\rangle/NJ_{0})$.
The conclusion is that the factorization (\ref{eq:sepN}) breaks down,
and that there exists a subsystem $\rho^{(i)}$ with a total spin
greater than $J_{0}$. This implies a block of a least $n_{0}=2J_{0}$
mutually entangled atoms. The experiments of Gross et al \citep{gross2010nonlinear}
measured the spin squeezing in a multi-well Bose Einstein condensate
to infer the multi-particle entanglement involving at least $\sim100$
atoms. Here, two hyperfine states of Rb act as the two modes of a
nonlinear interferometer. A similar multi-particle entanglement was
inferred by Riedel et al \citep{riedel2010atom}, using atom-chip
based interferometry.

Tura et al extended the approach that uses collective operators to
infer the existence of Bell correlations within the ensemble of atoms
\citep{tura2014detecting,tura2015nonlocality}. This was measured
by Schmied et al \citep{Schmied2016Bell} for a Bose Einstein condensate,
and by Engelsen et al \citep{engelsen2017bell} for ultra-cold but
not Bose-condensed atoms, at higher temperatures. There was however
no assessment of the collective number of atoms mutually sharing the
Bell nonlocality. These measurements also involved the collective
atomic spin-squeezing parameter.

In fact, a debate had arisen around how to interpret entanglement
criteria when applied to the atoms of a Bose Einstein condensate (BEC)
\citep{zanardi2002quantum,wiseman2003entanglement,barnum2004subsystem,killoran2014extracting}.
In a BEC, the atoms are identical bosonic particles which are indistinguishable
and which obey the symmetrization principle. Super-selection rules
apply for massive particles that exclude the possibility of superpositions
of states with different atom number in a single mode. A resolution
was put forward by Killoran, Crammer and Plenio \citep{killoran2014extracting},
who gave a connection between the so-called particle entanglement
and mode entanglement approaches. Using criteria based on super-selection
rules, Cramer et al \citep{cramer2013spatial} were able to quantify
the large-scale entanglement of ultra-cold bosons in $10^{5}$ sites
of an optical lattice. Dalton et al \citep{dalton2014new,dalton2017quantumI,dalton2017quantumII,dalton2020tests}
adopted second quantization to derive conditions based on super-selection
rules for both entanglement and steering between the modes associated
with the two atomic levels. This allowed the conclusion that the experimental
observation of interference in a two-mode BEC interferometer is sufficient
to imply entanglement \citep{dalton2014new} and steering \citep{opanchuk2019mesoscopic,dalton2020tests,rosales2018einstein}
between the modes, since the modes are distinguishable. Moreover,
the number of atoms collectively involved in the mode-entanglement
could be quantified, using the measurable fringe visibility or higher
moments \citep{rosales2018einstein,dalton2020tests}. A particular
atom interferometer experiment was examined \citep{Egorov_2011,egorov2013measurement},
to infer 40,000 atoms genuinely involved in the two-mode steering
\citep{opanchuk2019mesoscopic}.

In 2015, Islam et al used quantum interference to directly measure
the amount of entanglement in a lattice of ultra-cold bosonic atoms
\citep{islam2015measuring}. The measure was based on entanglement
entropy. Experiments have now demonstrated macroscopic superpositions
at the time-scale of everyday life \citep{kovachy2015quantum}, an
entanglement of $3000$ atoms\textcolor{red}{{} }with a non-positive
Wigner function \citep{mcconnell2015entanglement}, and 16 million
genuinely entangled atoms entangled through their electronic states
in a solid environment \citep{frowis2017experimental}. The work of
Frowis et al establishes the extraordinary level of genuine entanglement
based on an entanglement depth witness \citep{frowis2017experimental}.
In another approach, the quantum propagation of an attractive Bose
gas soliton was analyzed. This showed theoretically that such a system
would evolve dynamically to have nonlocal pair correlations, due to
the creation of a superposition of different types of fragments, caused
by quantum instabilities not present in the usual classical analysis
\citep{ng2019nonlocal}.

\subsection{EPR entanglement, steering and multipartite entanglement between
atomic groups}

In 2001, Julsgaard et al experimentally demonstrated entanglement
between two macroscopic spatially-separated ensembles of $\sim10^{12}$
Cesium atoms at room temperature \citep{Julsgaard:2001aa,krauter2011entanglement,muschik2011dissipatively,Muschik_2012}.
Their method identifies the spin $1/2$ system as the two-level atom
associated with an internal hyperfine atomic transition.  In \citep{Julsgaard:2001aa},
macroscopic spin operators are then defined for each ensemble, and
the correlations confirmed using variance criteria applied to collective
macroscopic spin observables, $\hat{J}_{X}$ and $\hat{J}_{Y}$. The
entanglement is created by first transmitting an off-resonant polarized
laser pulse through two atomic ensembles with opposite mean macroscopic
spins, in order to correlate the spins $\hat{J}_{X}$ of each ensemble
($i=1$, $2$). The process is then repeated with a second pulse to
correlate the atomic spins $\hat{J}_{Y}$ \citep{duan2000quantum,kuzmich2000atomic}.
The final simultaneous correlation of both $\hat{J}_{X}$ and $\hat{J}_{Y}$
for the ensembles gives the correlations necessary for an EPR entanglement
and (if strong enough) for an EPR paradox. The correlation is inferred
by the measurements of $\hat{J}_{X1}+\hat{J}_{X2}$ and $\hat{J}_{Y1}+\hat{J}_{Y2}$,
where $\hat{J}_{Xi}$ and $\hat{J}_{Yi}$ refer to the spins of the
ensemble labelled $i$. These measurements are made on the outputs
of the polarized pulses, which according to the theory have values
for the Stokes observables that are correlated with the spin sums.
The experiment reported entanglement between the ensembles using a
variance measure similar to type (\ref{eq:ent-cond}), but the correlation
did not satisfy (\ref{eq:epr-crit}) as necessary for an EPR paradox
(refer \citep{he2013towards}). 

To explore quantum correlations in the strictest sense, it is necessary
to obtain evidence of correlations where the values of the observables
at each spatially separated site are obtained by a local measurement,
as in the Bell tests. This motivated theoretical investigations which
analyzed how to achieve EPR-type quantum correlations at a mesoscopic
level, between groups of atoms \citep{fadel2020number,opanchuk2012dynamical,bar2011einstein,he2011planar,he2012einstein}.
In a step towards this goal, the experimental observation of the
entanglement between two distinct groups of atoms in a BEC was reported
by Gross et al \citep{Gross:2011aa}. Here, the correlations were
detected using the equivalent of an optical homodyne technique for
each system, as described for continuous variable measurements in
Section V. The atomic homodyne involved a second group of atoms that
form the local oscillator \citep{ferris2008detection}. In the atom-optics
equivalent to the photonic scheme, the beam splitter interaction that
combines the local oscillator with the signal field is carried out
with a Rabi rotation. Using atomic homodyne detection, the two-mode
squeezing criterion of type (\ref{eq:ent-cond}) allowed an inference
of entanglement between the two systems of atoms. While the two groups
were distinguishable, there was limited spatial separation. The stronger
correlations required for an EPR paradox and for steering were generated
experimentally by Peise et al \citep{peise2015satisfying}. The correlations
were verified using the atomic homodyne method and the EPR criterion
(\ref{eq:epr-crit}), although spatial separation of the atomic groups
was limited.

A significant advance came in 2018 from three experiments which confirmed
quantum correlations between the spatially separated atomic clouds
of a split Bose Einstein condensate (BEC). Entanglement, an EPR paradox
and EPR steering were detected between spatially separated groups
of several hundreds of atoms \citep{Matteo:2018aa,kunkel2018spatially,lange2018entanglement}.
Kunkel et al demonstrated entanglement and bipartite EPR steering
between the clouds of hundreds of Rb atoms in an expanding BEC \citep{kunkel2018spatially}.
The steering was measured using spatially resolved spin read-outs,
and properties such as monogamy of steering  were also investigated.
Kunkel et al certified the genuine multipartite entanglement of five
spatially separated mesoscopic groups of atoms, using witnesses constructed
by modifying techniques applied previously to continuous variable
systems \citep{teh2014criteria,van2003detecting}. Fadel et al \citep{Matteo:2018aa}
used high-resolution imaging to infer EPR steering based on the spin
correlations between spatially separated parts ($\sim$100 of atoms)
of a spin-squeezed Bose-Einstein condensate generated on an atom chip.
Variance criteria were also used to infer the correlations. Lange
et al \citep{lange2018entanglement} similarly demonstrated entanglement
between two spatially separated mesoscopic clouds of hundreds of Rb
atoms, obtained by splitting an ensemble of ultra-cold identical particles
prepared in a twin Fock state. The method of generation of entanglement
is analogous to that described in Section V, where a squeezed field
combines with a vacuum state on a beam splitter to create EPR entangled
outputs. These experiments give evidence of entanglement distributed
over several hundred atoms. It remains to rigorously quantify the
number of atoms genuinely entangled from each group, but a step in
this direction was provided in \citep{Reid2019_PRL}. Arguments validating
a large depth of entanglement can also be made based on the indistinguishability
of the atoms of the BEC \citep{Matteo:2018aa,lange2018entanglement,kunkel2018spatially}.
As of yet, it remains to demonstrate Bell nonlocal correlations between
spatially separated groups of atoms.

\section{Quantum correlations in optomechanics}

\textcolor{red}{}The question of the existence of quantum correlations
in systems that are macroscopic by mass is addressable in the field
of optomechanics. Quantum correlations in optomechanics are expected
to play a significant role in fundamental tests of quantum mechanics.
The idea that separated quantum systems may decohere was proposed
by Furry \citep{Furry_PR1936}, as a possible resolution of the Einstein-Podolsky-Rosen
paradox \citep{Einstein:1935}. Spontaneous decoherence is not observed
for low-mass systems like photons \citep{Tittel_1998_PRL}, electrons
\citep{Hensen_2015_Nature} or atoms \citep{Shin2019Bell}, where
entanglement has been verified for separated masses. The idea that
gravitational effects \citep{bassi2017gravitational} may be involved
in causing quantum decoherence \citep{diosi1984gravitation,penrose1996gravity,pikovski2015universal}
or changes in commutation relations \citep{Pikovski:2012aa}, has
led to substantial interest in quantum superpositions of more massive
objects than atoms \citep{marshall2003towards}. Experimental success
in cooling massive optomechanical systems to their quantum ground
state \citep{Teufel_Nature2011} has resulted in the generation of
entanglement in optomechanics. The simplest theoretical schemes to
generate quantum correlations in optomechanics produce entanglement
between the optical and mechanical subsystems in a single optomechanical
system. Entanglement between two separated optomechanical systems
requires a more complex approach, involving at least two mechanical
oscillator subsystems. Since one of the motivating factors in this
work is to test for the combined effects of quantum mechanics and
gravity, experiments that combine both entangled massive oscillators
and a controllable degree of spatial separation appear to be of greatest
interest. Recently there has been an interest in entanglement resulting
directly from gravitational interactions, as a possible direct test
of quantum gravity, explained below.

Typical experimental masses in cryogenic optomechanical experiments
are $m\sim50pg$, using aluminum cantilevers or capacitor plates for
the mechanical sub-system, together with a superconducting microwave
LC circuit for the 'optical' component. The number of atoms is therefore
of order $n_{a}\sim10^{12}$, which means that these devices are both
macroscopic and massive. Even larger optomechanical systems exist
in the form of the LIGO gravitational wave detectors, with mirror
masses of over $10kg$, so that $n_{a}\sim10^{26}$. However, these
are generally at room temperatures, making it difficult to observe
quantum effects.

\subsection{Entanglement between modes in an optomechanical system}

We first review the generation of entanglement within an optomechanical
system. These schemes typically rely upon the interaction Hamiltonian
between the optical $a$ and mechanical $b$ modes due to radiation
pressure. The fundamental Hamiltonian has the form\textcolor{red}{{}
}\citep{Pace1993,Law_PRA1994,Mancini_PRL2002}
\begin{equation}
H/\hbar=\omega_{0}\hat{a}^{\dagger}\hat{a}+\omega_{m}\hat{b}^{\dagger}\hat{b}+\hbar\chi\hat{a}^{\dagger}\hat{a}\left(\hat{b}+\hat{b}^{\dagger}\right),
\end{equation}
where $\omega_{0},\omega_{m}$ are the optical and mechanical oscillator
resonance frequencies respectively, while $\chi$ is the nonlinear
coupling strength between modes $a$ and $b$. Here, $\hat{a}$ and
$\hat{b}$ are the boson destruction operators for the optical and
mechanical modes respectively. The Hamiltonian generates a unitary
transformation operator that contains a nonlinear Kerr term $\left(\hat{a}^{\dagger}\hat{a}\right)^{2}$
\citep{Mancini_PRA1997,Bose_PRA1997}. Mancini et al \citep{Mancini_PRA1997}
and Bose et al \citep{Bose_PRA1997} obtained different entangled
states between the optical and mechanical modes, depending on specific
times in the system evolution. Similar couplings have been engineered
between two optical modes in a superconducting device \citep{Rodriguez2021Cooling},
but these are less interesting from the viewpoint of gravitational
effects. Typical experimental values in cryogenic nano-mechanical
systems are $\omega_{0}/2\pi\sim10GHz$, $\omega_{m}/2\pi\sim10MHz$
and $\chi/2\pi\sim100Hz$ \citep{Palomaki_Science2013}. These are
microwave electromagnetic frequencies, and the systems are cooled
to temperatures of around $T\sim10mK$ to reduce thermal excitations
of the oscillator. There is also some optical and mechanical damping
due to reservoirs, causing optical and mechanical decays at typical
rates of $\gamma_{0}/2\pi\sim500kHz,\gamma_{m}/2\pi\sim50Hz$, respectively.
In the weak coupling regime, where $\chi$ is small compared to the
optical damping rates, the coupling between the optical and mechanical
modes is enhanced by an external driving field. Depending on the driving
field frequency $\omega_{d}$, the optomechanical system behaves differently.

For a driving field that has a frequency such that $\omega_{d}=\omega_{m}+\omega_{o}$,
where $\omega_{m}$ and $\omega_{o}$ are the mechanical and optical
mode frequencies respectively, the driving field is said to be blue
detuned, defined by the detuning parameter $\Delta\equiv\omega_{o}-\omega_{d}=-\omega_{m}$.
In this case the blue-detuned optomechanical system has an effective
interaction Hamiltonian 
\begin{equation}
H_{int}^{b}=\hbar g\left(\hat{a}\hat{b}+\hat{a}^{\dagger}\hat{b}^{\dagger}\right)\label{eq:blue-det}
\end{equation}
where $g=\chi\sqrt{N}$ is the effective coupling strength due to
the driving field, for an internal stored microwave photon number
of $N$. This Hamiltonian is known to generate entanglement between
$a$ and $b$ modes \citep{Reid:1989}, and here is the physical source
of the entanglement between the optical and mechanical modes. The
dynamics from this effective optomechanical Hamiltonian in the presence
of noise and losses was studied by Vitali et al \citep{Vitali_PRL2007,Vitali_PRA2007},
Genes et al \citep{Genes_PRA2008}, and Hofer et al \citep{hofer2011quantum}
using linearized Langevin equations. Hofer et al and Vanner et al
analyzed pulsed schemes for the generation of non-classical states
\citep{vanner2011pulsed,hofer2011quantum}.

The scheme of Hofer et al is based on the description given above,
where a blue-detuned pulse first entangles the optical and mechanical
modes \citep{hofer2011quantum}. The entanglement verification process
requires the readout of the mechanical mode, which is is achieved
by applying a red-detuned pulse that transfers the mechanical state
to an optical state where measurements are made using optical homodyne.
A red-detuned driving field ($\Delta=\omega_{o}-\omega_{d}=\omega_{m}$)
leads to an effective interaction Hamiltonian 
\begin{equation}
H_{int}^{r}=\hbar g\left(\hat{a}\hat{b}^{\dagger}+\hat{a}^{\dagger}\hat{b}\right)\label{eq:red-det}
\end{equation}
which was shown to enable the transfer of a quantum state between
the optical and mechanical modes by Zhang, Peng and Braunstein \citep{Zhang_PRA2003}.
In the theory of Hofer et al, dissipation and noise are included and
the optomechanical system evolves according to the Langevin equations
(in the rotating wave approximation)
\begin{align}
\dot{\hat{a}}_{c} & =-\kappa\hat{a}_{c}-ig\hat{a}_{m}^{\dagger}-\sqrt{2\kappa}\hat{a}_{in}\nonumber \\
\dot{\hat{a}}_{m} & =-ig\hat{a}_{c}^{\dagger}\,\label{eq:lang}
\end{align}
where $\hat{a}_{c},\hat{a}_{m}$ are the boson operators for cavity
optical and mechanical modes respectively, $g$ is the effective optomechanical
coupling strength, $\kappa$ is the cavity decay rate, and $\hat{a}_{in}$
contains the quantum noise entering the cavity. In the limit of large
cavity decay rate $\kappa\gg g$, the adiabatic approximation allows
the cavity optical and mechanical modes to have the following expressions:
\begin{align}
\hat{a}_{c}\left(t\right) & \approx-i\frac{g}{\kappa}\hat{a}_{m}^{\dagger}\left(t\right)-\sqrt{\frac{2}{\kappa}}\hat{a}_{in}\left(t\right)\nonumber \\
\hat{a}_{m}\left(t\right) & \approx e^{Gt}\hat{a}_{m}\left(0\right)+i\sqrt{2G}e^{Gt}\intop_{0}^{t}e^{-Gs}\hat{a}_{in}^{\dagger}\left(s\right)ds\,\label{eq:lang-ae}
\end{align}
where $G=g^{2}/\kappa$. In particular, the cavity optical mode satisfies
the input-output relation, given by $\hat{a}_{out}=\sqrt{2\kappa}\hat{a}_{c}+\hat{a}_{in}$,
where $\hat{a}_{out}$ is the output of the cavity \citep{Gardiner_PRA1985}.
This relation is useful as it is the output field from the cavity
that is usually being measured. Hofer et al also defined normalized
temporal light modes: $A_{in}=\sqrt{\frac{2G}{1-e^{-2G\tau}}}\intop_{0}^{\tau}e^{-Gt}\hat{a}_{in}\left(t\right)dt$
and $A_{out}=\sqrt{\frac{2G}{e^{2G\tau}-1}}\intop_{0}^{\tau}e^{Gt}\hat{a}_{out}\left(t\right)dt$
and the mechanical modes $B_{in}=\hat{a}_{m}\left(0\right)$, $B_{out}=\hat{a}_{m}\left(\tau\right)$,
showing that it is the quadrature amplitudes of these modes that become
entangled. They showed that for $G\tau\rightarrow\infty$, the mechanical
state is perfectly transferred to the optical mode, apart from a phase
shift.\textcolor{teal}{}

Entanglement between the optical and mechanical modes in an optomechanical
system was demonstrated experimentally by Palomaki et al \citep{Palomaki_Science2013},
following the scheme of Hofer et al. They realized the optomechanical
system using an electromechanical circuit where an LC oscillator corresponds
to the optical mode and one of the capacitor plates is moveable, behaving
like a mechanical mode. The entanglement between modes is then generated
using the interaction Hamiltonian (\ref{eq:blue-det}), with a blue-detuned
microwave field. The experiment measured the quadrature amplitudes
of the entangled optical $\left(\hat{X}_{1},\hat{P}_{1}\right)$ and
mechanical modes $\left(\hat{X}_{2},\hat{P}_{2}\right)$, where $\hat{X}_{i}=\hat{a}_{i}+\hat{a}_{i}^{\dagger}$
and $\hat{P}_{i}=(\hat{a}_{i}-\hat{a}_{i}^{\dagger})/i$. The measured
statistical moments of these quadrature amplitudes allow the inseparability
parameter $\Delta_{sum}$ to be determined, as defined in (\ref{eq:Simon_Duan-1})
with $g=g'=1$. The stronger criterion (\ref{eq:ent-cond}) allowing
$g$, $g'\neq1$, although not measured in the Palomaki experiment,
was predicted from simulations to a give a more sensitive measure
\citep{KiesewetterPhysRevA_2014}.

Since linearization can fail when there are strong laser driving fields,
quantum phase-space simulations without any linearization approximations
were carried out by Kiesewetter et al \citep{KiesewetterPhysRevA_2014}.
The theory uses the exact positive-P representation \citep{Drummond_Gardiner_PositivePRep}
to transform the nonlinear quantum master equation into stochastic
equations, which can be numerically simulated. This allows the full
multi-mode output fields to be calculated, giving excellent quantitative
agreement with the pulsed optomechanical entanglement experiment of
Palomaki et al and justifying the linearization regime. Nonlinear
effects were studied in Teh et al \citep{teh2017simulation}, and
can be significant due to the strong pump fields that are often used
in experiments.

\subsection{Entanglement between optomechanical systems}

Different methods can be used to entangle two massive systems, but
the common feature is to entangle by interacting the masses with an
optical field. For example, in the scheme of Hofer et al, the outgoing
red pulse can be propagated through a second oscillator, and the state
of the pulse transferred onto the second oscillator, thus entangling
both oscillators.

In 2018, two experiments reported entanglement between oscillators
\citep{Riedinger_Nature2018,Ockeloen-Korppi_Nature2018}. Discrete
variable entanglement between two mechanical oscillators separated
by $\sim20$ cm was demonstrated in the experiment of Riedinger et
al \citep{Riedinger_Nature2018}. Here, a pump field is sent into
one of the two optomechanical systems, which creates a phonon-photon
pair in one of the systems. The photon leaks out of the optomechanical
system and is subsequently sent into a beam splitter and detected.
As the whole process does not provide information on which optomechanical
system the phonon-photon pair is created, the detection of the photon
at the beam splitter output heralds the mechanical mode into a superposition
state of one phonon in mechanical oscillator $A$ or $B$, given by
$|\Psi\rangle=\left(|1\rangle_{A}|0\rangle_{B}+|0\rangle_{A}|1\rangle_{B}\right)/\sqrt{2}$.
Here $|1\rangle_{A}|0\rangle_{B}$ is a state with a single excitation
in the mechanical oscillator $A$, with the mechanical oscillator
$B$ is in its ground state; and $|0\rangle_{A}|1\rangle_{B}$ is
similarly defined. An entanglement witness involving second-order
coherences was used to certify the entanglement.

Ockeloen-Korppi et al demonstrated entanglement between two electromechanical
systems \citep{Ockeloen-Korppi_Nature2018}. The method was based
on the idea of reservoir engineering to prepare the two cavity-coupled
mechanical oscillators into a steady state that is entangled, as proposed
by Woolley and Clerk \citep{Woolley_PRA2014}, Tan, Li and Meystre
\citep{Tan_PRA2013}, and Wang and Clerk \citep{Wang_PRL2013}. Here,
two mechanical modes with frequencies $\omega_{m,1}$ and $\omega_{m,2}$
are coupled to a single microwave cavity mode with frequency $\omega_{c}$.
Two driving fields are applied such that the effective interaction
Hamiltonian has the form 
\begin{equation}
H_{eff}=g_{+}\left[\left(\hat{a}+\hat{b}\right)\hat{c}+H.c.\right]+g_{-}\left[\left(\hat{a}+\hat{b}\right)\hat{c}^{\dagger}+H.c.\right]\,,
\end{equation}
which consists of a sum of the quantum state transfer term and entanglement
generation term. H.c. refers to hermitian conjugate. By tuning the
amplitude of the driving fields, a two-mode squeezed state was generated
between the two mechanical modes. \textcolor{red}{}Barzanjeh et al
also carried out an experiment in a similar electromechanical setting
using two microwave fields \citep{Barzanjeh_Nature2019}. The entanglement
is in the continuous variable quadratures of the modes, and the inseparability
parameter (\ref{eq:Simon_Duan-1}) with $g=g'$ is used to verify
the entanglement.

Entanglement between two optomechanical systems has also been recently
reported in the electromechanical experiment by Kotler et al \citep{Kotler_Science2021}.
Based on the theory similar to previous works \citep{Mancini_PRL2002,Tan_PRA2013,Wang_PRL2013,Li:2015aa},
a blue-detuned pulse is first applied to entangle the cavity with
a drum, producing a quantum correlated photon-phonon pair. A red-detuned
pulse is subsequently applied to transfer the photon state to a phonon
state in a second drum, and hence realizing the generation of entanglement
between the phonon-phonon pair.

More recent work by Mercier de Lepinay et al \citep{Mercier-de-Lepinay_Science2021}
generated a two-mode squeezed state using four driving fields. This
method has the feature that it involves a 'quantum-mechanics free
subsystem', a manifold where commutation relations are nearly zero.
This is closely related to planar spin squeezing \citep{he2011planar,he2012entanglement},
which has been achieved in macroscopic atomic spin systems \citep{puentes2013planar}.

The entanglement of an optomechanical system with an atomic spin ensemble
was realized by Thomas et al \citep{Thomas_Nature2021}, based on
the proposal by Hammerer et al \citep{Hammerer_PRL2009}. The light-spin
interaction is first established by sending a polarized light beam
through a sample of $10^{9}$ atoms with a collective macroscopic
spin along a certain direction, similar to the experiment of Julsgaard
et al \citep{Julsgaard:2001aa}. The propagated light is then sent
into a cavity that contains a mechanical dielectric membrane, where
the mechanical mode interacts with the light so that the mechanical
mode becomes entangled with the atomic ensemble.

Alternatively, one can use a direct state transfer from an optical
to mechanical system, using the red-detuned Hamiltonian (\ref{eq:red-det})
\citep{Zhang_PRA2003}. The red-detuned Hamiltonian (for vacuum inputs)
does not generate entanglement, and an external source of optical
entanglement is therefore required. The two entangled light fields
generated by parametric oscillation can be sent to two spatially separated
optomechanical systems, where the entangled optical modes are transferred
to the mechanical modes via the red-detuned effective Hamiltonian.
This has the advantage in principle that because the optical entanglement
has been generated for large spatial distances, one may be able to
obtain an arbitrary separation, for tests of massive entanglement
at different distances, as in proposals to test Furry's hypothesis
\citep{Kiesewetter_PRL2017}. This type of scheme was first studied
using linearization and a steady state approach by Zhang et al \citep{Zhang_PRA2003}.
However, to study the possible dynamics of gravitational decoherence,
a pulsed entanglement approach is important \citep{vanner2011pulsed},
and a full dynamical study without a linearization approximation was
numerically carried out by Kiesewetter et al \citep{Kiesewetter_PRL2017}.
A similar treatment gave a proposal to transfer a cat state from an
optical to mechanical mode \citep{teh2018creation}, which could in
principle be extended to generate entangled cat states in optomechanics.
Vanner has proposed another mechanism to generate cat states, using
a conditional pulsed measurement scheme \citep{vanner2011selective}.
A further analysis has been given by Hoff et al \citep{Hoff_PRL2016}.

\subsection{EPR steering}

A proposal to realize the correlations of an EPR paradox through radiation
pressure in optomechanics was first put forward by Giovannetti et
al \citep{giovannetti2001radiation}. For a pulsed system, the correlations
of the EPR paradox and of EPR steering were investigated by He and
Reid \citep{He_PRA2013}. These authors first considered the generation
of entanglement between the optical and mechanical modes of a single
optomechanical system as considered by Hofer et al. The entangled
state is characterized by a squeezing parameter $r$ that is proportional
to the coupling strength between the optical and mechanical modes.
In order to quantify the steering of the entangled state, they use
the steering criterion (\ref{eq:epr-crit}), which becomes 
\begin{align}
E_{m|c} & =\Delta\left(\hat{X}_{m}-g_{x}\hat{P}_{c}\right)\Delta\left(\hat{P}_{m}+g_{p}\hat{X}_{c}\right)<1\,\label{eq:steer}
\end{align}
as given by (\ref{eq:steer-gh}). Here, $\hat{X}_{m},\hat{P}_{m}$
are the quadratures of the mechanical mode, and $\hat{X}_{c},\hat{P}_{c}$
are the quadratures of the optical mode. The $g_{x},g_{p}$ are real
numbers that can be chosen to minimize $E_{m|c}$. Steering of the
mechanical mode by optical mode is confirmed when $E_{m|c}<1$. The
presence of thermal noise degrades the quantum correlation. Using
this steering criterion, which is necessary and sufficient for a two-mode
Gaussian system, the authors evaluated the minimal squeezing strength
required to show steering for a given thermal occupation number $n_{0}$.
They found that the required squeezing strength for steering does
not grow indefinitely with $n_{0}$ but asymptotically approaches
$r=0.5\text{ln}2$ as $n_{0}\rightarrow\infty$. On the other hand,
using the steering criterion $E_{c|m}<1$ for the steering of optical
mode by the mechanical mode, no such minimum squeezing strength is
required to demonstrate steering, and there is always steering of
the optical mode by the mechanical mode as long as $r\neq0$. The
authors argued that the steering of the mechanical mode was of interest,
because then the ``elements of reality'' considered by Einstein-Podolsky-Rosen
related to the massive object, rather than to the field. The model
used by the authors is essentially that of a two-mode squeezed state
with asymmetric reservoirs for the two modes, and illustrated the
sudden death of EPR steering that occurs with a certain threshold
amount of thermal noise on the steered system \citep{rosales2015decoherence}.
A multimode model appropriate for a pulsed treatment was put forward
by Kiesewetter et al and supported these predictions \citep{KiesewetterPhysRevA_2014}.

Next, the authors of propose to entangle two oscillators by first
entangling the mechanical mode of mechanical oscillator $M1$ and
cavity optical mode as before. The cavity optical mode is then transferred
to the mechanical mode of a second mechanical oscillator $M2$, and
hence entangling $M1$ and $M2$. They study the steering in the entanglement
of two optomechanical systems as a function of thermal noise using
the steering criterion of the form Eq. (\ref{eq:steer}) and provide
the squeezing strength threshold required to observe steering. The
same steering criterion is used in the work of Kiesewetter et al \citep{Kiesewetter_PRL2017}
where steering is studied for different mechanical modes storage times,
as well as for thermal noise.

Other proposals using the steering criterion are given by Sun et al
\citep{Sun_NJP2017} in a different setting. In that work, a dielectric
membrane is placed in a cavity that divides the cavity into two independent
cavity modes. Two pump fields enter these cavity modes and create
entanglement among the two cavity and mechanical modes.  By varying
the phase difference between the two pump fields, the degree of bipartite
entanglement between the mechanical mode and one of the optical mode
varies. This is then extended to the case of steering where the condition
on the phase difference required to show steering is established. 

For Gaussian states, the criterion (\ref{eq:epr-crit}), and hence
(\ref{eq:steer}) with the optimally selected values of $g$ and $g'$,
has been shown to be necessary and sufficient for steering in two-mode
Gaussian systems \citep{Jones_PRA2007}. This means it is possible
to construct a measure of such steering. A Gaussian steering quantifier
$\mathcal{G}^{A\rightarrow B}$ that quantifies the steerability of
mode $B$ by mode $A$ was put forward by Kogias et al \citep{Kogias_PRL2015},
and has the form 
\begin{align}
\mathcal{G}^{A\rightarrow B}\left(\sigma_{AB}\right) & =\text{max}\left\{ 0,\frac{1}{2}\text{ln}\frac{\text{det}A}{\text{det}\sigma_{AB}}\right\} \,\label{eq:Gaussian_steer}
\end{align}
where $\sigma_{AB}$ is the covariance matrix of the bipartite system
$AB$, and $\text{det}A$ is the determinant of the covariance matrix
of the subsystem $A$. This steering quantifier is shown to be related
to the steering parameter $E_{B|A}$ in Eq. (\ref{eq:steer}) via
the expression $(E_{B|A})_{opt}=e^{-2\mathcal{G}^{A\rightarrow B}}$,
where here $(E_{B|A})_{opt}$ is the value of $E_{B|A}$ for optimally
chosen $g$ and $g'$ as given by (\ref{eq:steer-gh}). However, the
Gaussian quantifier only applies under the assumption of Gaussian
states, whereas the condition (\ref{eq:epr-crit}) (and hence (\ref{eq:steer}))
holds for all states, as a witness to EPR steering and as a one-sided
device-independent witness to entanglement \citep{Opanchuk_PRA2014}.
While the steering parameter $E_{B|A}$ has the advantage of clear
operational interpretations, the steering quantifier $\mathcal{G}^{A\rightarrow B}$
allows some mathematical properties such as convexity, additivity
and monotonicity under quantum operations to be readily proven \citep{Kogias_PRL2015}.
This quantifier is studied in the work of Tan and Zhan \citep{Tan_PRA2019},
and El Qars et al \citep{El-Qars_EPJ2017}. Zhong et al use the EPR
parameter $E_{B|A}$ and the Gaussian steering quantifier to study
one-way EPR steering between two macroscopic magnons located in optically
driven cavities \citep{zhong2021one}.

\subsection{Bell nonlocality}

Schemes that generate Bell nonlocality in the optomechanical systems
may follow closely the schemes that generate entanglement. The difference
lies in the quantum correlation verification process. The variance
entanglement criteria used for entanglement verification cannot be
applied to demonstrate Bell nonlocality. Rather, a CHSH inequality
$\left|B\right|=\left|E\left(a,b\right)-E\left(a,b^{'}\right)+E\left(a^{'},b\right)+E\left(a^{'},b^{'}\right)\right|\leq2$
is to be checked, where $\left|B\right|>2$ is required to demonstrate
Bell nonlocality. Here, $E$ is a correlation function and it is a
function of different settings $a,a^{'},b,b^{'}$.

In the theoretical work by Vivoli et al \citep{Vivoli_PRL2016}, Hofer,
Lehnert and Hammerer \citep{Hofer_PRL2016}, and Manninen et al \citep{Manninen_PRA2018},
entanglement is generated between the mechanical and optical modes
using the blue-detuned driving field as described in the previous
section. The difference in those works lies in the verification of
Bell nonlocality of the entangled state generated. Vivoli et al and
Hofer et al propose to coherently displace the optical fields before
measuring the photons, similar to earlier continuous variable Bell
approaches. The amplitudes of coherent displacement $\alpha,\alpha^{'},\beta,\beta^{'}$
constitute the different settings in the correlation function $E\left(\alpha,\beta\right)$,
which are then used to check against the CHSH inequality.

Yet another possible choice of measurement settings is considered
in the work of Manninen et al \citep{Manninen_PRA2018}. In that work,
quadrature phases of the modes are measured using homodyne detection
and the different settings are the phases of the local oscillator
in the homodyne detection scheme. This measurement scheme has been
experimentally shown to violate a Bell inequality in an optical system
\citep{Thearle_PRL2018}. Finally, although not a Bell test per se,
we mention that other quantum paradoxes, including a delayed choice
wave-particle duality experiment and a Leggett-Garg test of macro-realism,
have been theoretically proposed for mechanical resonators \citep{qin2019proposal,asadian2014probing}.

\subsection{Experimental optomechanical Bell test}

The first experimental Bell test involving an optomechanical system
has been carried out by Marinkovic et al \citep{Marinkovic_PRL2018}.
Here, the origin of the entanglement involves two-particle interference
between four photonic modes as in the earlier photonic interferometric
proposals \citep{Horne_PRL1989,Reid1986}. This experiment measures
coincidences in two detectors using photon counting and is thus a
discrete variable Bell test, unlike the theoretical schemes in the
previous section, where continuous variables are measured. The physical
system of Marinkovic et al involves two nano-mechanical resonators
with $10^{10}$ atoms, whose entanglement is mediated by photons.
As discussed previously, a blue-detuned pulse is used to generate
entanglement between the optical and mechanical modes. 

However, different from other schemes, this pulse is not sent directly
into the optomechanical system. Rather, the pulse is first sent into
an interferometer with a beam splitter where the output from it is
then sent into either one of the identical optomechanical systems
that is located in each arm of the interferometer. An electro-optical
modulator is present in one of the interferometer arms to induce a
phase difference $\phi_{b}$ between the two arms. The optomechanical
system that receives the pulse will have entanglement between its
optical and mechanical modes where a cavity photon and phonon are
created. Up to this point, an entangled state between the optical
and mechanical modes is generated in one of the optomechanical systems
in this interferometer.

In the optomechanical system that contains the entangled state, the
cavity photon leaks out of the cavity and goes through a beam splitter
before being detected by photon detectors. The photon detection implies
the existence of a single phonon, while the beam splitter before the
photon detection erases the information on where the detected photon
is originated from. This puts the state of the whole system into a
superposition of single phonon state in one optomechanical system
or the other. In other words, the photon detection heralds an entangled
state between the mechanical modes of two optomechanical systems. 

A red-detuned pulse is also sent into the interferometer some time
after the blue-detuned pulse. Similarly, the phase shift $\phi_{r}$
in one of the interferometer arms is controlled by an electro-optical
modulator. This red-detuned pulse transfers the mechanical state into
the cavity optical state that leaks out of the optomechanical system,
which is then measured just as in the case for blue-detuned pulse.

The observables measured are the number of coincidences $n_{ij}$
in the two detectors when both the blue and red detuned pulses are
sent into the interferometer. Here, $i,j$ $\left(i,j=1,2\right)$
correspond to the detection when the blue and red detuned pulses are
sent, respectively. For instance, $n_{12}$ is the number of times
the blue drive triggers a photon detection at detector $1$ and a
subsequent detection at detector $2$ by the red drive. The correlation
function $E\left(\phi_{b},\phi_{r}\right)$ is related to these coincidences
by
\begin{align}
E\left(\phi_{b},\phi_{r}\right) & =\frac{n_{11}+n_{22}-n_{12}-n_{21}}{n_{11}+n_{22}+n_{12}+n_{21}}\,
\end{align}
where $\phi_{b}$ and $\phi_{r}$ are the phase difference acquired
in the arm of the interferometer when the blue and red detuned pulses
are sent, respectively. Marinkovic et al obtain $B=2.174_{-0.042}^{+0.041}$
and show a Bell violation by more than $4$ standard deviations. While
a significant step forward, this test requires a fair-sampling assumption,
and also does not use spatially separated detectors. Moreover, the
measurement settings are for the phases of the photon part of the
interferometer, rather than the mechanical oscillator amplitudes.
The settings are selected prior to the photon entering the mechanical
oscillator, which suggests that the hidden variables being tested
relate to the photon rather than oscillator fields.

\subsection{Gravitational quantum entanglement}

In the proposals discussed and experimentally demonstrated above,
the entanglement is generated through optomechanical interactions.
The entanglement of two oscillators is achieved by interacting with
an optical field, the quantum nature of which is essential to the
mechanism. These methods may potentially test for gravitationally
induced decoherence, owing to the presence of a massive test particle.

A more direct test of the quantum features of gravity would come from
the direct gravitational coupling of two massive test particles, as
proposed by Marletto and Vedral \citep{marletto2017gravitationally},
and Bose et al \citep{bose2017spin}. There have been a number of
related proposals, including \citep{AlBalushi2018,Krisnanda2020}.
The basic idea is to create a superposition state in an optical field,
and transfer this to a massive mirror, which can interact gravitationally
with a second massive mirror. The result is a quantum entanglement
of two mirrors obtained through gravitational interactions, and hence
implicitly involving a superposition of two distinct metric tensors,
thus providing evidence for quantum gravity.  This is a quantum version
of the Cavendish experiment \citep{cavendish1798}, which measured
the gravitational constant $G$, and hence the density of the earth.

Due to the extreme weakness of gravitational forces, the original
experiment was by no means trivial. Generating quantum entanglement
purely through gravitational interactions is considerably more difficult,
and has not been achieved as yet. However, with improvements in technology
since 1798, one may hope that a quantum Cavendish experiment is not
impossible in future. This might require test masses as large as the
kilogram-mass LIGO mirrors, or even a space-based experiment. There
are a number of technical challenges, including the elimination of
unwanted non-gravitational interactions, as well as the problem of
decoherence caused by external microgravity sources. Recently, the
first step was achieved, of measuring entanglement of large, LIGO-scale
masses with an optical field \citep{Yu2020}. 

\section{Conclusions}

This review summarizes the developments in our understanding of macroscopic
quantum correlations since the original papers of Einstein, Podolsky
and Rosen (EPR) and Bell. The original papers pertained to just two
systems of one particle each, and the correlations between the two
systems ruled out all local hidden variable theories. Contrary to
what might have been expected at the time of EPR and Bell, quantum
mechanics has since been shown to predict such correlations for higher
spin systems comprising many particles $-$ either with many particles
at just two locations, or with single particles at a large number
of locations, or with many particles at multiple sites. Furthermore,
it was found that the difference between the predictions of quantum
mechanics and local hidden variables theories can (in a certain context)
be more extreme as systems become larger. While the larger systems
become more sensitive to decoherence which will erase the difference,
this is in principle controllable. In fact, all pure entangled states
can exhibit Bell nonlocality. Counterintuitively, strong Bell violations
can be predicted in the presence of a macroscopic coarse-graining
of measurement outcomes.

Experiments have so far supported the predictions of quantum mechanics
in the mesoscopic regime. Higher dimensional Bell inequalities have
been violated in photonic systems, and Einstein-Podolsky-Rosen paradoxes
have been confirmed for high optical fluxes incident on detectors.
In a step towards demonstrating the nonlocality of a cat state, the
quantum coherence of superpositions of coherent states well separated
in phase space has been measured, for microwave fields in a cavity.
The genuine multipartite nonlocality associated with multipartite
photonic states has also been verified.

Moreover, there is evidence of quantum correlations in massive systems.
Experiments confirm the existence of multi-particle Bell correlations
inferred within a Bose-Einstein condensate, and the genuine entanglement
of tens of ions in a trap and of millions of atoms in a solid, certified
by rigorous theoretical methods. EPR-type entanglement has been detected
for spatially-separated propagating atoms, and for ensembles of atoms
at room temperature. EPR correlations in the form of a rigorous paradox
(and multipartite entanglement) has been demonstrated for the atomic
clouds of a split Bose-Einstein condensate, where each cloud contains
hundreds of atoms. These are the first steps towards a rigorous demonstration
of Bell correlations between massive systems, and between freely propagating
massive particles. As we approach more rigorous testing of the quantum
correlations in a macroscopic regime, there is a potential for fundamental
tests of quantum mechanics. This is especially true for tests involving
more massive objects. To date, there has been no Bell test involving
the position and momentum of well-separated massive particles or objects,
and no Bell test where the hidden variables are directly associated
with spatially separated macroscopic objects.

Leggett and Garg in 1985 explained the possibility of testing for
the incompatibility between macroscopic realism and quantum mechanics.
This is not necessarily achieved by confirming Bell correlations in
macroscopic systems. Leggett and Garg argued such tests might be carried
out for dynamically evolving macroscopic superposition states. Recently,
Leggett and Garg's definition of macroscopic realism (macro-realism)
has been tested for macroscopic superconducting qubits, with results
supporting the quantum predictions that counter macrorealism. While
loopholes remain, there is now an expanding number of theoretical
proposals, including for optomechanical Leggett-Garg tests and for
Bell-Leggett-Garg tests involving dynamically evolving cat states
at separated sites.
\begin{acknowledgments}
This work was funded through the Australian Research Council Discovery
Project scheme under Grants DP180102470 and DP190101480. The authors
also wish to thank NTT Research for their financial and technical
support.
\end{acknowledgments}

\bibliographystyle{elsarticle-num}
\bibliography{Quantum-correlations}

\end{document}